\pdfoutput=1

\documentclass[12pt]{article}
\usepackage{graphicx}
\usepackage{amsmath}
\usepackage{amssymb}
\usepackage{hyperref}
\usepackage[numbers]{natbib}

\usepackage{multirow}

\newcommand{\V}[1]{\mathbf{#1}}
\newcommand{\M}[1]{\mathbf{#1}}
\newcommand{\Vg}[1]{\boldsymbol{#1}}

\newcommand{\T}[1]{#1^{\text{T}} }

\newcommand{\dpart}[2]{\frac{\partial #1}{\partial #2}}

\newcommand{\Var}[1]{\text{Var} \left[ #1 \right] }
\newcommand{\Exp}[1]{\text{E} \left[ #1 \right]}

\usepackage{xcolor}

\usepackage{subfig}
\usepackage{ mathrsfs } % to get bold faced script font
\DeclareMathAlphabet{\mathscrbf}{OMS}{mdugm}{b}{n}

\begin{document}

\title{Robust topology optimisation of microstructural details without length scale separation - using a spectral coarse basis preconditioner}

\author{Joe Alexandersen\footnote{E-mail: joealex@mek.dtu.dk}, Boyan S. Lazarov \\ \small\textit{Department of Mechanical Engineering, Solid Mechanics,} \\ \small\textit{ Technical University of Denmark, Nils Koppels All\'e 404,} \\ \small\textit{DK-2800 Kgs. Lyngby, Denmark} }

\date{23rd of September 2014}

\maketitle

\begin{abstract}
This paper applies topology optimisation to the design of structures with periodic microstructural details without length scale separation, i.e. considering the complete macroscopic structure and its response, while resolving all microstructural details, as compared to the often used homogenisation approach. The approach takes boundary conditions into account and ensures connected and macroscopically optimised microstructures regardless of the difference in micro- and macroscopic length scales. This results in microstructures tailored for specific applications rather than specific properties.

Dealing with the complete macroscopic structure and its response is computationally challenging as very fine discretisations are needed in order to resolve all microstructural details. Therefore, this article shows the benefits of applying a contrast-independent spectral preconditioner based on the multiscale finite element method (MsFEM) to large structures with fully-resolved microstructural details.

The density-based topology optimisation approach combined with a Heaviside projection filter and a stochastic robust formulation is used on various problems, with both periodic and layered microstructures. The presented approach is shown to allow for the topology optimisation of very large problems in \textsc{Matlab}, specifically a problem with 26 million displacement degrees of freedom in 26 hours using a single computational thread.


\end{abstract}

\section{Introduction} \label{sec:intro}

The systematic design of novel materials with extremal properties is possible by applying topology optimisation to the design of material microstructures. The usual approach is to consider the homogenised properties of a single unit cell. This completely decouples the problem, as one does not consider the macroscopic response at all. One may then optimise a periodic microstructure to achieve a certain effective property, such as directional stiffness  \citep{Sigmund1994,Sigmund1995}, negative Poisson's ratio or thermal expansion coefficient \citep{Sigmund1996}, enhanced poroelastic pressure-coupling \citep{Andreasen2012}, among many others. 

Topology optimisation \cite{Bendsoee2003} is an iterative design process which distributes material in a design domain by minimising a selected objective functional, e.g. compliance, material weight or Poisson's ratio, while satisfying a set of constraints. The material distribution is represented by a design field which takes the value one if the point is occupied with material and zero if not. In order to utilise gradient-based optimisation techniques, the design field is allowed to take intermediate values. At each iteration step, the design field is updated using the gradients of the objective and constraint functionals. The discretisation determines the resolution of the optimisation process, as well as its computational cost. Therefore, in order to avoid excessive computations, the design is often limited to a single mechanical element with homogeneous material properties  or a single periodic cell. 

An alternative multiscale approach to the topological design of material unit cells can be taken by introducing the homogenised macroscopic response in the optimisation process. This is actually the origin of the homogenisation approach to topology optimisation presented in the seminal paper by Bends\o e and Kikuchi \citep{Bendsoee1988} and later by Bends\o e \citep{Bendsoee1989}. A hierarchical approach is taken in \citep{Rodrigues2002}, where the macroscopic density and microstructure is designed iteratively. This approach has been applied to bone modelling \citep{Coelho2009} and is well suited for parallel computations due to the decoupled homogenisation cells \citep{Coelho2011a}. Similar methodologies have recently been proposed for nonlinear elasticity \citep{Nakshatrala2013,Xia2014}. 

The lack of connectivity between the varying microstructural cells is a common problem of the above two-scale optimisation approaches. This may be neglectable when the microstructures are infinitely small, but when considering current manufacturability, a finite size must be attributed to the microstructures. Ensuring connected microstructures has been sparsely treated in the literature, but good examples include \citep{Schury2012} in the context of free material optimisation (FMO) and \citep{Radman2013} in the context of functionally graded materials (FGM).

The convergence of periodic structures towards homogenised microstructures is investigated in \citep{Xie2012,Zuo2013}. These papers show that an optimised periodic structure, with finite geometric periodicity, converges to optimised material unit cells with an infinite geometric periodicity, obtained using homogenisation for a single microstructure repeated throughout the computational domain.

This paper restricts itself to a single-scale optimisation approach, where the periodic microstructural details filling a given domain are optimised for the macroscale response. The complete macroscopic structure and its response is considered, while resolving all microstructural details as compared to the homogenisation approach common to all of the above papers, except \citep{Xie2012,Zuo2013}. The approach takes boundary conditions into account and ensures connected and macroscopically optimised microstructures regardless of the difference in micro- and macroscopic length scales. Furthermore, the restriction to microstructural details controlled by the size of the coarse mesh and a global volume constraint, implicitly introduces a maximum length scale similar to the results presented in \cite{Guest2009}. Such a feature can be utilised for providing manufacturability of the design, e.g. in the structural optimisation of high-rise buildings, where the unit cell models a single building unit (room, office or apartment), or for imposing properties which are not directly related or not included in the physical model due to computational cost, e.g., requiring specified porosity in a scaffold design. 

Dealing with the complete macroscopic structure and its response is computationally challenging as very fine discretisations are needed in order to resolve all microstructural details. This work therefore explores the application of a contrast-independent spectral preconditioner based on the multiscale finite element method (MsFEM), as presented in \citep{Lazarov2014}, to large structures with fully-resolved microstructural details.
The origins of MsFEM can be traced back to \cite{Babuska1994,Babuska1983} where special multiscale basis functions are utilised for solving elliptic problems. The method has been extended further to be applicable to a wide range of linear and non-linear multiscale problems, e.g. \cite{Hou1997, Efendiev2009}. The main idea is to construct basis functions  which provide a good approximation of the solution on a coarse grid \cite{Efendiev2009}. For obtaining good accuracy, the coarse basis functions need to have similar oscillatory behaviour as the fine scale solution, which can be achieved by oversampling techniques \cite{Efendiev2009}. The application of MsFEM to linear elasticity and material jumps isolated in the interior of the coarse elements is reported in \cite{Buck2013}. This condition cannot be guaranteed in the topology optimisation process, which limits the applicability of the original MsFEM in topology optimisation. For oscillatory high-contrast material properties, an alternative approach is proposed in \cite{Galvis2010,Efendiev2011}. The method constructs several basis functions per coarse node, which are capable of representing the important features of the solution and the accuracy of the approximation is controlled by the dimension of the coarse space. 

The layout of the paper is as follows: Section 2 covers the theoretical background of the methods used; Section 3 demonstrates the capabilities of the MsFEM approach through numerical experiments; Section 4 covers the implementation details of the presented approach; Section 5 details numerical experiments with the approach; Section 6 goes into details with several microstructural design examples; and Section 7 covers a discussion and conclusions of the presented work.

\section{Theoretical background} \label{sec:theory}

\subsection{Linear elasticity} \label{sec:theory_linela}

The considered problem is static linear elasticity, which is governed by the Navier-Cauchy equation:
\begin{subequations} \label{eqd10001}
\begin{align}
- {\rm div}\, \Vg{\sigma}\left(\V{u}\right)&=\V{f}\\
\Vg{\sigma}\left(\V{u}\right)&=\M{C}:\Vg{\varepsilon}\left(\V{u}\right)
\end{align}
\end{subequations}
where $\Vg{\sigma}$ is the stress tensor, $\Vg{\varepsilon}$ is the linearised strain tensor, $\M{C}$ is the linear elastic stiffness tensor, $\V{u}$ denotes the displacement field and $\V{f}$ is the input to the system in the form of distributed or concentrated forces. The displacements and the input forces are vector quantities with number of elements equal to the dimensionality of the considered physical space. The considered problems  are two-dimensional, however, the presented approach is applicable to three dimensions without any significant amendments. The system occupies a bounded physical domain $\Omega\in \mathbb{R}^2$ with the boundary $\Gamma=\overline{\Gamma_{D_i} \cup \Gamma_{N_i} }$ being decomposed into two disjoint subsets for each component $i=1,2$. $\Gamma_{D_i}$ is the part of the boundary with prescribed zero displacement, $u_i=0$, and $\Gamma_{N_i}$ denotes the part with prescribed traction, $t_i$.  The stiffness tensor $\M{C}$ is assumed to be isotropic and has the following form $\M{C}\left(\V{x}\right)=E\left(\V{x}\right)\M{C}_0$ where $\M{C}_0$ is the stiffness tensor for a predefined Poisson's ratio, $\nu < 0.5$, and unit Young's modulus, $E_{0} = 1$. The Young's modulus $E\left(\V{x}\right)$ is spatially varying and bounded $E\left(\V{x}\right)\in \left[E_{\min}, E_{\max} \right]$. 

The variational formulation, see e.g. \cite{Braess2007}, of equation \eqref{eqd10001} is to find $\V{u}\in V_0$ such that:
\begin{equation}
a\left(\V{u},\V{v}\right)=l\left(\V{v} \right) {\text{ for all }} \V{v}\in V_0
\end{equation}
where $V_0=\left\{\V{v} \in \left[H^1\left(\Omega\right)\right]^2 \left| \, v_i=0 {\text{ on }} \Gamma_{D_i}, i=1,2 \right.  \right\} \subset V=\left[ H^1 \left(\Omega \right)\right]^2$ and the bilinear form $a$ and linear form $l$ are defined as:
\begin{subequations} 
\begin{align}
a\left( \V{u},\V{v}\right) &= \int_\Omega \left(\M{C}:\Vg{\varepsilon}\left(\V{u}\right) \right):\Vg{\varepsilon}\left(\V{v}\right) {\text d}\V{x}\\
l\left(\V{v}\right) &= \int_\Omega \left(\V{f}\cdot \V{v}\right){\text d}\V{x} + \int_{{\Gamma}_N} \left(\V{t}\cdot \V{v}\right) {\text d} s
\end{align}
\end{subequations} 
The weak formulation is discretised using the finite element space $V_h\subset V_0$ with vector-valued shape functions defined on a uniform mesh $\mathcal{T}^h$. The discretization leads to a linear system of equations of the form:
\begin{equation}
\label{eqd10002}
\M{K} \V{u}=\V{f}
\end{equation}
where $\M{K}$ is the so-called stiffness matrix, the vector $\V{u}$ consists of the nodal displacements and the vector $\V{f}$ contains the nodal forces applied to the discrete problem. 

\subsection{Topology optimisation}

Topology optimisation is a material distribution method that seeks to find an optimised structure for a given physical problem with respect to a certain objective functional subject to design constraints \citep{Bendsoee2003}.
The physical problem, in this work linear elasticity, is discretised using the finite element method, as described in section \ref{sec:theory_linela}. The design is parametrised by attributing each finite element with a relative density, $\rho_e$, which determines whether the element is solid, $\rho_{e} = 1$, or void, $\rho_{e} = 0$. In order to solve the problem using gradient-based optimisation, the binary density variables are relaxed to continuous variables, $\rho_{e} \in \left[ 0, 1 \right]$.

The Young's modulus of each element is made to be dependent on the element relative density using the modified SIMP approach \citep{Bendsoee2003} and thus the element stiffness matrix can be formulated as:
\begin{equation} \label{eq:elementstiffness}
\M{K}_{e} \, (\rho_{e}) = \left( E_{min} + (E_{max}-E_{min}){\rho_{e}}^{p} \right) \M{K}_{0}
\end{equation}
where $E_{min}$ is the minimum Young's modulus attributed to void areas in order to avoid ill-conditioning of the stiffness matrix, $E_{max}$ is the maximum Young's modulus attributed to solid areas, $\rho_{e}$ is the element relative density, $p$ is a penalisation parameter and $\M{K}_{0}$ is the stiffness matrix for an element with a Young's modulus of 1.

In this work, we consider the minimum compliance problem under a constraint on the volume fraction of solid material. The optimisation problem is posed as follows:
\begin{align} \label{eq:topopt_prob}
\underset{ \Vg{\rho} \in \mathbb{R}^{ n_{\! d} } }{\text{minimise: }} & f\negthinspace \left( \Vg{{\rho}}, \V{u} \right) = \T{\V{f}} \V{u} \nonumber\\
\text{subject to: } & g\negthinspace \left( \Vg{{\rho}} \right) = \frac{ \T{\Vg{\rho}}\V{v} }{ v_{f} \T{\V{e}}\V{v} } - 1 \leq 0  \\
 & \Vg{\mathscrbf{R}}\negmedspace \left( \Vg{{\rho}}, \V{u} \right) = \V{0} \nonumber \\
 & 0 \leq \rho_{e} \leq 1 \,\,\,\, \text{ for } e = 1,...,n_{d}  \nonumber
\end{align}
where $f$ is the compliance functional, $g$ is the volume constraint functional, $\Vg{\rho}$ is a vector of the $n_{d}$-number of design variables, $\V{v}$ is a vector of the $n_{d}$-number of element volumes,  $\Vg{e}$ is a $n_{d}$-entry long vector of ones and $\Vg{\mathscrbf{R}}\negmedspace \left( \Vg{{\rho}}, \V{u} \right) = \M{K}\negthinspace \left( \Vg{{\rho}} \right) \V{u} - \V{f}$ is the residual of the discretised system of equations, . The optimisation problem is solved using the nested formulation, where the discretised system of equations for the state field is solved separately from the design problem.

\subsection{Computational issues of topology optimisation}

Topology optimisation is an iterative process and requires the solution of a large system of equations for the state field, equation \eqref{eqd10002}, at every design iteration. Solving the linear systems of equations can often account for more than 99\% of the total computational time \cite{Aage2013}. Therefore, the solution cost of the optimisation procedure depends strongly on the time necessary for solving the state equations.
Realistic mechanical problems with microstructural scales comparable to the macroscale require very fine discretisations to capture the microstructural scales, making direct solvers prohibitive even in two dimensions. An alternative is to utilise iterative solvers, where the convergence and the solution cost is controlled mainly by the applied preconditioner. Classical preconditioning techniques, such as incomplete factorisation, sparse approximate inverse or stationary iterative methods, cannot provide a mesh-independent number of iterations. Furthermore, for topology optimisation problems with high contrast between the material parameters, ${E_{max}}/{E_{min}}$, the number of iterations increases with increasing contrast \cite{Aage2013, Amir2014}.
Geometric multigrid techniques \cite{Vassilevski2008} can provide mesh-independent solution for smooth material properties, however, similar to the classical preconditioners, the number of iterations depends on the contrast when the mesh is not aligned with the different materials.

\subsection{Multiscale finite element (MsFEM) coarse basis for linear elasticity} \label{sec:theory_msfem}

An effective alternative to geometric multigrid and classical preconditioners was demonstrated in \cite{Galvis2010, Efendiev2011}. The method is a multiscale approach providing the solution of the diffusion equation with strongly heterogeneous coefficients. The solution cost is independent of the contrast of the diffusion coefficients and is significantly reduced by using coarse space approximations which contain important features of the solution. The approach has been extended and applied in topology optimisation of linear elastic problems in \cite{Lazarov2014} and the main steps are outlined below in the context of structures with periodic microstructural details.

	\begin{figure}
	\centering
	\includegraphics[width=0.6\textwidth]{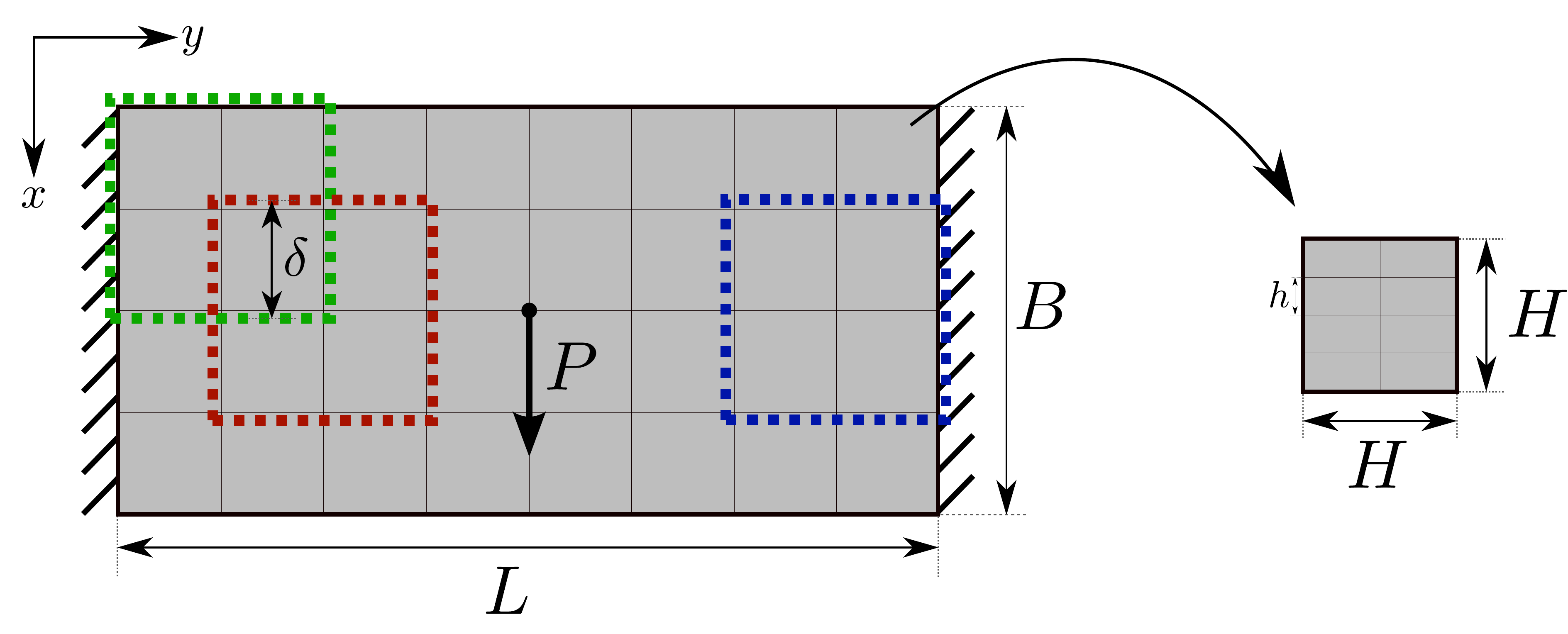}
	\caption{Illustration and dimensions of a double-clamped beam subjected to a concentrated load in the centre. The beam is made up of a single periodic microstructure with square unit cells. The fine mesh, coarse mesh and agglomerates are illustrated.} \label{fig:msfem_numexp_illustration}
	\end{figure}

The fine mesh $\mathcal{T}^{h}$, utilised for discretising equation \eqref{eqd10001} in section \ref{sec:theory_linela}, is assumed to be obtained by a refinement of a coarser one $\mathcal{T}^{H}=\left\{K_j\right\}_{j=1}^{N_c}$, where $K$ denotes a coarse mesh cell and $N_c$ is the number of coarse nodes. In the context of microstructural design, the coarse cells represent the unit cells and the fine cells the discretisation of the unit cells, which is illustrated in figure \ref{fig:msfem_numexp_illustration}.

The nodes of the coarse mesh are denoted as $\left\{\V{y}_i\right\}_{i=1}^{N_c}$ and the neighbourhood of node $\V{y}_i$ is defined as:
\begin{equation}
\omega_i=\bigcup\left\{\overline{K_j}\in\mathcal{T}^{H}; \V{y}_i\in\overline{K_j} \right\}
\end{equation}
which is illustrated in figure \ref{fig:msfem_numexp_illustration} using dotted lines. These neighbourhoods will be denoted as agglomerates, since they can be viewed as a group of coarse elements agglomerated together with an overlap $\delta$ \cite{Efendiev2011b}.

A set of coarse basis functions, $\left\{\Vg{\phi}_{i,j}, j=1\dots N_{i,j} \right\}$, defined with respect to $\mathcal{T}^{h}$, is introduced for each node $\V{y}_i$ in the coarse mesh with support on $\omega_i$, where $N_{i,j}$ is the number of basis functions for the \textit{i}'th coarse node. The coarse basis functions are based on the modes obtained by solving the following local generalised eigenvalue problem on each agglomerate $\omega_i$:
\begin{equation} \label{eq:continuous_aggeigenprob}
-\rm{div}\left(  \V{C}\left(\V{x}\right):\Vg{\varepsilon}\left(\V{u}\right)\right) = \lambda E\left(\V{x}\right) \V{u},\quad \V{x}\in\omega_i
\end{equation}
The local eigenvalue problem is discretised using the fine mesh, which leads to the following discrete generalised eigenvalue problem in matrix-vector form:
\begin{equation} \label{eq:discrete_aggeigenprob}
\M{K}_{\omega_i} \Vg{\psi}^{\omega_i}_{j}=\lambda^{\omega_i}_{j} \M{M}_{\omega_i} \Vg{\psi}^{\omega_i}_{j}
\end{equation}
where $\M{K}_{\omega_i}$ is the stiffness matrix, $\M{M}_{\omega_i}$ is a stiffness-based mass matrix, $\Vg{\psi}^{\omega_i}_{j}$ is the \textit{j}'th eigenvector and $\lambda^{\omega_i}_{j}$ is the \textit{j}'th eigenvalue, all for the \textit{i}'th agglomerate, $\omega_i$. It is important to note that the local matrices take the physical boundary conditions of the full problem into account and the coarse basis thus automatically fulfils the specified boundary conditions.

The eigenvalues are ordered as $\lambda_1^{\omega_i}\leq\lambda_2^{\omega_i}\leq\dots\leq\lambda_j^{\omega_i}\leq...$ and the first several eigenvectors corresponding to eigenvalues smaller than a prescribed global threshold, $\lambda_\Omega$, are selected to form the coarse basis. The coarse basis functions, represented on the fine grid, are defined as $\Vg{\phi}_{i,j} = \Vg{\psi}^{\omega_i}_{j} \chi_i$; that is they are constructed by multiplying the eigenfunctions, $\Vg{\psi}^{\omega_i}_{j}$, with a partition of unity, $\left\{\chi_i\right\}_{i=1}^{N_c}$, subordinated to $\omega_i$ such that $\chi_i\in H^1\left( \Omega \right)$ and $\left|\nabla \chi_i\right|\leq 1/H, i=1,\dots, N_c$, and $H$ is the characteristic length of a coarse element $K$. Therefore, the set of coarse basis functions $\left\{\Vg{{\phi}}_{i,j}\right\}$ associated with node $\V{y}_i$ is defined as the fine space finite element interpolant of $\left\{\Vg{\psi}_j^{\omega_i}\left(\V{x}\right)\chi_i\left(\V{x}\right) , j=1\dots N_i \right\}$, see e.g. \cite{Efendiev2012a}. For each $\omega_i$, $N_i$ is determined as the number of eigenvalues smaller than a globally selected threshold value, $\lambda_\Omega$.

The solution in the coarse space is sought as $\V{u}_a=\sum_{i,j} c_{i,j} \Vg{\phi}_{i,j}$, where $c_{i,j}$ is the coefficient of the \textit{j}'th coarse basis function for the \textit{i}'th agglomerate, $\omega_i$. A coarse discretisation of the variational formulation is given as $\M{K}_c \V{u}_c=\V{f}_c$, where the coarse stiffness matrix $\M{K}_c$ and the coarse right-hand side $\V{f}_c$ are obtained as:
\begin{equation} \label{eq:coarse_projection}
\M{K}_c=\M{R}_c \M{K} \T{{\M{R}_c}}, \quad \V{f}_c=\M{R}_c \V{f}
\end{equation}
The vector $\V{u}_c$ contains the coefficients $c_{i,j}$ and $\T{{\M{R}_c}}=\left[\Vg{\phi}_1, \Vg{\phi}_2,\dots,\Vg{\phi}_{N_t} \right]$, where $N_t=\sum_{i=1}^{N_c} N_i$, is a matrix describing the mapping from the coarse to the fine space and consisting of nodal values of the coarse basis functions in the fine space. An approximation of the nodal solution in the fine space is obtained as $\V{u}_a=\T{{\M{R}_c}} \V{u}_c$.
 
\subsubsection{Periodic microstructure} \label{sec:theory_periodic}

Building the MsFEM basis for general linear elastic systems requires solving the eigenproblem, equation \eqref{eq:discrete_aggeigenprob}, for each agglomerate and as discussed in \cite{Lazarov2014} this process is computationally very expensive. It is important to note that when dealing with periodic microstructural details, the computation of eigenfunctions is limited to a small amount of unique agglomerates. Figure \ref{fig:msfem_numexp_illustration} shows the three unique agglomerates for a double-clamped beam with a periodic microstructure. Due to the imposed periodicity of the microstructure, all agglomerates along the left-hand side are the same (constrained in both $x$- and $y$-direction at the left-hand side), all agglomerates along the right-hand side are the same (constrained in both $x$- and $y$-direction at the right-hand side) and all internal agglomerates are the same (unconstrained and allowing rigid body motion). This means that significant time can be saved on the computation of the local spectral bases.
 
\subsubsection{Application as multigrid-like preconditioner}

MsFEM can be applied as a solver in topology optimisation, however, if the coarse solver does not provide an accurate enough approximation of the fine-scale solution, the optimisation will results in sub-optimal designs with disconnected features \cite{Lazarov2014}. An alternative is proposed in \cite{Galvis2010}, where the basis is utilised in a two-level additive Schwarz preconditioner for the iterative solution of the fine scale solver. The preconditioner is scalable and results in contrast independent number of iterations. Here, the coarse basis is simply applied as a coarse grid solve in a multigrid-like preconditioner \cite{Efendiev2011b} for the iterative solution of the fine-scale system of equations, equation \eqref{eqd10002}. The residual of the fine-scale system is restricted to the coarse basis and the correction is approximated using a MsFEM coarse-scale solve and projected back to the fine space. In the current work, the MsFEM coarse-scale solve is combined with a single pre- and post-smoothing using symmetric Gauss-Seidel and used as a preconditioner for the generalised minimal residual (GMRES) iterative method \cite{Saad1986}. The linear elasticity problem, as well as the preconditioner, is symmetric and positive definite and, hence, the preconditioned conjugate gradient (PCG) method could be used. However, the authors' experience is that GMRES performs better compared to PCG for the presented problems and thus GMRES is used throughout.

\section{Demonstration of the MsFEM preconditioner} \label{sec:msfem}

\subsection{Numerical experiments} \label{sec:msfem_numexp}	

In order to demonstrate the performance and applicability of MsFEM for topology optimisation problems, a test problem is formulated. The test problem is a short beam clamped at both ends and subjected to a concentrated load in the centre, as shown in figure \ref{fig:msfem_numexp_illustration}. The height of the beam is set to $B = 1$, the length to $L=2$ and the size of the force is set to $P = 0.01$. The beam is made up of 8 by 16 square coarse cells, with an edge length of $H = 1/8$, containing the same microstructure. Each coarse cell is discretised with 20 by 20 linear finite elements, leading to a full macroscopic mesh of 160 by 320. The problem is investigated for varying values of the eigenvalue threshold, $\lambda_{\Omega}$, and MsFEM is applied both as a solver and as a preconditioner in combination with GMRES.

\begin{figure}
	\centering
		\begin{tabular}{cc}
			\subfloat[``Final'']{\includegraphics[trim= 390 90 390 65, clip, width=0.13\textwidth]{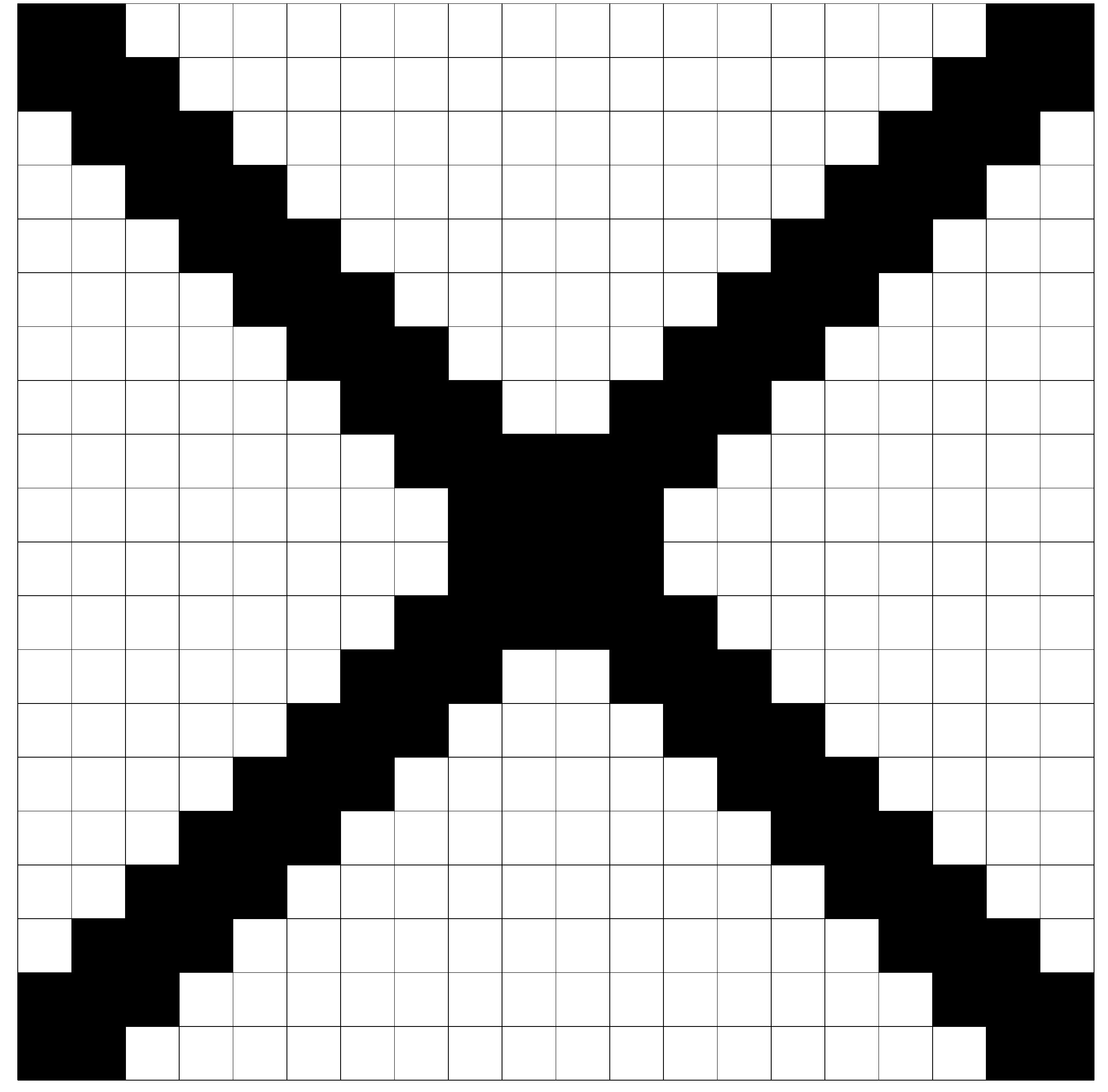} \label{fig:cross_microandmacro-a}}
			&
			\addtocounter{subfigure}{1}
			\multirow{2}{*}[1.8cm]{
			\subfloat[Macrostructure: ``Intermediate'']{\includegraphics[width=0.6\textwidth]{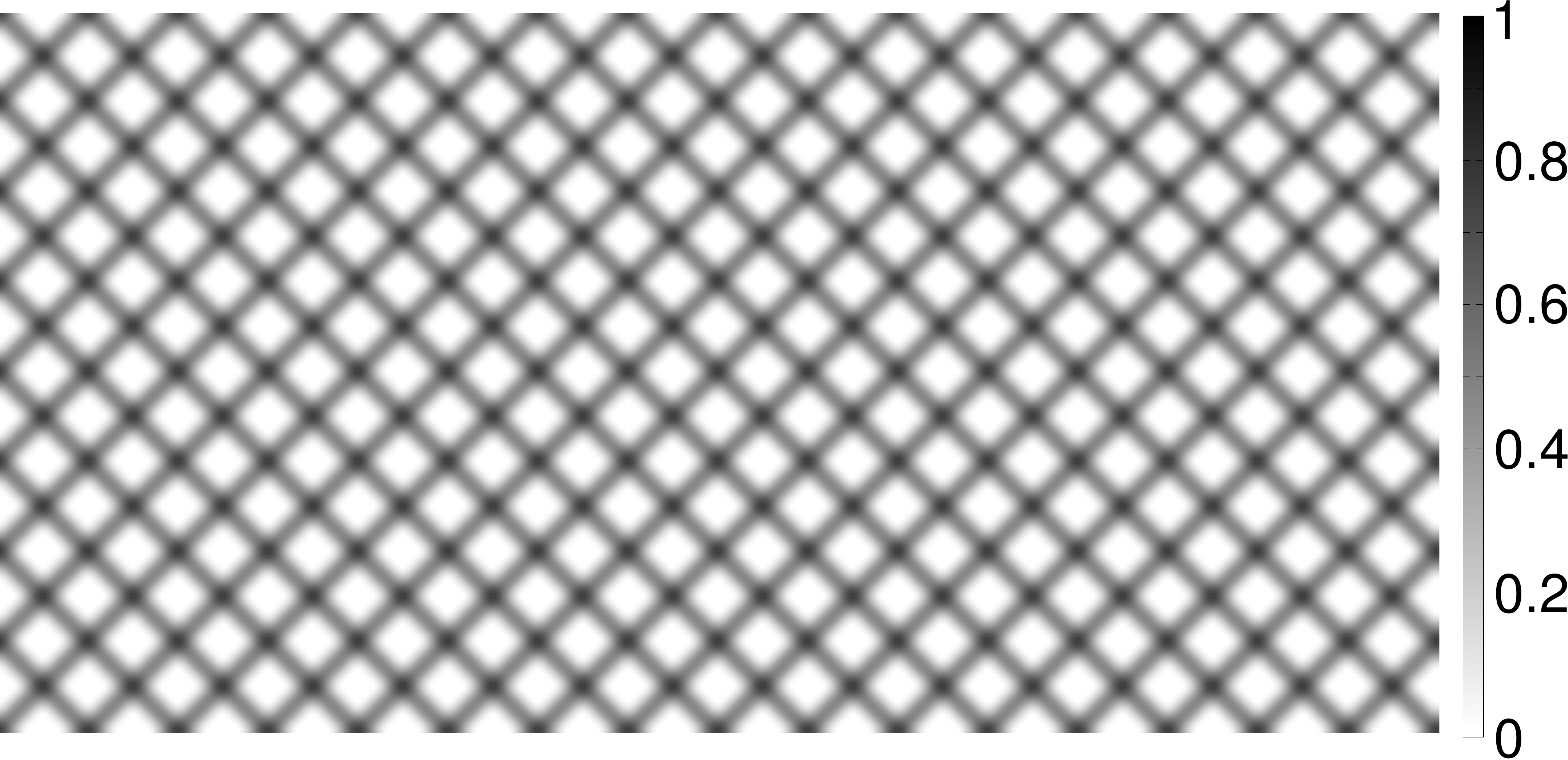}			 \label{fig:cross_microandmacro-c}}
			}
			\\
			\addtocounter{subfigure}{-2}
			\subfloat[``Intermediate'']{\includegraphics[trim= 390 90 390 65, clip, width=0.13\textwidth]{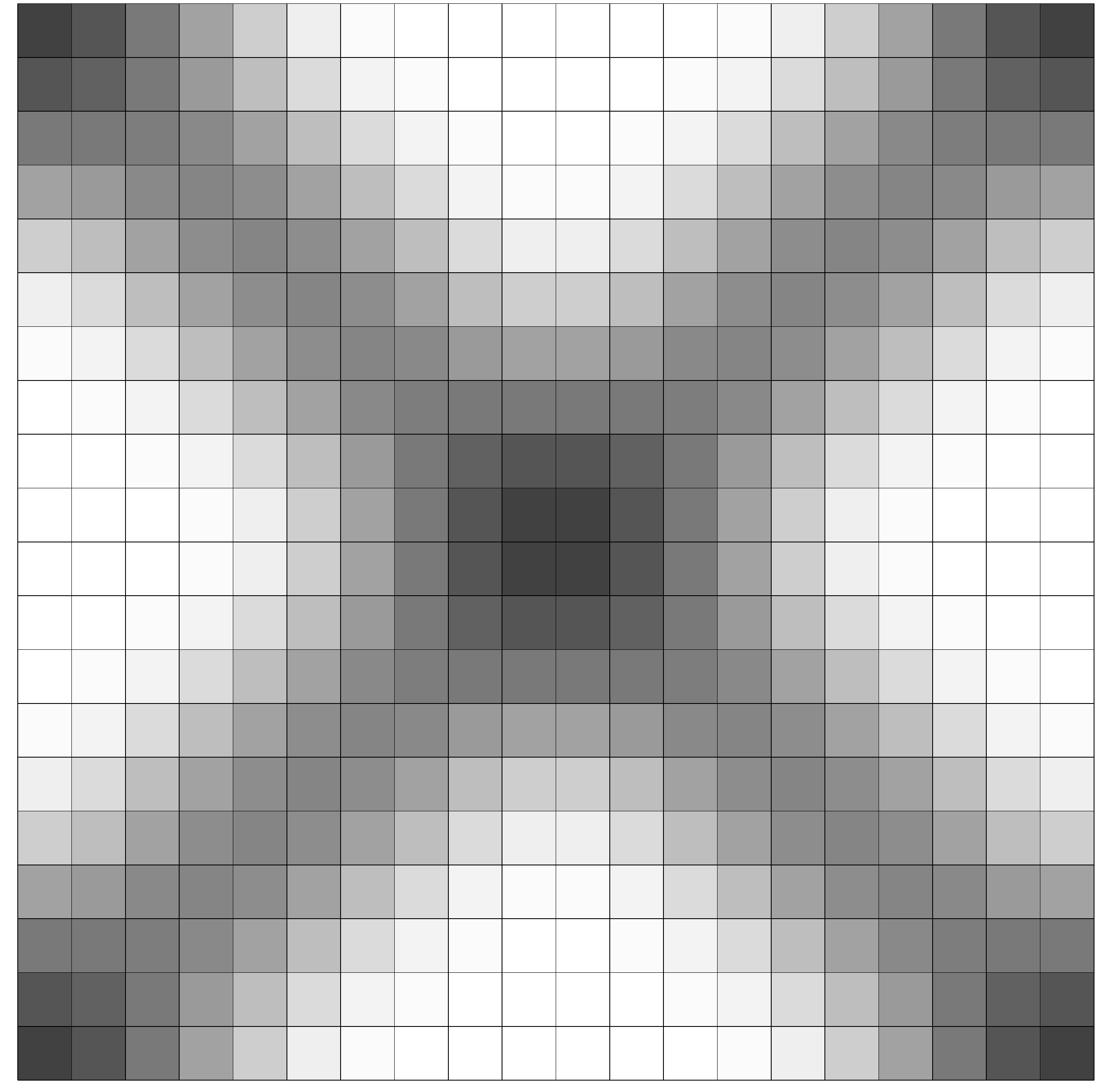} \label{fig:cross_microandmacro-b}}
			&
		\end{tabular}
	\caption{The investigated cross structure with a unit cell discretised by 20x20 elements. The macrostructure is made up of 8 by 16 unit cells, which is shown in two versions: (a) final 0-1 design and (b) the corresponding ``intermediate'' design generated by filtering the first. Subfigure (c) shows the macrostructure for the ``intermediate'' design.} \label{fig:cross_microandmacro}
\end{figure}
Figure \ref{fig:cross_microandmacro} shows the structure under consideration for these numerical experiments. The microstructure is chosen as a cross structure and is considered in two versions, the first of which is the clear 0-1 design shown in figure \ref{fig:cross_microandmacro-a}. This design is representative of a final design of a topology optimisation process using a Heaviside projection as will be described in section \ref{sec:implem_filtering}. The structure is also considered in a smeared out version, as can be seen in figure \ref{fig:cross_microandmacro-b}, which is representative of an intermediate design during the optimisation process where the design usually contains significant amounts of intermediate stiffnesses. The smeared out design is obtained using a density filter, as will be described in section \ref{sec:topopt_robust}, with a filter radius of 4 times the element side length. The macrostructure for the intermediate design case is shown in figure \ref{fig:cross_microandmacro-c}. The structure is analysed with a SIMP penalisation parameter of p=3. Similar trends are observed for the two structures and thus only the results for the filtered version are shown.
	
	\begin{figure}[t]
	\centering
	\subfloat[Eigenvalue threshold, $\lambda_{\Omega}$]{\includegraphics[width=0.5\textwidth]{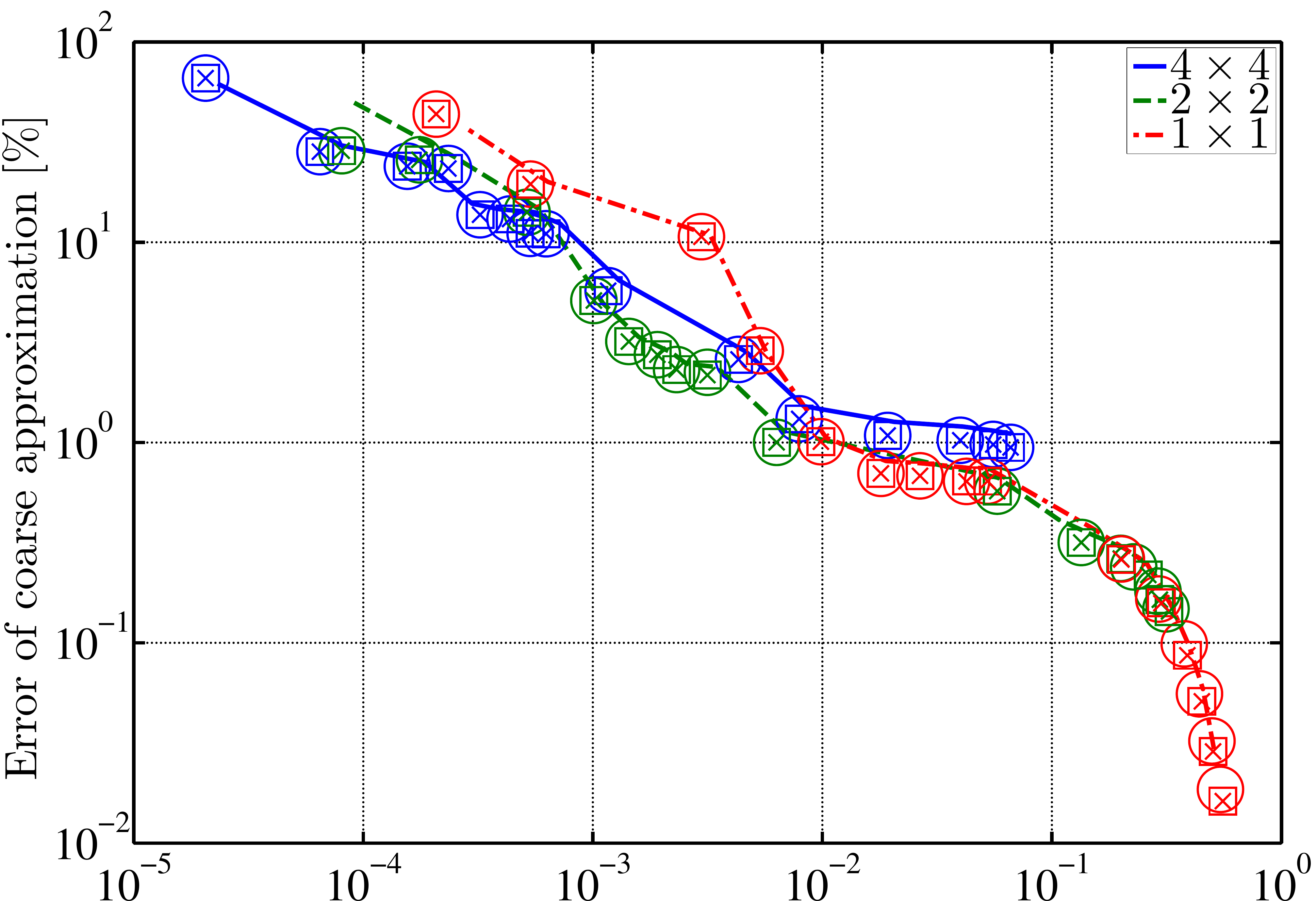}
	\label{fig:crossfilter4_ConvOfError-a}}
	\hspace*{0.02\textwidth}
	\subfloat[Number of coarse DOF]{\includegraphics[height=0.33\textwidth]{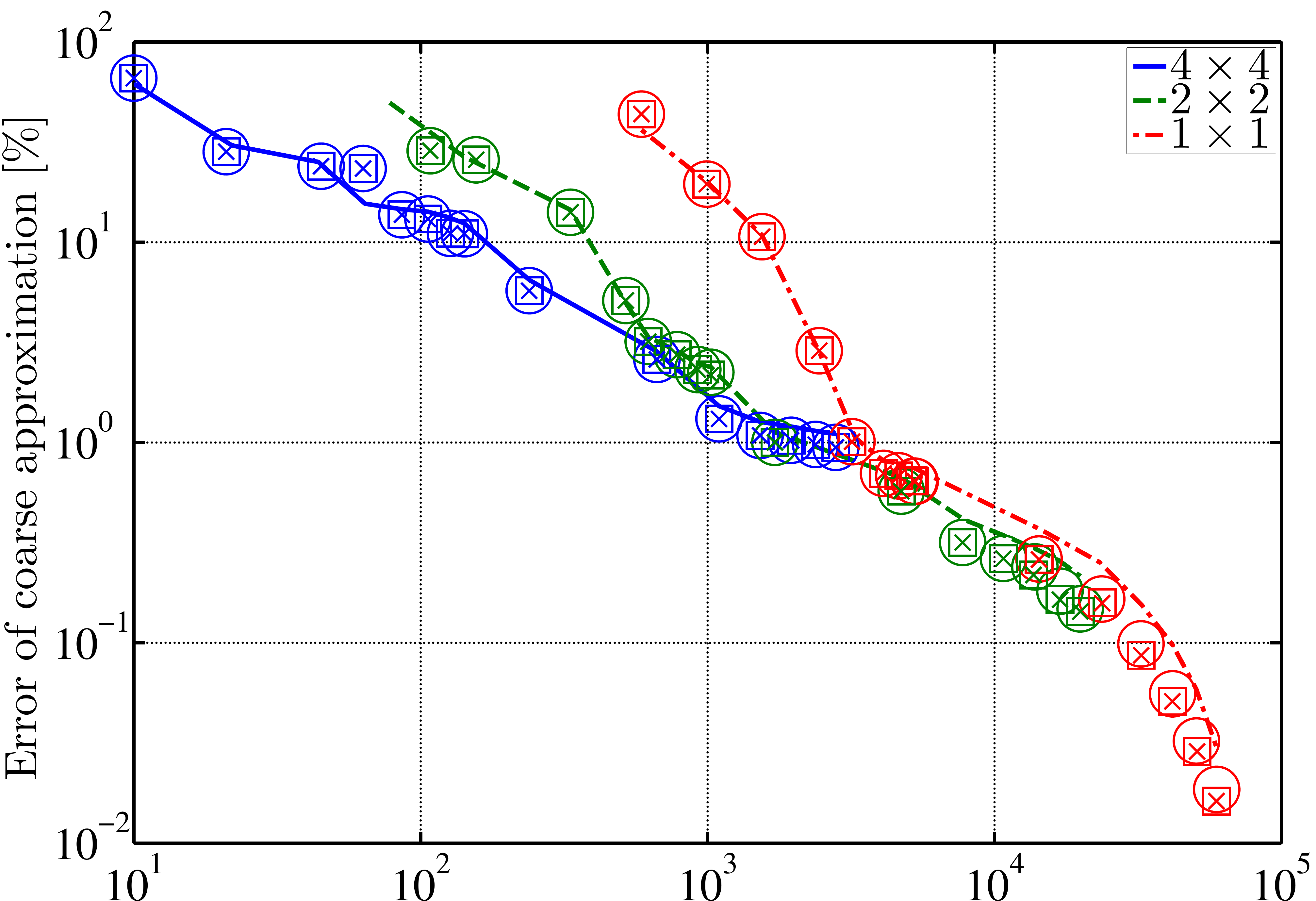}
	\label{fig:crossfilter4_ConvOfError-b}}
	\caption{Relative error in the energy norm for a single MsFEM solve of the filtered cross structure. The error is shown as a function of (a) varying eigenvalue threshold and (b) the corresponding number of coarse degrees of freedom. Colours denote different agglomerate sizes, see legend, and symbols denote different contrasts: lines; $E_{min} = 10^{-3}$, circles; $E_{min} = 10^{-6}$, squares; $E_{min} = 10^{-9}$ and crosses; $E_{min} = 10^{-12}$.} \label{fig:crossfilter4_ConvOfError}
	\end{figure}
	Figure \ref{fig:crossfilter4_ConvOfError} shows the relative error in the energy norm, $\left\lVert \V{u} - \V{u}_{a} \right\rVert_{\M{K}} / \left\lVert \V{u} \right\rVert_{\M{K}} = \T{(\V{u}-\V{u}_{a})}\M{K}(\V{u}-\V{u}_{a}) / \T{\V{u}}\M{K}\V{u}$, of the solution obtained from a single coarse-scale MsFEM solve for the intermediate design case. The analysis is done for a constant maximum Young's modulus, $E_{max} = 1$ and a set of minimum Young's moduli, $E_{min}\in \left\lbrace 10^{-3},10^{-6},10^{-9},10^{-12} \right\rbrace$, as well as for different sizes of the agglomerates, denoted by the numbers of unit cells covered by each coarse cell, $\left\lbrace 1\times1, 2\times2, 4\times 4 \right\rbrace$.
It can be seen that the approximated solution converges with respect to both the eigenvalue threshold and the corresponding number of degrees of freedom. Even though small differences are observed between $E_{min} = 10^{-3}$ and the lower, it can clearly be seen that the behaviour is contrast-independent for $E_{min} \in \left\{ 10^{-6},10^{-9},10^{-12} \right\}$. It can be seen that the energy norm error follows a common trend, in relation to the eigenvalue threshold, independent of the agglomerate size, similar to the scalar case \cite{Efendiev2011}. Clear differences can be seen in the lower end of the datasets for the error in relation to the number of coarse degrees of freedom. The curves do, however, converge towards a common trend as the number of degrees of freedom is increased.

	As can be seen from the error plots, a very large set of coarse basis functions need to be used to obtain a good accuracy of a single MsFEM solve. This results in a very large computational cost, both for solving the generalised eigenvalue problems but also for solving the coarse-scale system. This is why the MsFEM basis is applied as a preconditioner to the full system instead.
	
	\begin{figure}[t]
	\centering
	\subfloat[Eigenvalue threshold, $\lambda_{\Omega}$]{\includegraphics[width=0.5\textwidth]{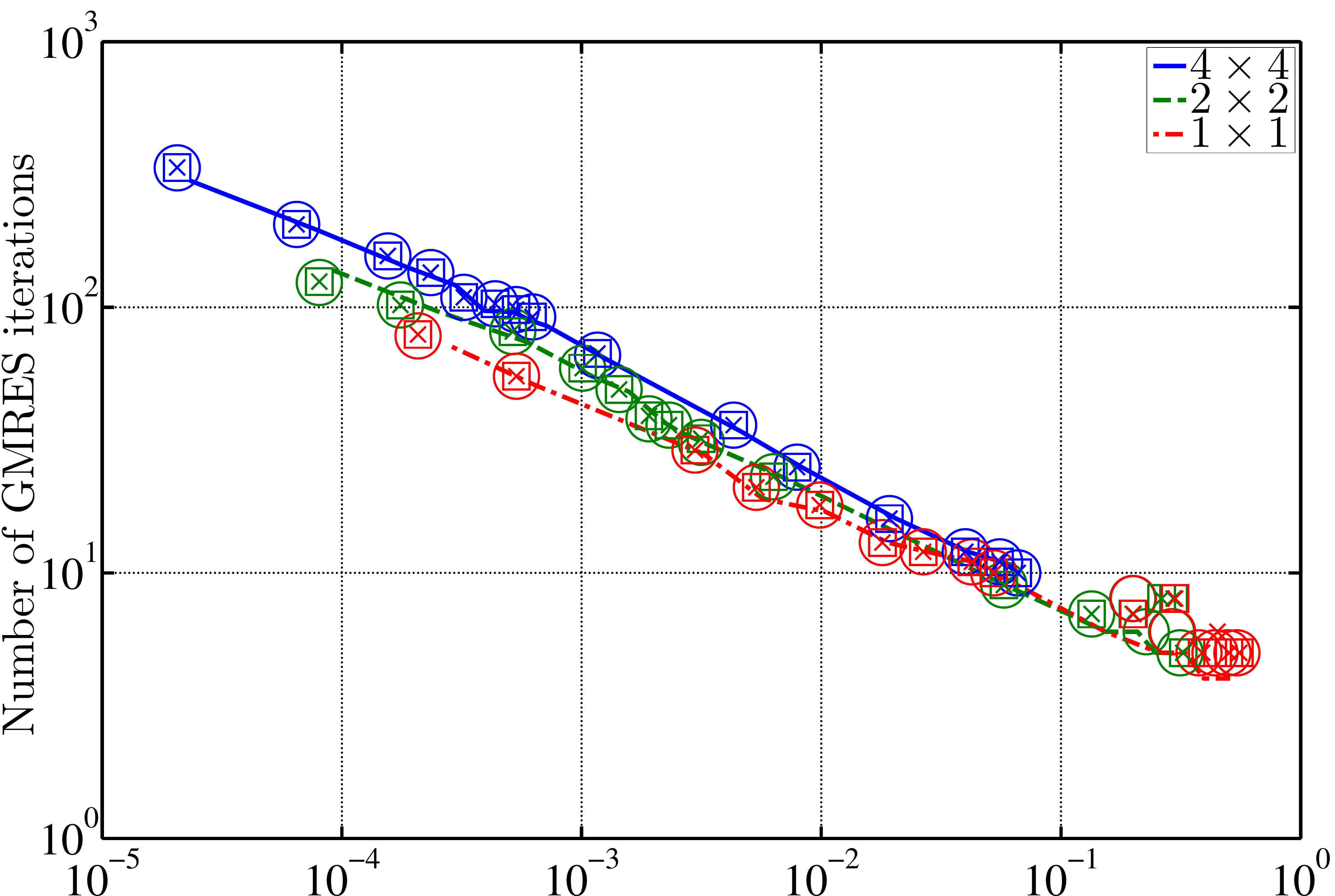}
	\label{fig:crossfilter4_ConvOfIter-a}}
	\hspace*{0.02\textwidth}
	\subfloat[Number of coarse DOF]{\includegraphics[width=0.5\textwidth]{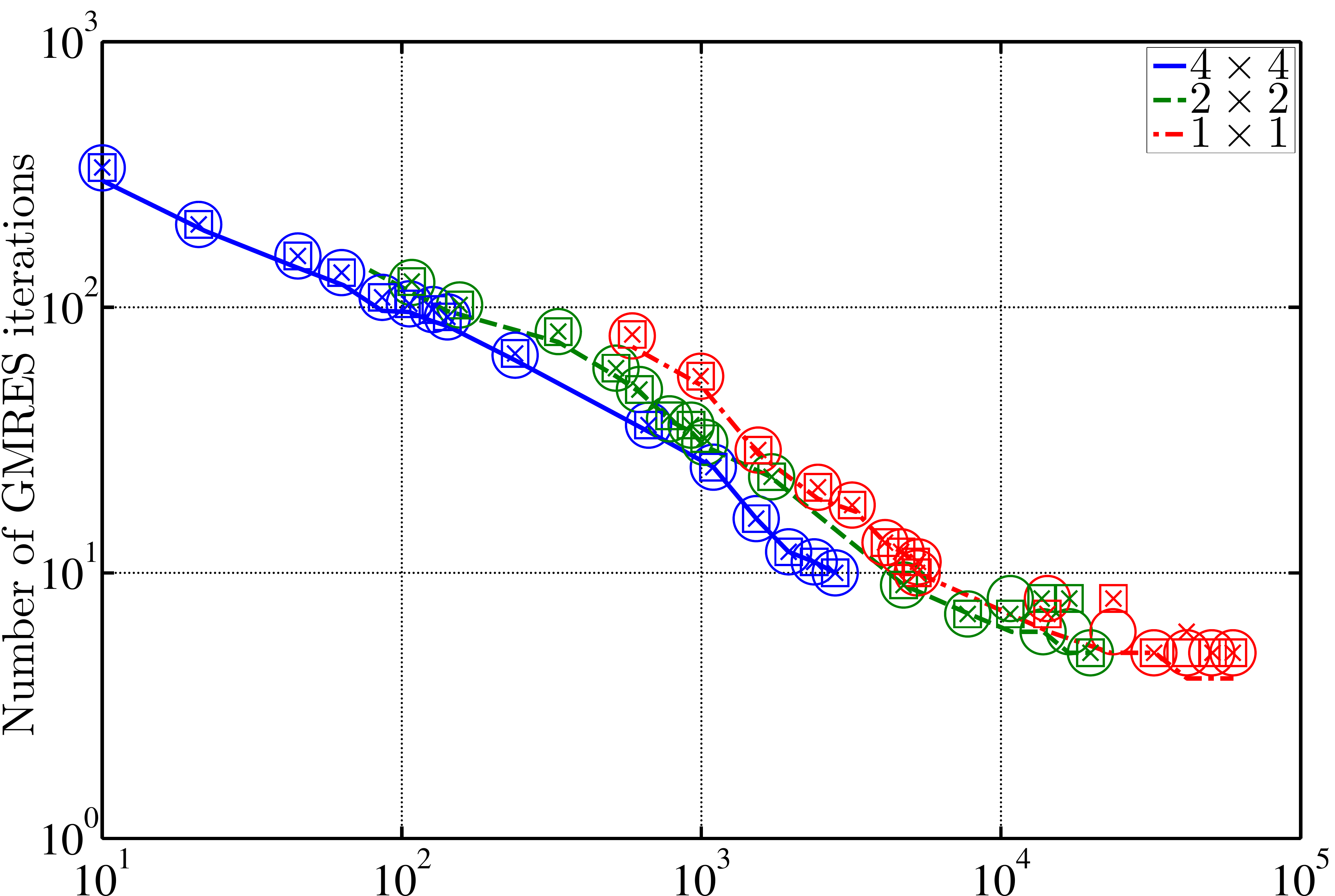}
	\label{fig:crossfilter4_ConvOfIter-b}}
	\caption{Number of GMRES iterations until convergence for the filtered cross structure. The number of iterations is shown as a function of (a) varying eigenvalue threshold and (b) the corresponding number of coarse degrees of freedom. See figure \ref{fig:crossfilter4_ConvOfError} for explanation of symbols and colours.} \label{fig:crossfilter4_ConvOfIter}
	\end{figure}
	Figure \ref{fig:crossfilter4_ConvOfIter} shows the number of GMRES iterations needed to converge to a precision of $\varepsilon_{rel} = 10^{-8}$ relative to the preconditioned norm of the forcing vector. As expected, the number of GMRES iterations decreases as the eigenvalue threshold is increased and the basis is enlarged. Besides the results showing clear contrast-independence, they also show that the convergence of the GMRES solver is not significantly affected by the size of the agglomerate. More specifically, by looking at figure \ref{fig:crossfilter4_ConvOfIter-a} it can be seen that by choosing the same eigenvalue threshold, the number of GMRES iterations needed to converge is very close to the same.

	\subsection{Diagonal weighting matrix}	
	
Instead of computing the stiffness-based mass matrix for a given agglomerate, as specified by equation \eqref{eq:continuous_aggeigenprob}, it is possible to simply use the diagonal of the agglomerate stiffness matrix as the weighting matrix of the generalised eigenproblem, equation \eqref{eq:discrete_aggeigenprob}. This is because the two weighting matrices are spectrally equivalent and $\lambda_{\text{diag}(\M{K})} \approx C h^{2} \lambda_{\M{M}}$ for some constant C \citep{Vassilevski2008, Brezina2012}. This is interesting from a computational point of view, as the diagonal leads to a cheaper solution procedure; both in that the stiffness-based mass matrix is not assembled, and in that the matrix-products involved in solving the eigenproblems become cheaper. The numerically observed convergence behaviour is very similar to that seen in figure \ref{fig:crossfilter4_ConvOfError} and is thus not shown. The diagonal of the agglomerate stiffness matrix is used as the weighting matrix throughout the rest of this paper.
				
	\subsection{Spectral basis computational cost} \label{sec:msfem_compcost}
	 
Considering the total computational time taken for the projection operation, equation \eqref{eq:coarse_projection}, and the iterative solution, there exists an optimal basis size that provides the fastest solution process. Many factors influence this time, such as the sparsity of the projection matrix and also the number of iterations taken to reach convergence. It should also be noted that the times are not the same for the three agglomerate sizes, because the larger the agglomerate, the larger the support area and the denser the projection matrix becomes. Thus, it is clear that increasing the size of the basis does not necessarily lead to a faster solution even though it results in fewer GMRES iterations, as each of these become more expensive along with the projection operation. It can thus be concluded that it cannot pay to increase the size of the agglomerate, as it does not appear to influence the error nor the convergence of the iterative solver, figure \ref{fig:crossfilter4_ConvOfIter}, as well as leading to a more expensive solution procedure due to larger eigenproblems and denser projection matrices.

\subsection{Note on length scale separation}

The main assumption in standard homogenisation is that the macroscale medium consists of an infinite number of periodic unit cells. The homogenised solution is an asymptotic solution, which is valid only when the cell characteristic length $H$ is orders of magnitude smaller than the macroscopic problem scale, i.e., $H << L$. For cell characteristic lengths in the intermediate range, $H < L$, the homogenised solution can differ significantly from the true solution. Also, the homogenised material properties cannot account for the boundary conditions and any load variations of the fine scale problem.

Furthermore, in the design of functionally graded materials, i.e. materials with spatially varying properties, the periodicity condition is often violated. The transition from one cell type to another cell type is applied abruptly within a distance comparable to $H$. In the above mentioned cases, the proposed MsFEM-GMRES approach will provide the physical response without any simplifications for relatively low computational cost. Hence, it will be superior, but more costly, compared to the standard homogenisation for optimisation problems, where localised constraints (loading, boundary conditions, surface effects) significantly affect the design topology.

	\begin{figure}
	\centering
	\subfloat[Double-clamped beam with concentrated load]{
	\hspace*{0.01\textwidth}\includegraphics[height=0.2\textwidth]{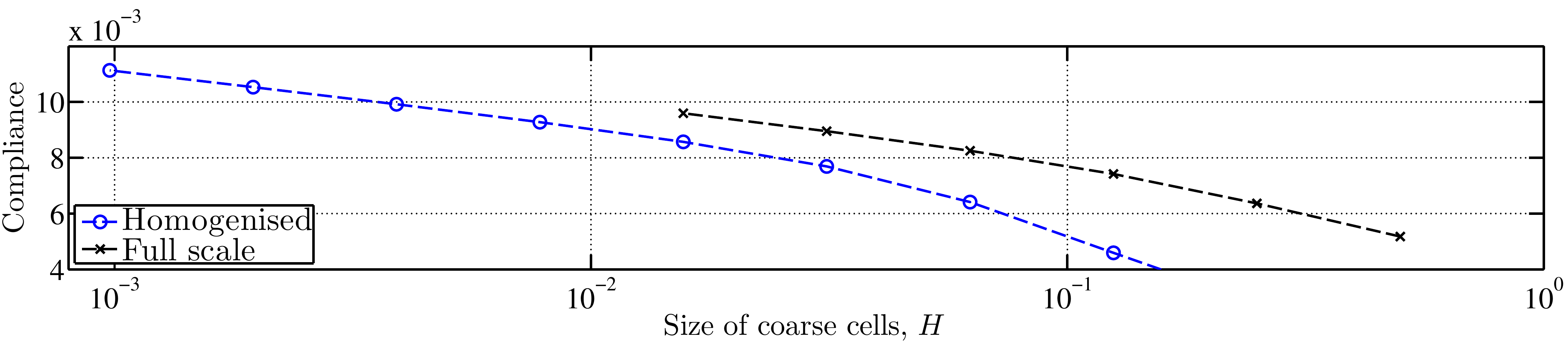}
	\label{fig:crossnofilter_ConvToHomog} }
	\\
	\subfloat[Cantilever beam with distributed load]{
	\includegraphics[height=0.192\textwidth]{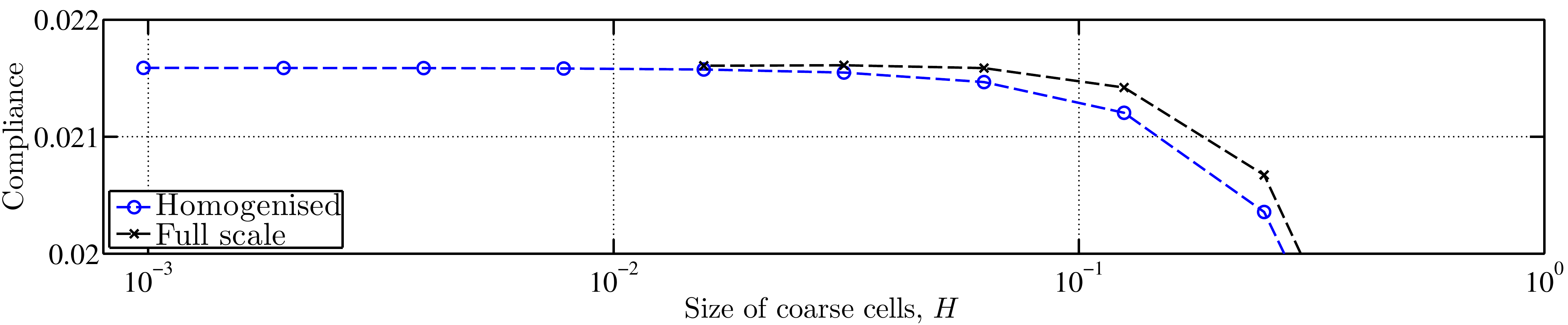}
	\label{fig:cantidist_ConvToHomog} }
	\caption{Convergence of compliance with respect to the size of the coarse cell using both homogenisation and full scale analysis using MsFEM-GMRES.} \label{fig:ConvToHomog}
	\end{figure}	
	Figure \ref{fig:ConvToHomog} shows the change of compliance with respect to the size of the coarse cell using both homogenisation and full-scale analysis using MsFEM-GMRES. The compliance is shown both for the double-clamped beam with a concentrated load in the centre, shown in figure \ref{fig:msfem_numexp_illustration}, as well as a cantilever with a distributed load at the end, as will be introduced in section \ref{sec:results_cantidist}. It should be noted that convergence is not expected for the case with a point-load, which is also observed for both methods in figure \ref{fig:crossnofilter_ConvToHomog}.
	It is important to note that any observed convergence is slightly different for the two solution types. For the full-scale analysis, the size of the microstructure decreases when decreasing the size of the coarse cells. The convergence of compliance with respect to microstructure size is assumed to be dominant over the increased finite element resolution of the macroscopic domain.
	For the homogenised case, decreasing the size of the coarse cells solely increases the resolution of the finite element discretisation of the homogenised problem. That is, the microstructure size can be seen as constant and is assumed to be small enough to fulfil the assumptions of homogenisation. 
	Thus, figure \ref{fig:ConvToHomog} illustrates that the macroscopic compliance of structures with finite size periodic microstructural details converges towards that estimated by homogenisation. However, it also emphasises that if one attributes a large finite size to a microstructure designed using homogenisation, one may be introducing a substantial error in the macroscopic response. This is especially pronounced for problems with localised behaviour as the illustrated double-clamped beam with a concentrated load in the centre, figure \ref{fig:crossnofilter_ConvToHomog}.

\section{Implementation details} \label{sec:implem}

\subsection{Topology optimisation of microstructural details} \label{sec:implem_micro}

Referring to figure \ref{fig:msfem_numexp_illustration}, the design domain is made up of repeated unit cells covering the entire domain. The unit cells contain a periodic microstructure that is discretised with a certain number of finite elements and each of these are assigned a relative density which is controlled by a corresponding design variable. Due to the imposed periodicity of the design, the same design variable controls the relative density of several fine-scale elements in the macroscopic problem. The total sensitivity for a given design variable is thus calculated by summing up local sensitivities of all fine-scale elements which it controls.

It is important to emphasise that the state field is solved on the macroscopic level resolving all microstructural details. The objective functional is thus the fine-scale macroscopic compliance. By considering the full structure, one ensures that all microstructural details will be connected in order to provide a beneficial macroscopic behaviour, as disconnected members would be detrimental to the compliance objective. 

One of the aims of this paper is to investigate the performance of the spectral basis preconditioner, as described in section \ref{sec:theory_msfem}, and the performance is thus compared to reference results. For calculating reference results, the equation system is solved using the standard backslash (\texttt{mldivide}) in \textsc{Matlab} and these are then compared to the results obtained when using the spectral preconditioner based on MsFEM combined with the standard \textsc{Matlab} implementation of the GMRES iterative method (\texttt{gmres}). 

The optimisation problem is solved using the method of moving asymptotes (MMA) \citep{Svanberg1987}, of which the used \textsc{Matlab} implementation is courtesy of Krister Svanberg.

\subsection{Adjoint sensitivity analysis}

In order to use a gradient-based optimisation method, such as MMA, one needs the sensitivities, or gradients, of the objective functional and any constraint functionals. These sensitivities are easily found for the structural compliance using the discrete adjoint method \cite{Michaleris1994}. Because the problem is self-adjoint, the sensitivity for the $e$'th element simply becomes:
\begin{equation} \label{eq:elementsensitivity}
\dpart{f}{\rho_{e}} = - \V{u}_{e} \dpart{\M{K}_{e}}{\rho_{e}} \V{u}_{e}
\end{equation}
where $\V{u}_{e}$ is the element nodal displacements and $\M{K}_{e}$ is the element stiffness matrix.
The derivative of the element stiffness matrix with respect to the element density is easily found as:
\begin{equation}
\dpart{\M{K}_{e}}{\rho_{e}} = p (E_{max}-E_{min})\rho_{e}^{p-1} \M{K}_{0}
\end{equation}

\subsection{Heaviside projection filtering} \label{sec:implem_filtering}

Heaviside projection filtering \cite{Guest2004,Wang2010} is applied in order to facilitate the formation of crisply-defined topologies, or 0-1 topologies, without intermediate densities. First, the design variables are filtered using density filtering \cite{Bourdin2001} and subsequently the filtered density variables are passed through a smoothed Heaviside function in order to project the densities below or above a given threshold to 0 or 1, respectively:
\begin{equation}
\bar{\tilde{\rho}}_{e} = \frac{ \tanh (\beta \eta) + \tanh (\beta ( \tilde{\rho}_{e} - \eta ) ) }{ \tanh (\beta \eta) + \tanh (\beta (1 - \eta)) }
\end{equation}
where $\eta \in [0; 1]$ determines the projection threshold and $\beta > 0$ determines the sharpness of the approximation.

After the design variables, $\Vg{\rho}$, have been filtered, $\tilde{\Vg{\rho}}$, and projected, $\Vg{\bar{\tilde{\rho}}}$, one has obtained the actual physical densities. These physical densities are used for the calculation of the element stiffness matrix, equation \eqref{eq:elementstiffness}, and therefore, the sensitivities, equation \eqref{eq:elementsensitivity}, are corrected using the chain rule, as detailed in e.g. \cite{Sigmund2007}.

\begin{figure}
\centering
\subfloat[Periodic filtering]{\includegraphics[height=0.25\textwidth]{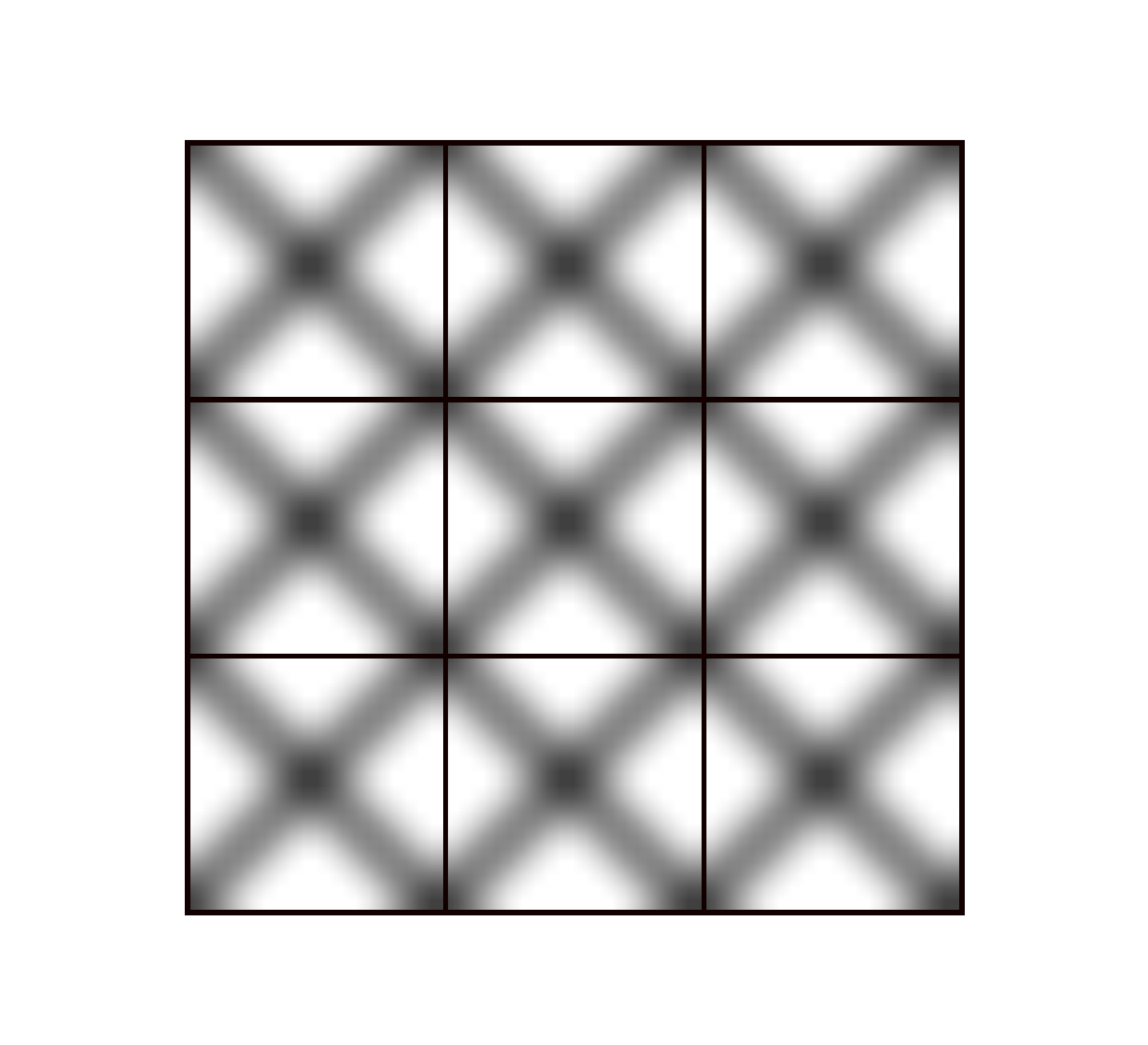}\label{fig:filtering_periodic}}
\subfloat[Multiple blocks]{\includegraphics[height=0.25\textwidth]{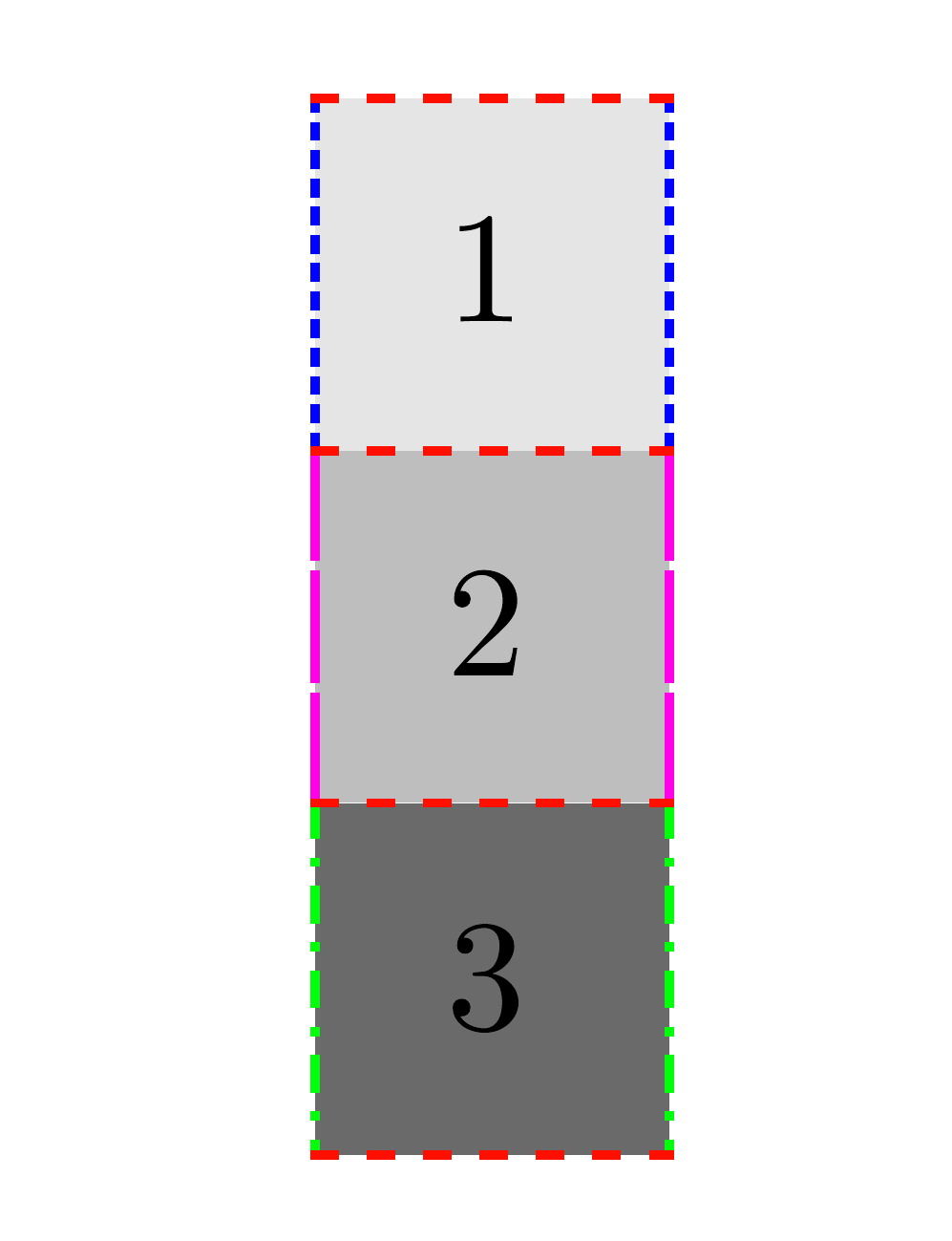}\label{fig:filtering_threeblock}}
\caption{Illustration showing the filtering procedures: (a) single periodic microstructure where the filtering is performed on a $3\times3$ block - (b) multiple blocks where periodicity and connectivity constraints imposed on the filtered density field. The colours and line styles indicate which edges are coupled together explicitly by collapsing their degrees of freedom.} \label{fig:filtering}
\end{figure}
The standard neighbourhood-based filtering approach, see e.g. \cite{Sigmund2007}, is used for all designs in this paper, except for the problems containing multiple microstructures, as will be described in section \ref{sec:topopt_multiblock}. For these problems, the PDE-based density filter, as introduced in \citep{Lazarov2011}, is used due to the fact that connectivity and periodicity boundary conditions are easily introduced. For the single microstructure designs, periodicity is introduced by applying the standard filter on a $3\times3$ block of the microstructural design as illustrated by figure \ref{fig:filtering_periodic}.

\subsection{Robust topology optimisation} \label{sec:topopt_robust}

All manufacturing methods introduce sources of uncertainty in the final realised structure, the most prevalent of which is an uncertainty in the manufactured geometry. This is caused by either too little or too much material being removed or added during the manufacturing process. The possible variability in the manufactured structures can be taken into account by considering several design realisations during the optimisation process \citep{Wang2010}. 

The optimisation problem is cast in a stochastic robust formulation \cite{Lazarov2012a}, in which the mean and variance of the macroscopic compliance is considered. The different realisations of the design are obtained by using several projections using different threshold values simulating uniform over- and under-material deposition or removal. The optimisation problem is posed as follows:
\begin{align} \label{eq:robust_prob}
\underset{ \Vg{\rho} \in \mathbb{R}^{ n_{\! d} } }{\text{minimise: }} & f\negthinspace \left( \Vg{{\rho}}, \V{u} \right) = \Exp{\T{\V{f}} \V{u}} + \kappa \sqrt{ \Var{\T{\V{f}} \V{u}} } \nonumber\\
\text{subject to: } & g\negthinspace \left( \Vg{{\rho}} \right) = \frac{ \Exp{\T{\bar{\tilde{\Vg{\rho}}}}\V{v}} }{ v_{f} \T{\V{e}}\V{v} } - 1 \leq 0  \\
 & \Vg{\mathscrbf{R}}\negmedspace \left( \Vg{{\rho}}, \V{u} \right) = \V{0} \nonumber \\
 & 0 \leq \rho_{i} \leq 1 \,\,\,\, \text{ for } i = 1,...,n_{d}  \nonumber
\end{align}
where $\Exp{\cdot}$ and $\Var{\cdot}$ denotes the expected value and variance of a given quantity, respectively, and $\kappa$ is a weighting factor for the inclusion of the variance in the objective. The sensitivities are computed as described in \cite{Lazarov2012a}.

Throughout this paper, the weighting factor, $\kappa$, is set to 1.0 and three design realisations are used with equal weights, specifically the set $\eta \in \left\lbrace 0.3, 0.5, 0.7\right\rbrace$. The final designs are verified using a large set of uniformly distributed points, $\eta \in \left[0.3,0.7\right]$.

\subsubsection{Constant $\beta$ approach for moderate $\beta$}

A constant $\beta$ approach as presented in \citep{Guest2011} is adopted in order to eliminate changes in the allowable minimum length scale during the optimisation process. By having a constant $\beta$, controlling the slope of the projection, in combination with a constant filter radius and constant thresholds, one ensures a constant minimum length scale. This ensures that small features that cannot be supported by the mesh and length scale do not appear during the optimisation process. Using the classical $\beta$-continuation, these smaller features are progressively removed during the optimisation process and convergence issues can be encountered along the way. The approach relies on simple changes to the size of the initial asymptotes in the MMA algorithm, which is set to $\frac{0.5}{1+\beta}$. No external move limits are imposed and thus the full control of the updates are left in the hands of MMA. Even though one removes the need to use a continuation approach on the $\beta$-parameter by making these modifications, it is observed to be beneficial to perform continuation of the SIMP penalisation parameter in order to get high quality solutions, as also noted in the original paper \citep{Guest2011}. If starting from a homogeneous intermediate guess, the constant beta approach retains the possibility to relatively freely form topologies. However, if starting from a random guess, many areas will be projected to solid and void and the constant beta approach then in many ways resembles some form of level set approach.

For the single design results presented in this work, a continuation approach is used on the SIMP penalisation parameter, increasing it in steps of $p \in \left\lbrace  1,2,3,4,5  \right\rbrace $ when the maximum relative change in objective function reaches below $0.001$ or when the number of iterations for the same parameter size reaches 50. After the last step, the optimisation is stopped when a maximum number of iterations is reached or the maximum relative change criteria is met.

Unfortunately, when combined with the robust topology optimisation formulation, one is somewhat restricted in the maximum size of $\beta$. The size of $\beta$ needs to be moderate in order to ensure the ability of the optimiser to form a good initial topology from both homogeneous and random starting distributions. The important thing is for the smooth parts of the shifted approximative Heaviside functions to overlap, or rather that the parts with non-negligible sensitivities overlap, and this sets an upper limit for the initial $\beta$ for the optimisation to proceed in practice. Thus, the robust topology optimised designs in this paper have been obtained from a three step continuation approach. During the initial stage where the topology is forming, the SIMP penalisation parameter is set to $p_{0} = 1.5$ and $\beta$ is set to a moderate value of $\beta_{0} \in [ 8, 32 ]$ depending on the filter radius - the larger the filter radius relative to the element size, the larger $\beta$ is needed to provide the same level of intermediate densities \cite{Guest2004}. During the next stage, the SIMP penalisation parameter is raised to $p_{1} \in [ 3.0, 5.0 ]$ depending on the problem - some microstructural design problems require higher penalisation than others to form 0-1 topologies. For the final stage of the optimisation, the $\beta$ parameter is increased to $\beta_{1} \in [ 32, 128 ]$, again depending on the filter radius. Even though one has not completely eliminated the continuation approaches, the fact that $\beta$ is already set to a moderately high value from the beginning reduces the likelihood of unsupportable features forming during the optimisation process. The next step is introduced after 100 iterations or when the maximum relative change in objective function reaches below $0.001$. After the last step, the optimisation is stopped when a maximum number of iterations is reached or the maximum relative change criteria is met.

\subsection{Multiple blocks} \label{sec:topopt_multiblock}

Restricting the design to be made up of a single periodic microstructure is quite a severe constraint on the design freedom. The obtained designs performance will be far from the unconstrained optimal design. This is due to the fact that the same microstructure is placed throughout all parts of the design domain and the optimised microstructure will thus be one that satisfies optimality in an average sense, but where the parts with the largest sensitivities will dominate the optimisation procedure.
This restriction can be partially alleviated by partitioning the design domain into multiple subdomains, within which a unique locally periodic microstructure is designed.

Smoothly connected microstructures, both within the subdomains and across their interfaces, are obtained by considering the full macroscopic problem and by imposing periodicity and connectivity constraints on the optimised design. This is achieved by using the PDE-based density filter \cite{Lazarov2011}, which easily allows for imposing such constraints on the filtered density field by collapsing the corresponding degrees of freedom in the discrete filter problem.
Figure \ref{fig:filtering_threeblock} illustrates the periodicity and connectivity constraints imposed on the filtered density field for a design with three layers of locally periodic microstructures. The colours indicate which edges are coupled together explicitly by collapsing their degrees of freedom. The top and bottom (red) of all cells are to be connected ensuring both periodicity, in the vertical direction, within the block and connectivity across the interfaces. The left and right sides (blue, green and magenta) of each cell type are to be connected ensuring periodicity within the blocks in the horisontal direction.

The increase in the number of microstructures present in the design, expands the number of unique agglomerates to consider when calculating the eigenmodes for the spectral basis. For instance, considering a double-clamped beam where the design is made up of three layers with each their microstructure, the number of agglomerates increases to 15. There is 3 sets to accommodate the different boundary conditions, as discussed for the single microstructure case in section \ref{sec:theory_periodic}, within which there now exists 5 unique agglomerate types, more specifically one for each layer and one for each interface.

\section{Numerical experiments} \label{sec:numexp}

\subsection{Basis re-utilisation} \label{sec:numexp_basisreuse}

	It is rather trivial to argue that while the design is evolving slowly, which is prevalent during the first and final stages of the presented optimisation procedure, it is possible to reuse the previously computed spectral basis for the iterative solution. 	
	\begin{figure}[t]
	\centering
	\subfloat[Iteration 1]{\includegraphics[trim= 390 90 390 65, clip, width=0.15\textwidth]{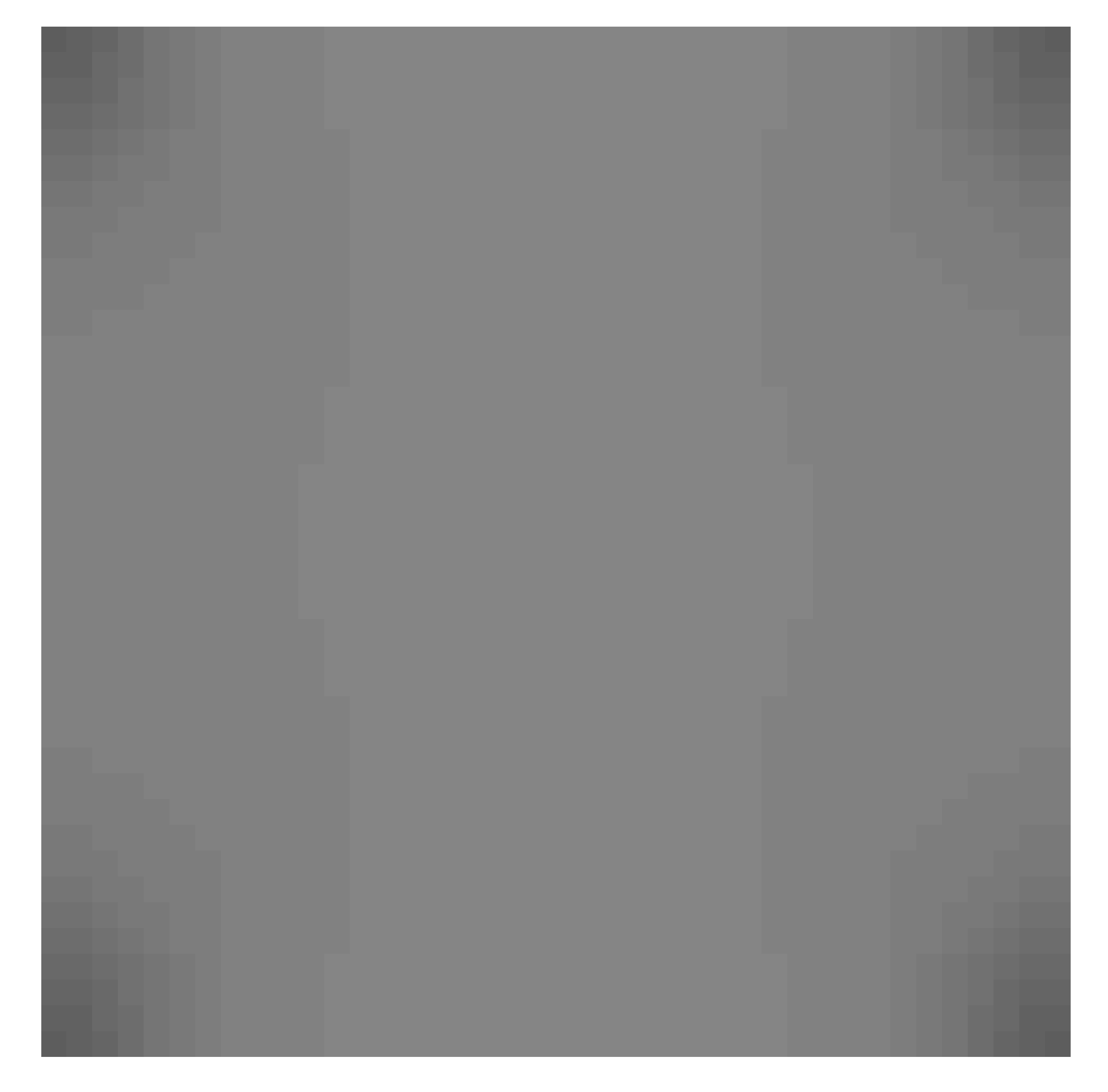}
	\label{fig:basisrecomp_desevolution-a}}
	\hspace*{0.01\textwidth}
	\subfloat[Iteration 10]{\includegraphics[trim= 390 90 390 65, clip, width=0.15\textwidth]{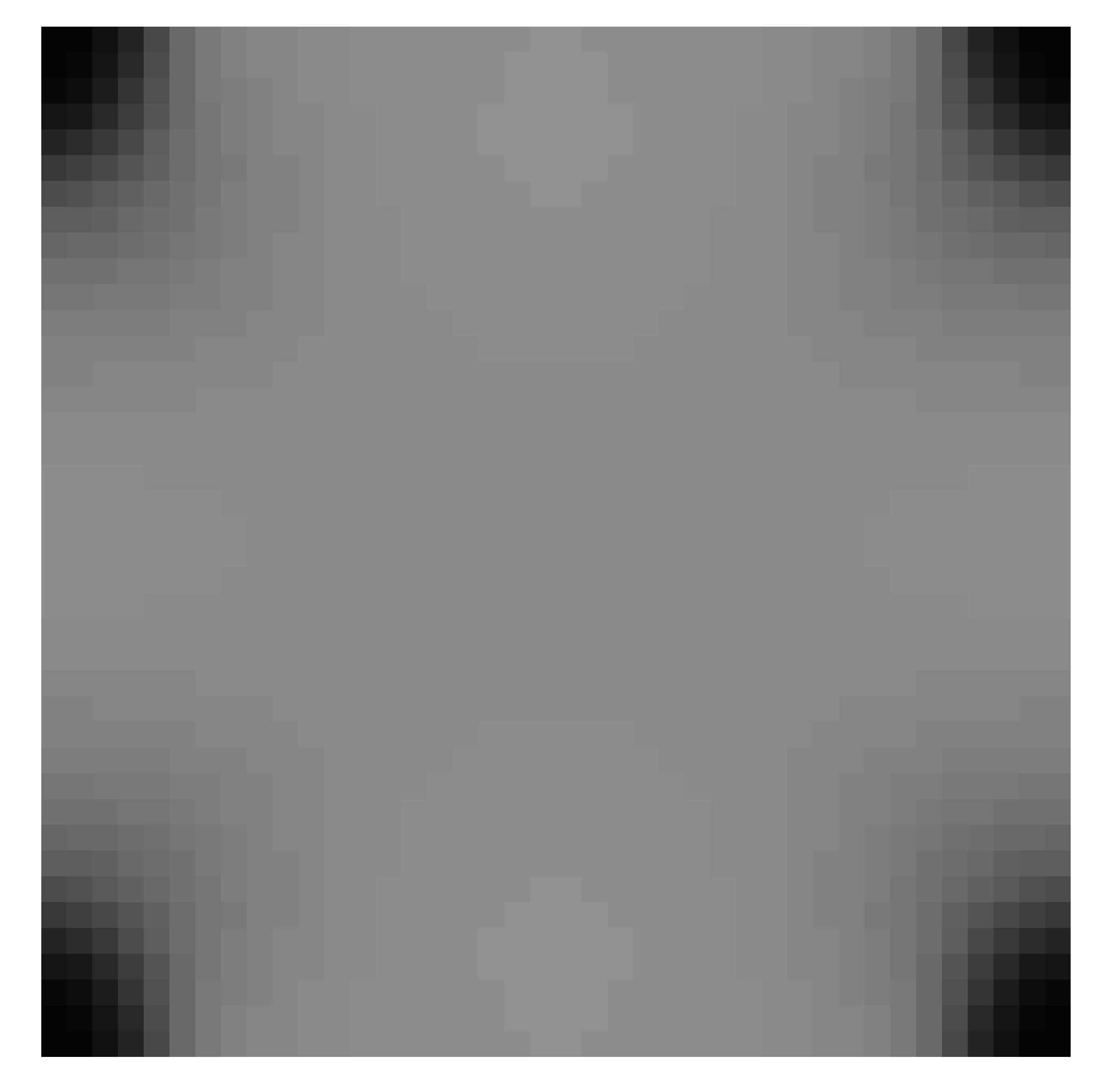}	
	\label{fig:basisrecomp_desevolution-b}}
	\hspace*{0.01\textwidth}
	\subfloat[Iteration 30]{\includegraphics[trim= 390 90 390 65, clip, width=0.15\textwidth]{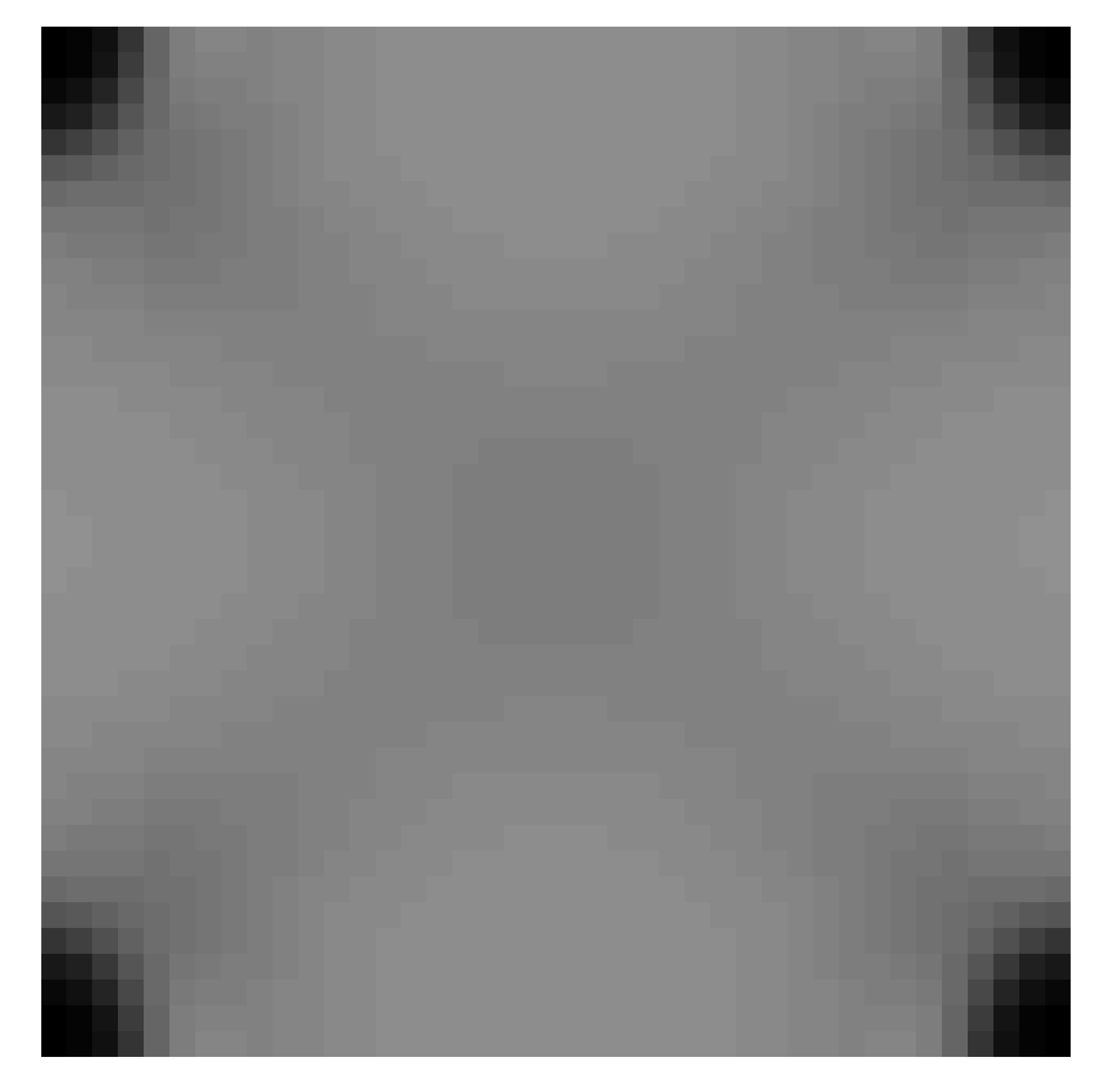}	
	\label{fig:basisrecomp_desevolution-c}}
	\hspace*{0.01\textwidth}
	\subfloat[Iteration 170]{\includegraphics[trim= 390 90 390 65, clip, width=0.15\textwidth]{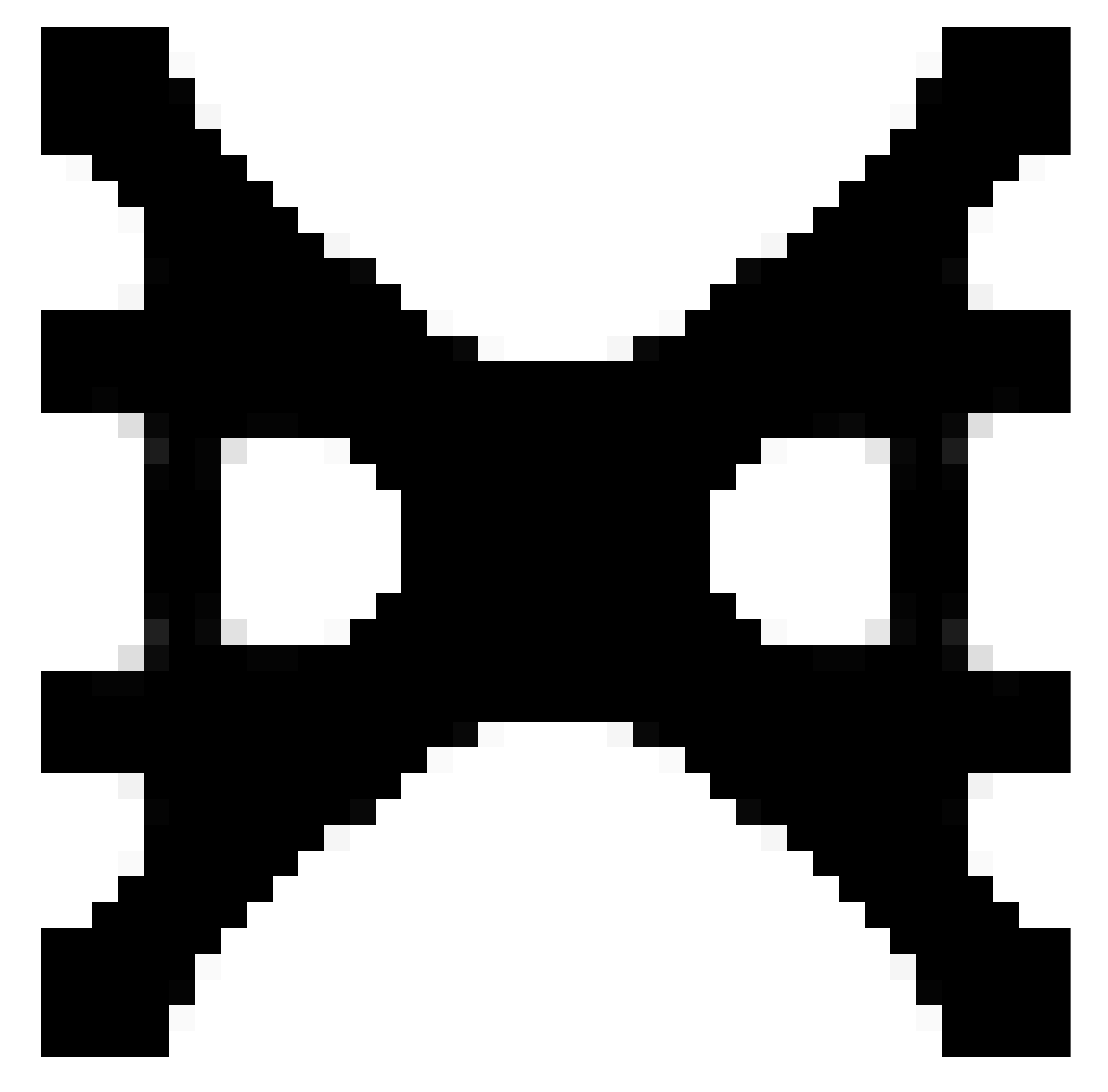}
	\label{fig:basisrecomp_desevolution-d}}
	\hspace*{0.01\textwidth}
	\subfloat[Iteration 190]{\includegraphics[trim= 390 90 390 65, clip, width=0.15\textwidth]{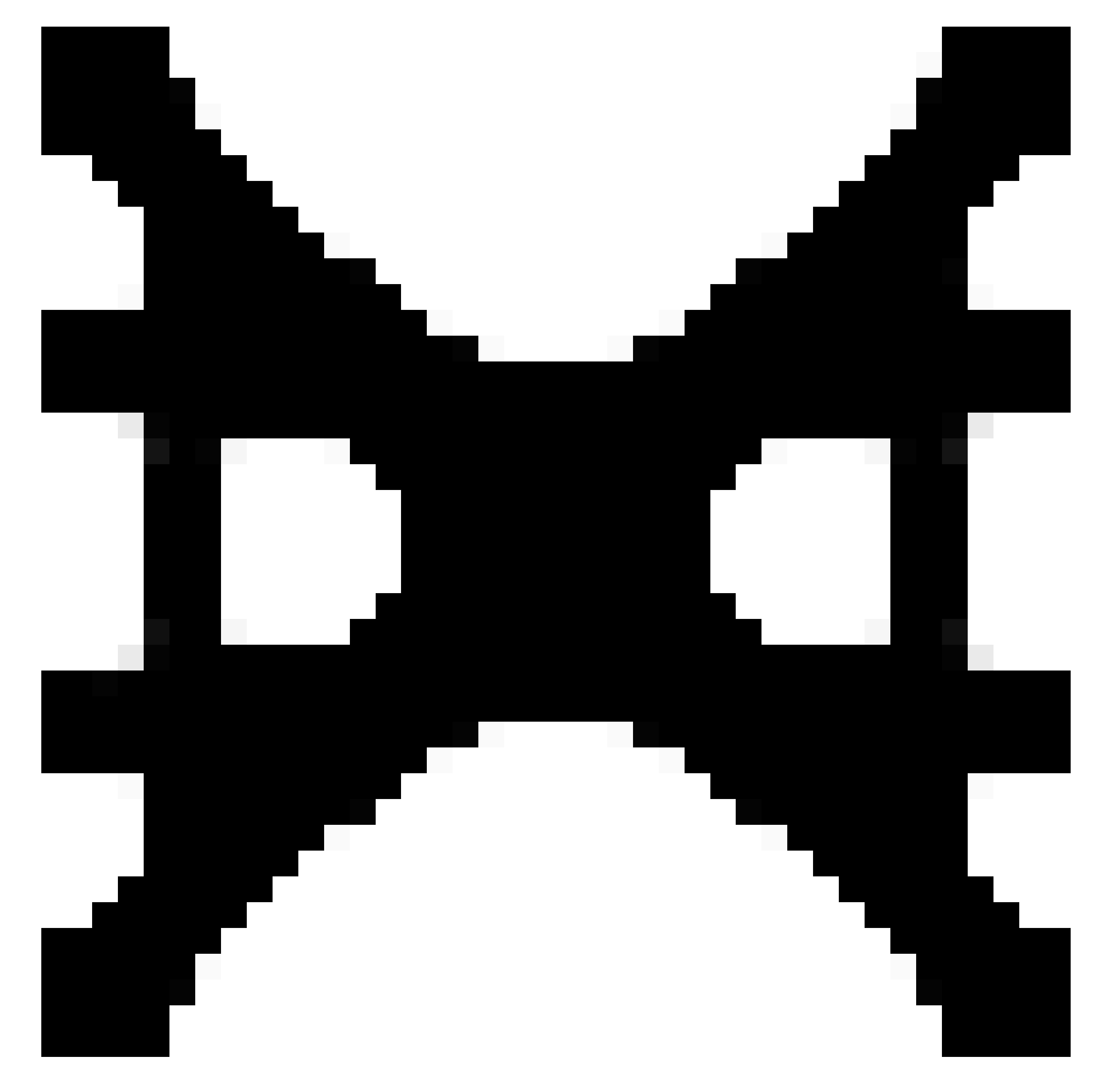}	
	\label{fig:basisrecomp_desevolution-e}}
	\hspace*{0.01\textwidth}
	\subfloat[Iteration 200]{\includegraphics[trim= 390 90 390 65, clip, width=0.15\textwidth]{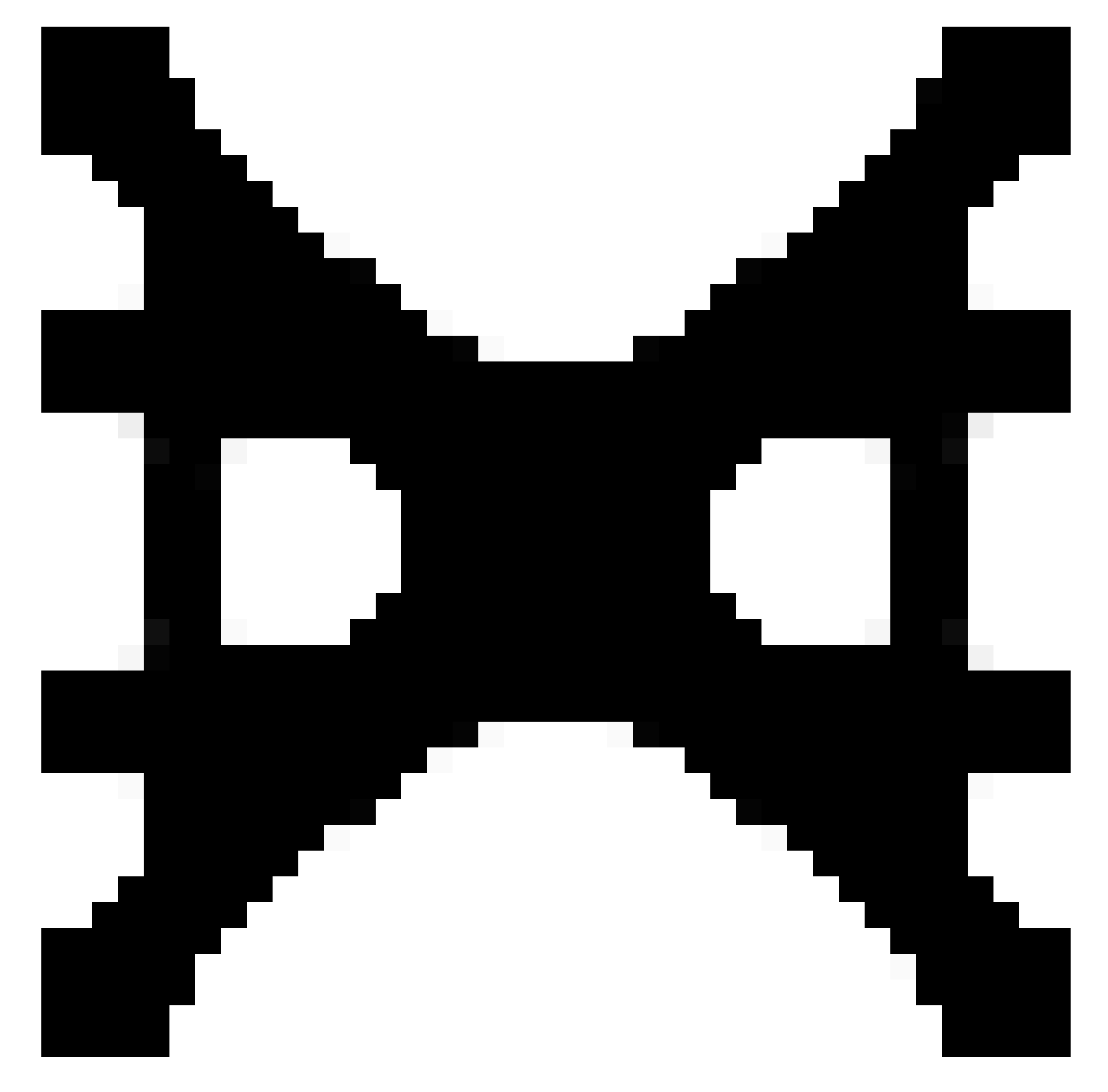}	
	\label{fig:basisrecomp_desevolution-f}}
	\caption{The design distributions for the first and final stages of the optimisation process; design iterations 1, 10, 30, 170, 190 and 200.} \label{fig:basisrecomp_desevolution}
	\end{figure}
	In order to test this hypothesis, the previously described double-clamped beam problem, as shown in figure \ref{fig:msfem_numexp_illustration}, is optimised using different strategies for updating the MsFEM basis during the optimisation process. The beam consists of 4 by 8 coarse cells within which the microstructure is discretised using 40 by 40 elements. The investigation is carried out for a constant $\beta = 64$ using a single realisation with threshold $\eta = 0.5$. The maximum number of design iterations is set to 200 in order to keep the total number of iterations the same for all the different strategies in order to conduct a fair comparison. 
	
Figure \ref{fig:basisrecomp_desevolution} illustrates the relatively slow development of the design during the first and final stages of the optimisation. It can be seen that the design is evolving quite slowly during the first iterations, figures \ref{fig:basisrecomp_desevolution-a} to \ref{fig:basisrecomp_desevolution-c}, due to the low SIMP penalisation value and this supports the arguments that the MsFEM basis does not need to be updated very frequently during this period. Furthermore, it is clearly shown that the design is hardly changing during the final iterations, figures \ref{fig:basisrecomp_desevolution-d} to \ref{fig:basisrecomp_desevolution-f}, where the topology has been determined and the final minor adjustments are being carried out.

	\subsection{Simple update scheme based on GMRES-iterations} \label{sec:numexp_basisupdate}

	In \citep{Lazarov2014} the spectral basis is updated for the agglomerates within which the design is changing more than a specified tolerance. Here a simple heuristic update scheme based on the number of GMRES iterations is used. The spectral basis is updated when the number of GMRES iterations changes more than a given percentage from the design iteration where the basis was previously computed.
	 
	 In order to further speed up the iterative solution of the linear systems of equations, the GMRES solver can be initialised with the displacement solution from the previous design iteration as the initial guess for the GMRES solver for the new design iteration, rather than starting from a zero-vector initial guess. When initialisation of GMRES is used, the heuristic basis update rule is modified so that if the number of GMRES iterations decreases after a basis update, the update threshold is updated to be with respect to the lowest number of GMRES iterations encountered after the previous basis computation.
	 
	 Furthermore, it has been shown in \citep{Amir2012,Amir2014} that the convergence criteria can be relaxed significantly during topology optimisation when using an iterative solver. The reason for this is that one does not necessarily require the displacement solution to be solved to full precision in order for the optimisation to progress correctly. This leads to significant reductions in computation time compared to solving to full precision. For the investigations presented in this section, the stopping tolerance is set to $\varepsilon_{rel} = 10^{-5}$ relative to the preconditioned norm of the forcing vector.
	
A metric that is correlated to the development of the design is the non-discreteness measure, $M_{nd}$, as introduced in \cite{Sigmund2007}:
\begin{equation}
M_{nd} = \frac{ \sum_{i=1}^{n_{d}} 4 \bar{\tilde{\rho}}_{i} (1-\bar{\tilde{\rho}}_{i}) }{ n_{d} }
\end{equation}
$M_{nd}$ is an indication of the amount of intermediate densities and thus an indication of the contrast of the stiffness distribution. When $M_{nd}$ is high, there is large amount of intermediate densities and when $M_{nd}$ is low, there is small amount of intermediate densities. For volume-constrained compliance minimisation where the design will never contain all 0 or all 1 densities, this is equivalent to saying: when $M_{nd}$ is high, there is low contrast and when $M_{nd}$ is low, there is high contrast.
	\begin{figure}
	\centering
\hspace*{-0.016\textwidth}
\includegraphics[width=0.91\textwidth]{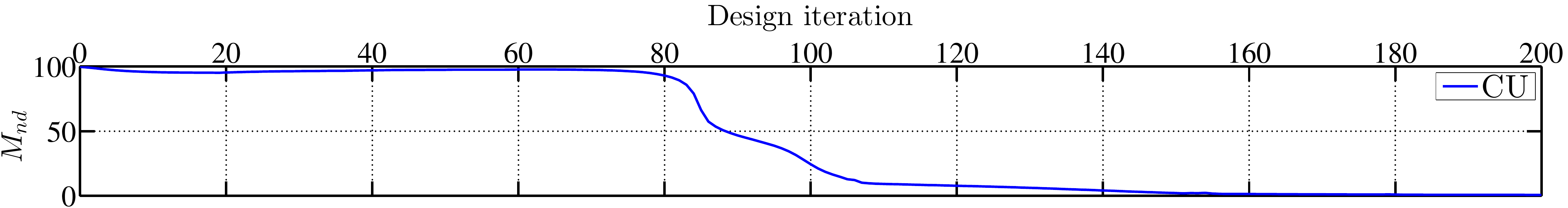} 
	\\
	\includegraphics[width=0.9\textwidth]{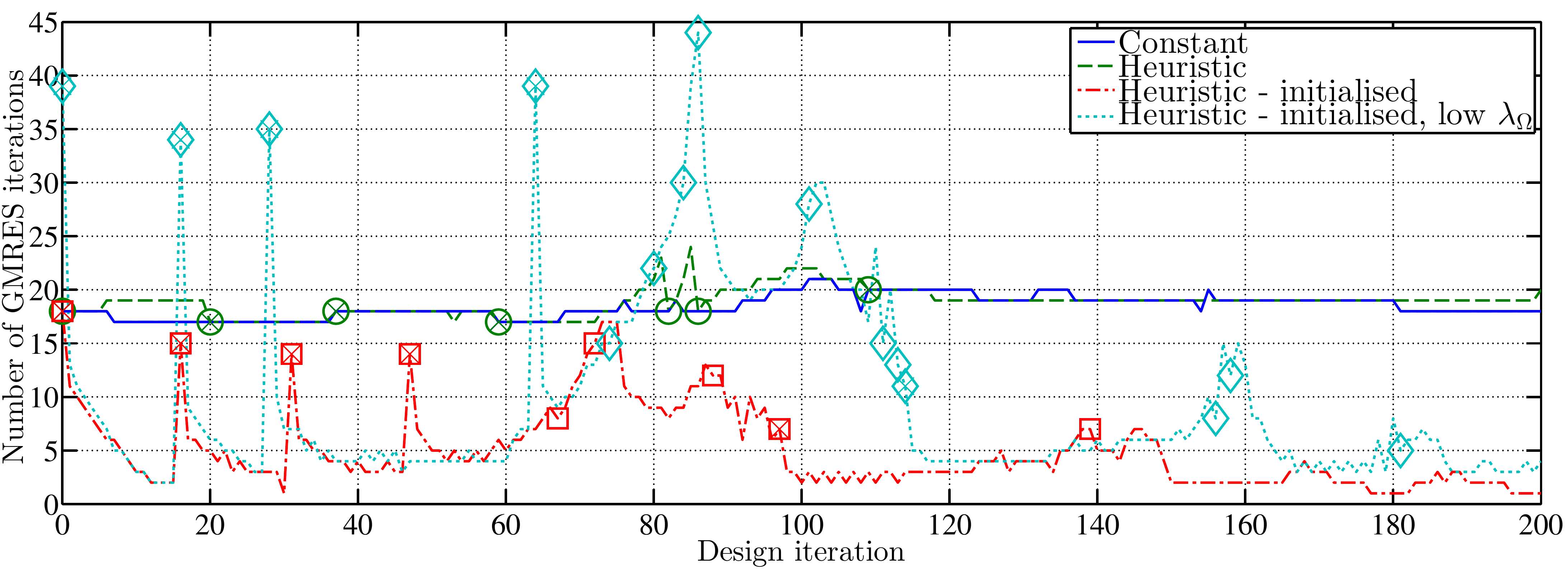}	
	\caption{Development of the non-discreteness measure, $M_{nd}$, and a comparison of the number of GMRES iterations for the different basis computation strategies for a relative tolerance of $\varepsilon_{rel} = 10^{-5}$. The lines show the number of GMRES iterations for the various strategies and the symbols show when the basis is updated. Crosses highlight forced updates due to change in continuation parameter.} \label{fig:basisrecomp_overall} 
	\end{figure}
The top plot of figure \ref{fig:basisrecomp_overall} shows how $M_{nd}$ changes during the optimisation process. It is clearly seen that the level of intermediate densities is relatively constant at the first and final stages of the optimisation process and that the contrast goes from low to high.
The bottom plot of figure \ref{fig:basisrecomp_overall} shows the number of GMRES iterations during the optimisation process for the described strategies. It can be seen that the number of GMRES iterations is more or less constant throughout the optimisation process when the MsFEM basis is recomputed at every design iteration (denoted by ``constant'' in the legend). Keeping the development of $M_{nd}$ in mind, this clearly supports the evidence that the MsFEM preconditioner is contrast-independent for a constant $\lambda_{\Omega}$. This is of course an expensive approach and results in a total time for the entire design process of 768.5 seconds, which is significantly higher as compared to the reference time of 306.5 seconds when using the direct solver.

When adopting the heuristic update rule, the number of GMRES iterations can be seen to remain very close to constant throughout the optimisation process (denoted by ``heuristic'' in the legend). Here it can be seen that the MsFEM basis is only computed a total of 7 times during the 200 design iteration optimisation process, yielding a significantly lowered total time for the entire design process of 444.0 seconds. It is interesting that of the 7 updates, the first is the initial basis computation and 4 are when the penalisation parameter, $p$, is updated which is a requirement made in the algorithm. 

However, this is still significantly slower than the reference design process using the direct solver. Thus, as suggested, the displacement solution from the previous design iteration is used as the initial guess for the GMRES solver for the new design iteration. This significantly decreases the number of iterations needed to obtain the required accuracy when the design is slowly evolving. This is shown in figure \ref{fig:basisrecomp_overall} (denoted by ``heuristic - initialised'' in the legend), where it can be seen that 4-10 GMRES iterations are needed for large parts of the optimisation process, namely the initial and final stages, whereas significantly more are needed in the middle stage where the topology is forming and the design is changing rapidly. 

By lowering the eigenvalue threshold, from $\lambda_{\Omega} = 6.69\times10^{-4}$ to $\lambda_{\Omega} = 4.64\times10^{-5}$, and thus decreasing the basis size, it is possible to decrease the computational work associated with the projection and a single GMRES iteration, as discussed in section \ref{sec:msfem_compcost}. As can be seen in figure \ref{fig:basisrecomp_overall} (denoted by ``heuristic - initialised, low $\lambda_{\Omega}$'' in legend), the required number of GMRES iterations generally increases as expected, however, despite this the resulting optimisation process only takes 298.7 seconds, which is 7.8 seconds faster than the reference time of 306.5 seconds when using the direct solver. This is an interesting observation and shows that the choice of the eigenvalue threshold, $\lambda_{\Omega}$, is hardly trivial. Further investigation into determining the eigenvalue threshold using error estimates is the subject of future research.

\begin{figure}
	\centering
		\begin{tabular}{cc}
			\subfloat[Direct solver - reference case]{\includegraphics[trim= 390 90 390 65, clip, width=0.16\textwidth]{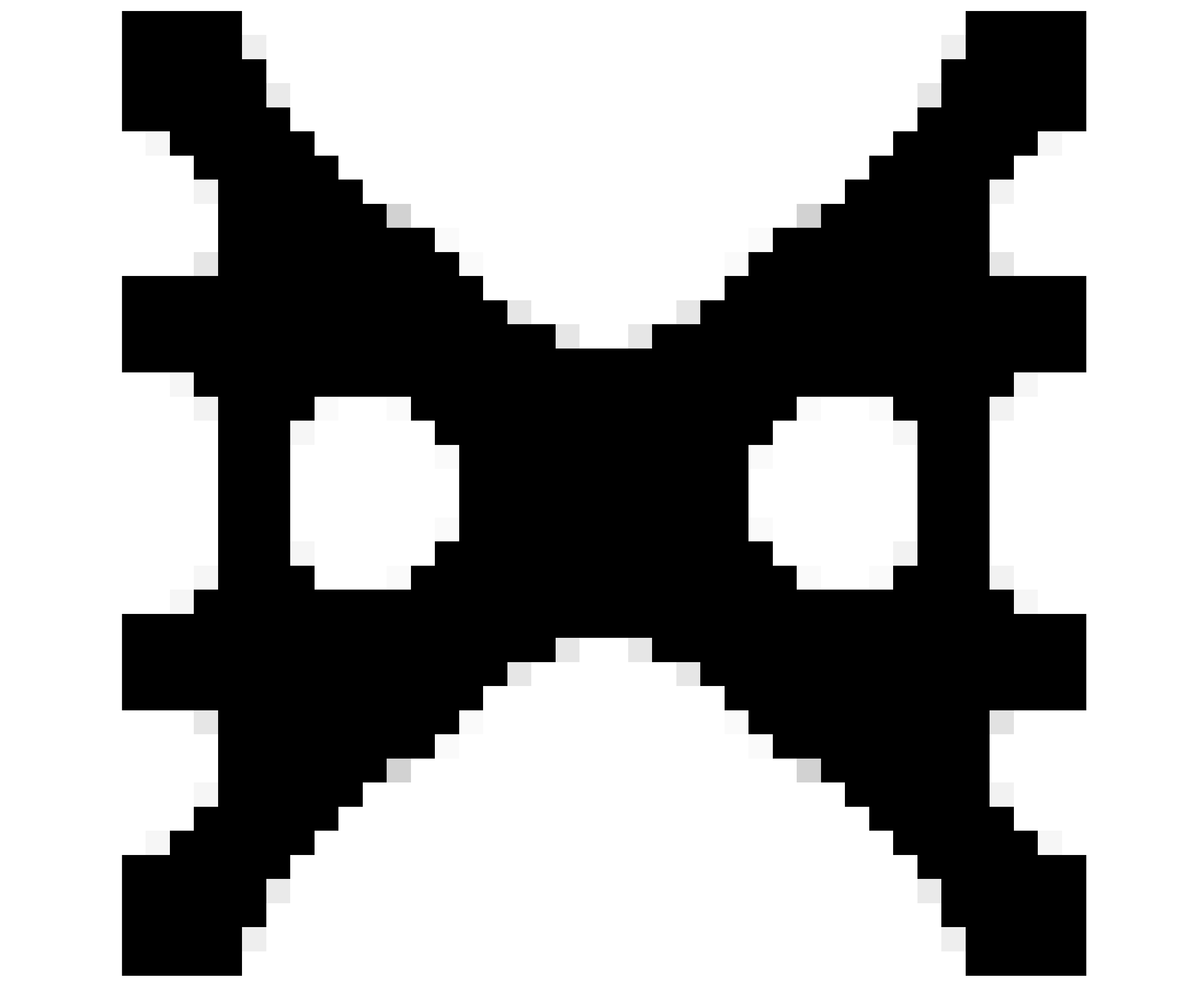}\label{fig:basisrecomp_microandmacro-a}}
			&
			\addtocounter{subfigure}{1}
			\multirow{2}{*}[1.6cm]{
			\subfloat[Deformed macrostructure: Direct solver - reference case]{
			\includegraphics[clip, width=0.6\textwidth]{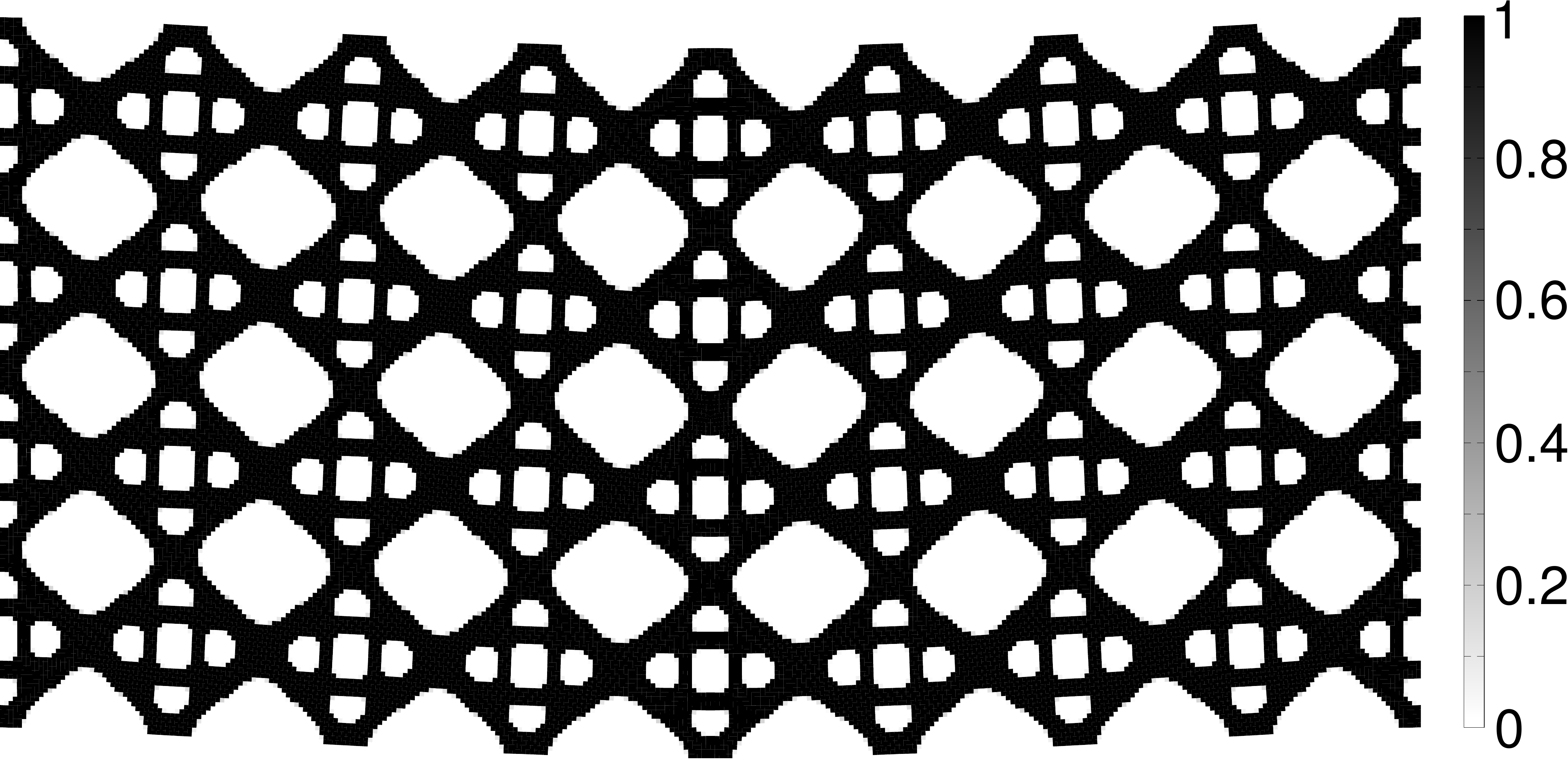}	\label{fig:basisrecomp_microandmacro-b}}
			}
			\\
			\addtocounter{subfigure}{-2}
			\subfloat[Iterative solver - relaxed convergence criteria and seeded initial guess]{\includegraphics[trim= 390 90 390 65, clip, width=0.16\textwidth]{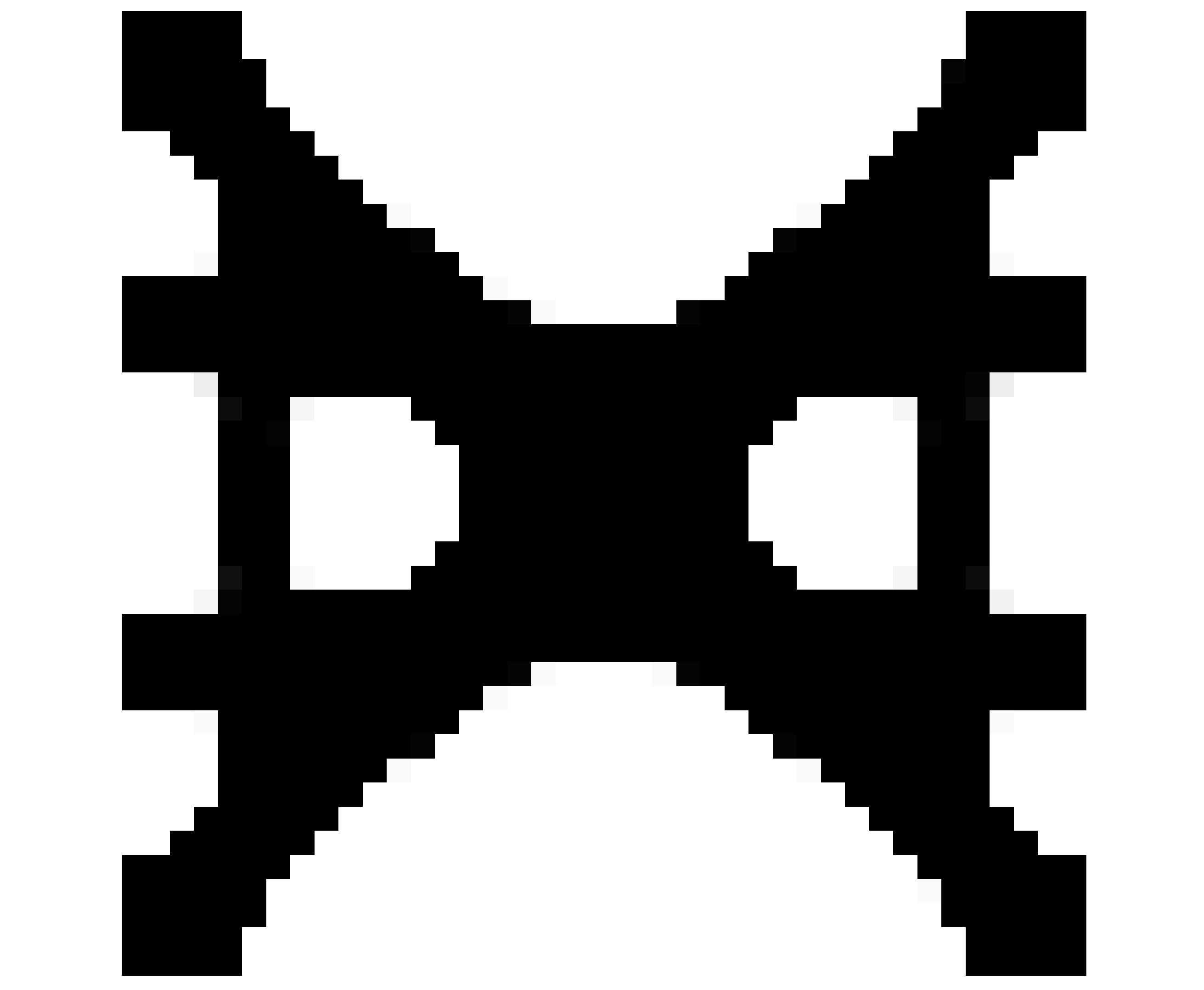}	\label{fig:basisrecomp_microandmacro-c}}
			&
		\end{tabular}
	\caption{Comparison of (a) the reference design by use of a direct solver and (b) the optimised design when using the proposed iterative method with relaxed convergence criteria. The deformed macrostructure is shown in (c) and is made up of 4 by 8 unit cells.} \label{fig:basisrecomp_microandmacro}
\end{figure}
All four final topologies are qualitatively very similar to the reference case and thus only a selected one is shown. The final topology for the optimisation process using the heuristic update rule combined with an initialised GMRES solver (``heuristic - initialised'') is shown in figure \ref{fig:basisrecomp_microandmacro-b} and can be seen to be very similar to the reference design obtained using the direct solver shown in figure \ref{fig:basisrecomp_microandmacro-a}, except for hard corners as compared to rounded corners. 
The final compliance values are as follows; reference using direct solver: $C_{end} = 1.0967\times 10^{-3}$, ``constant'': $C_{end} = 1.0812\times 10^{-3}$, ``heuristic'': $C_{end} = 1.0762\times 10^{-3}$, ``heuristic - initialised'': $C_{end} = 1.0649\times 10^{-3}$, ``heuristic - initialised, low $\lambda_{\Omega}$'': $C_{end} = 1.0787\times 10^{-3}$. It is very interesting to note that all designs obtained using the MsFEM-GMRES approach actually performs better than the reference case, specifically $1.4\%$,  $1.8\%$, $2.9\%$ and $1.6\%$, respectively, for the four cases. However, this is coincidental due to the non-convexity of the optimisation problem and can in no way be guaranteed in general.

	\subsection{Note on application to several thresholds for robust}
        Requiring robustness of the design response with respect to uniform erosion and dilation usually results in designs with similar topologies across the design realisations. This property can be utilised in the construction of the MsFEM preconditioner in order to further reduce the computational cost. Two strategies can easily be identified: 1) an offline building of the basis for the possible variations of the design and using it in an online optimisation process \cite{Efendiev2013a}, and 2) building a basis for a representative design realisation and re-utilising it in the solutions of all other realisations. The first strategy is computationally expensive in the preparatory part (offline computations) compared to the second one, however it is expected to provide faster online computations. The second strategy does not require any additional modifications of the algorithm for building the coarse basis and provides a fast solution for the online optimisation process, as demonstrated later. Therefore, the second strategy is utilised in the examples presented in section \ref{sec:results}.

The mass matrix in the eigenproblem defined, equation \eqref{eq:discrete_aggeigenprob}, acts as a filter which selects modes dominant within the solid region. As the dilated design realisation embeds all other design realisations, the basis for the dilated design will provide a good basis for all other design realisations. Therefore, the representative design is selected to be the dilated one. For small difference between the design realisations, which is the case here, the basis for the dilated design ensures a similar number of iterations for all solutions. If the difference between the design realisation is large, more rigorous selection criteria needs to be derived, which is left for future research.

\section{Microstructural design examples} \label{sec:results}

Throughout the microstructural design examples, the contrast in stiffness is constant and the Young's moduli of the solid and void are $E_{max} = 1$ and $E_{min} = 10^{-9}$, respectively.

\subsection{Double clamped beam with concentrated load} \label{sec:results_doubleclamp}

The first numerical example is the double clamped beam with a concentrated load, shown in figure \ref{fig:msfem_numexp_illustration}, as considered earlier in section \ref{sec:msfem_numexp}. The problem is first investigated using the standard single design topology optimisation formulation and secondly using the robust formulation. The problem is optimised for an increasing number of unit cells in order to investigate the convergence of the compliance, as well as to compare the speed of the MsFEM-GMRES iterative solver, combined with the proposed basis re-utilisation scheme, to using the standard \textsc{Matlab} direct solver. In order to make a fair comparison between the two, all of the timing runs have been executed on the same computer, where \textsc{Matlab} has been restricted to use a single computational thread. The computer is a Dell Precision T7500 with two Intel Xeon X5650 CPUs and 96GB memory, running CentOS 6.4 and \textsc{Matlab} 2013b.

\subsubsection{Single design topology optimisation}

The problem is first investigated using the standard single design topology optimisation formulation. The following parameters are used: $r_{min} = 4h$, $v_{f} = 0.5$, $\beta = 64$, $\eta = 0.5$, $\lambda_{\Omega} = 6.5\times10^{-4}$, $\varepsilon_{rel} = 10^{-6}$.
	
	\begin{figure}
	\centering
	\begin{tabular}{ccc}
		\subfloat[$M_{x} = 2$]{\includegraphics[clip, width=0.15\textwidth]{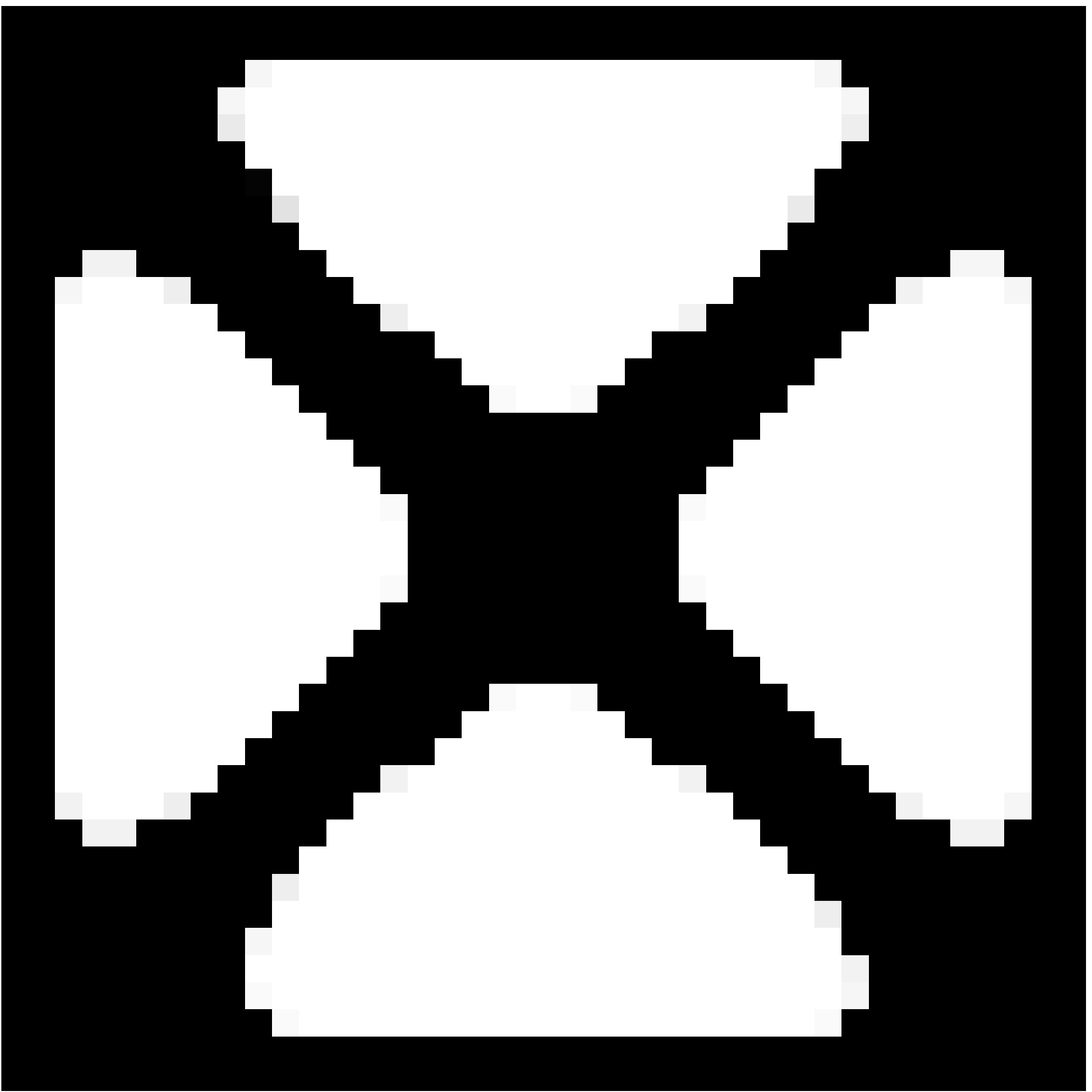}	\label{fig:doubleclamped_singleproj-a}}
		&
		\subfloat[$M_{x} = 4$]{\includegraphics[clip, width=0.15\textwidth]{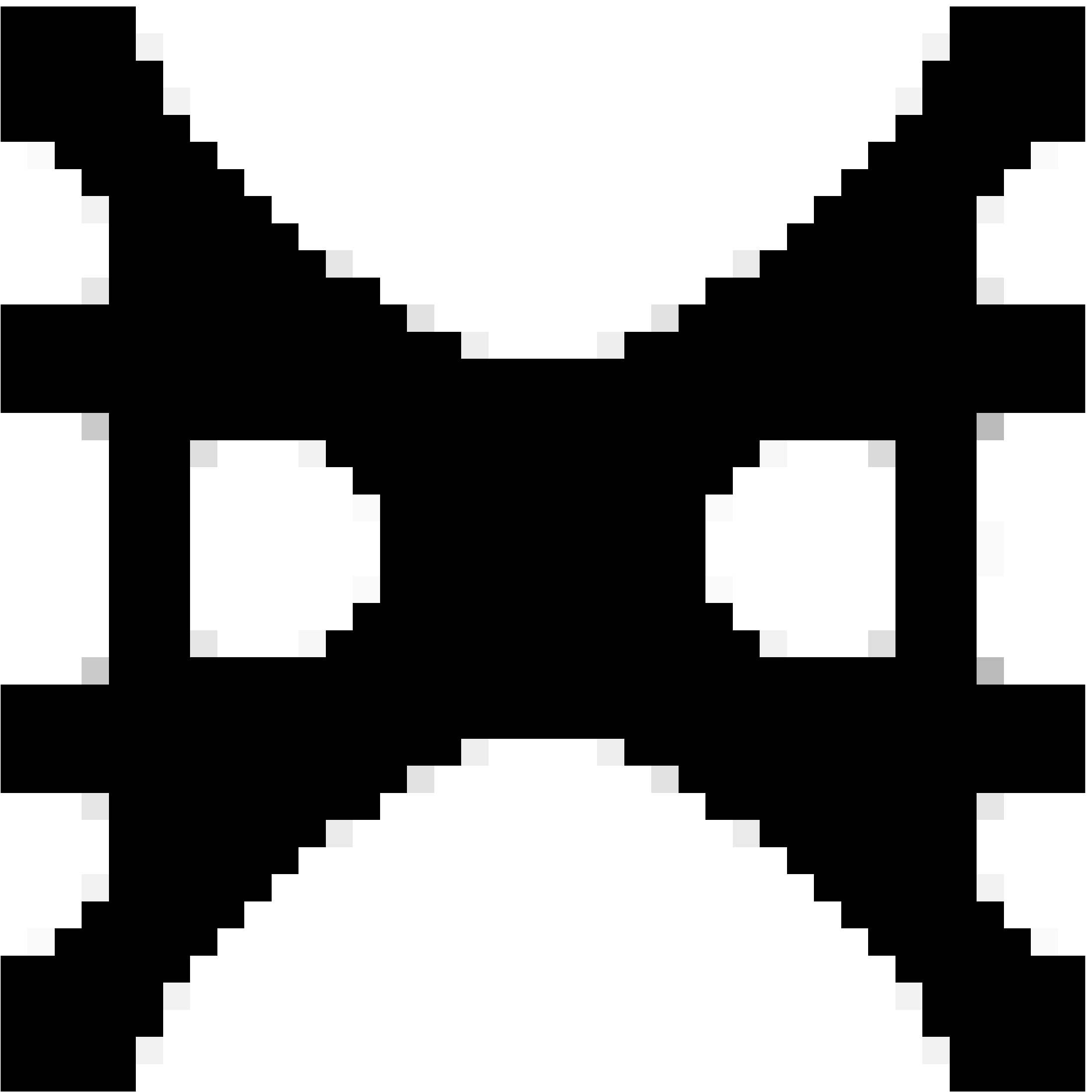}	\label{fig:doubleclamped_singleproj-b}}
		&
		\addtocounter{subfigure}{2}
		\multirow{2}{*}[2.3cm]{
		\subfloat[$M_{x} = 64$]{\includegraphics[clip, width=0.6\textwidth]{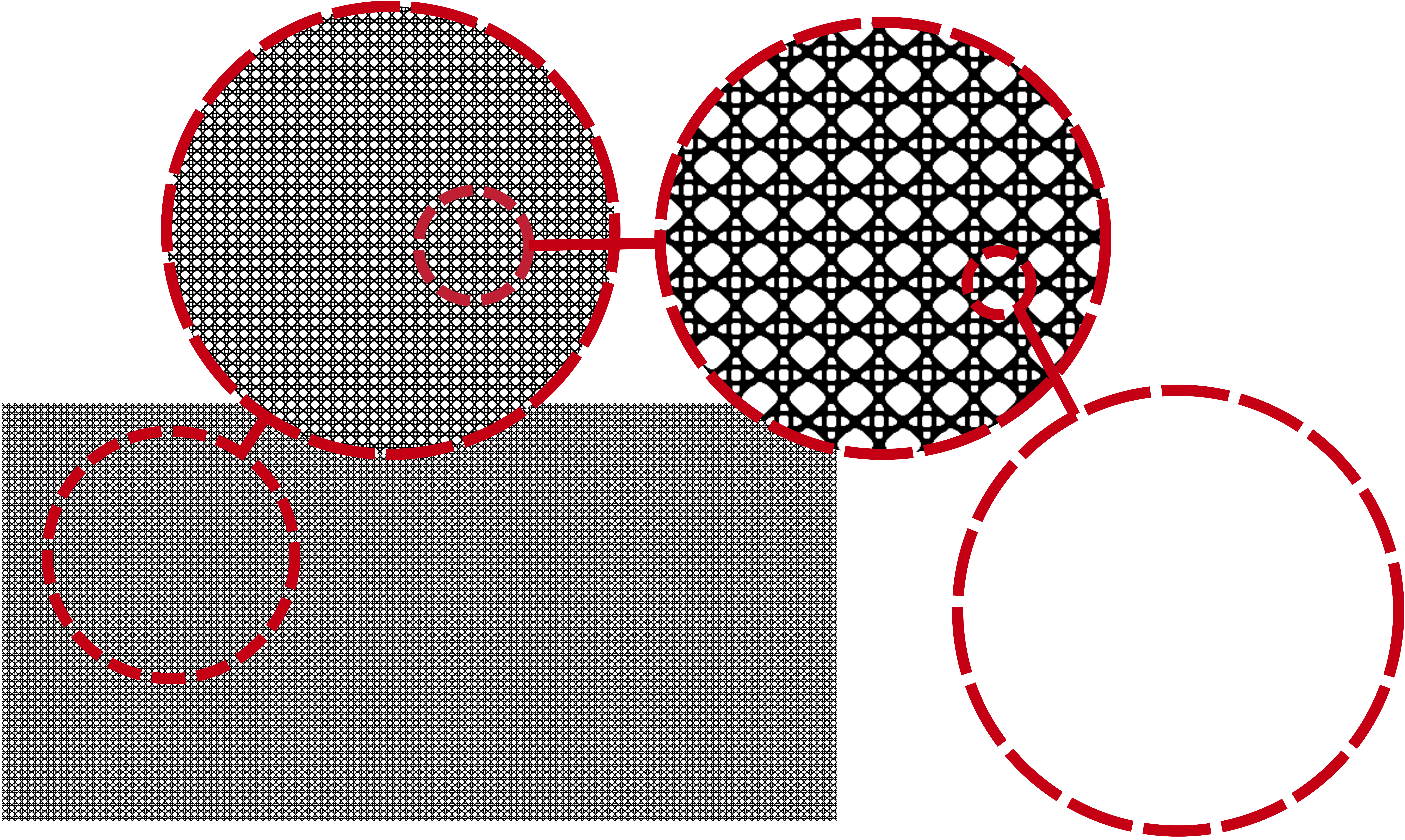}	\label{fig:doubleclamped_singleproj-e}}
		}
		\\
		\addtocounter{subfigure}{-3}
		\subfloat[$M_{x} = 8$]{\includegraphics[clip, width=0.15\textwidth]{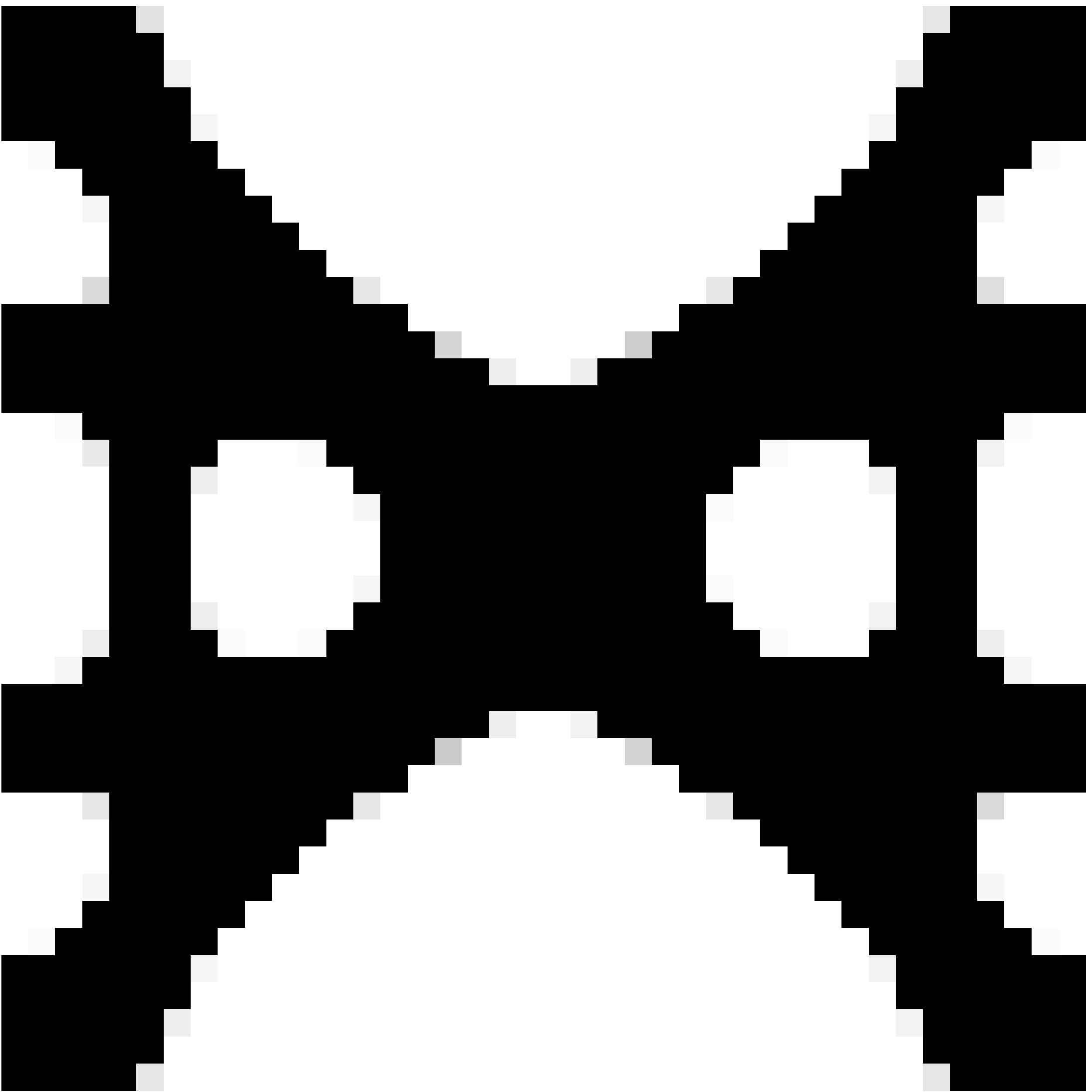}	\label{fig:doubleclamped_singleproj-c}}
		&
		\subfloat[$M_{x} = 16$]{\includegraphics[clip, width=0.15\textwidth]{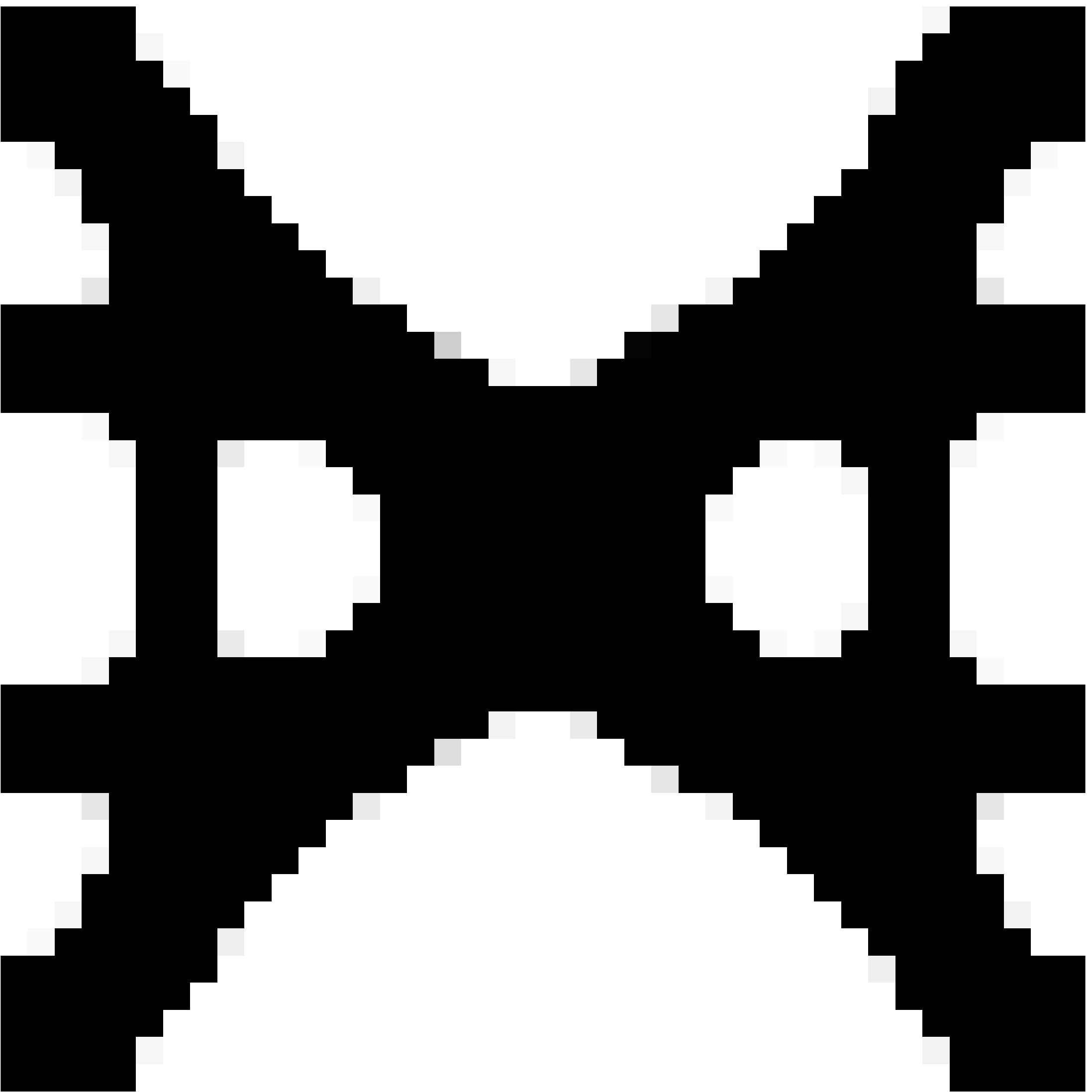}	\label{fig:doubleclamped_singleproj-d}}
		&
	\end{tabular}
	\caption{Optimised microstructures for the double-clamped beam with a concentrated load. The microstructure is shown for varying number of cells across the height, $M_{x} \in \{ 2,4,8,16 \} $, and the full macrostructure is shown for $M_{x} = 64$.} \label{fig:doubleclamped_singleproj}
	\end{figure}
Figure \ref{fig:doubleclamped_singleproj} shows the optimised periodic microstructures for the double-clamped beam with a concentrated load in the centre. The microstructure is shown for increasingly smaller unit cell sizes. The number of unit cells across the height of the beam is characterised by $M_{x} = B/H$, which will be used throughout the paper. It is observed that the microstructure converges very fast and the overall topology does not qualitatively change for $M_{x} > 4$, therefore only a select few have been shown out of the investigated set, $M_{x} \in \{ 2,4,6,8,12,16,32,64 \} $. This trend was also observed in \citep{Zuo2013} for the same problem, although the optimised designs differ. This difference in topology is likely due to a difference in the effective length scale of the current approach and that of \citep{Zuo2013}; the former using density filtering combined with a projection at $\eta = 0.5$ and the latter using BESO.

For the optimised structures shown in figure \ref{fig:doubleclamped_singleproj}, it is observed that the compliance increases as the number of unit cells is increased. This is attributed to the increased restriction of the design freedom imposed through the increased periodicity and is also observed in \citep{Zuo2013}. It is also seen that the MsFEM-GMRES approach gives qualitatively the same topologies as when using a direct solver, however, slight deviations are observed in the compliance values.

From the numerical experiments of this section, it is also observed that in terms of average time per design iteration, the direct solver approach is faster for smaller problems, but that the MsFEM-GMRES approach becomes faster for larger problems, $M_{x} > 6$, as expected. For $M_{x} = 16$ the observed average times are 59.44 and 83.41 seconds for the MsFEM-GMRES and direct approaches, respectively. The relative difference is $28.7\%$ and thus, the savings in time are quite substantial. Using a linear fit, the projected reduction in time for $M_{x} = 32$ and $M_{x} = 64$ is $41.0\%$ and for $54.3\%$, respectively.  However, for $M_{x} = 64$ it has been observed in practice that it takes approximately 26 hours for 200 design iterations. This yields a significantly lower average time per design iteration than the projected, yielding an expected reduction in time of $79.2\%$. Furthermore, a total time of 26 hours is quite impressive when considering that the optimisation was performed using \textsc{Matlab} restricted to a single computational thread and taking into account that for $M_{x} = 64$, the total number of elements is $13,\negthinspace107,\negthinspace200$ and the total number of degrees of freedom is $26,\negthinspace229,\negthinspace762$. It should be noted that this problem was also solvable using a direct solver on the same machine, but the time taken for a single design iteration was so immense that the job was not allowed to finish.

\subsubsection{Robust topology optimisation}

The problem is secondly investigated using the robust topology optimisation formulation. The following parameters are used: $r_{min} = 4h$, $v_{f} = 0.5$, $\beta_{0} = 16$, $\beta_{1} = 64$, $p_{1} = 5$, $\eta = 0.3$, $\lambda_{\Omega} = 6.5\times10^{-4}$, $\varepsilon_{rel} = 10^{-6}$.

	\begin{figure}[b]
	\centering
	\subfloat[$M_{x} = 2$]{\includegraphics[clip, width=0.16\textwidth]{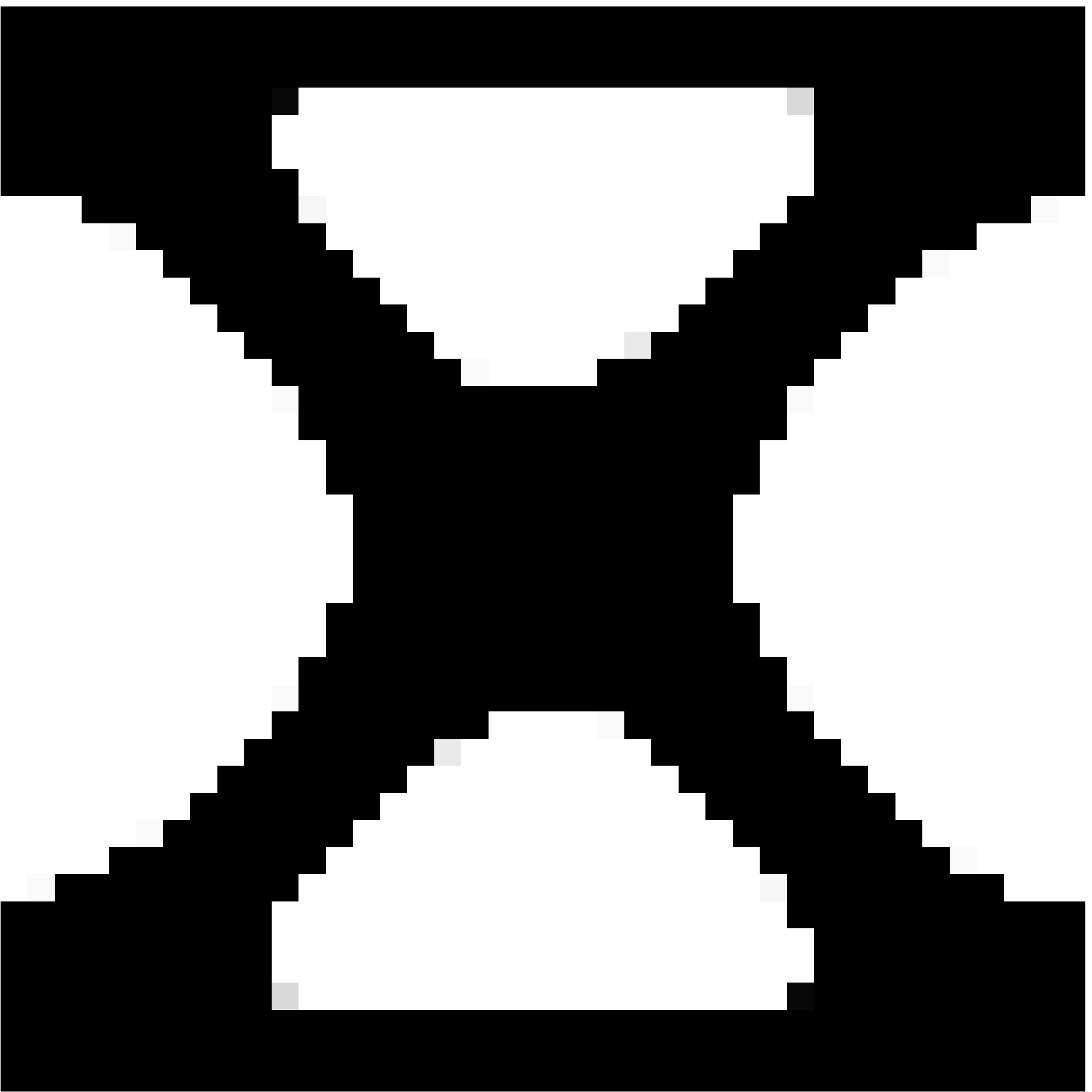}
	\label{fig:doubleclamped_robust-a}}
	\hspace*{0.04\textwidth}
	\subfloat[$M_{x} = 4$]{\includegraphics[clip, width=0.16\textwidth]{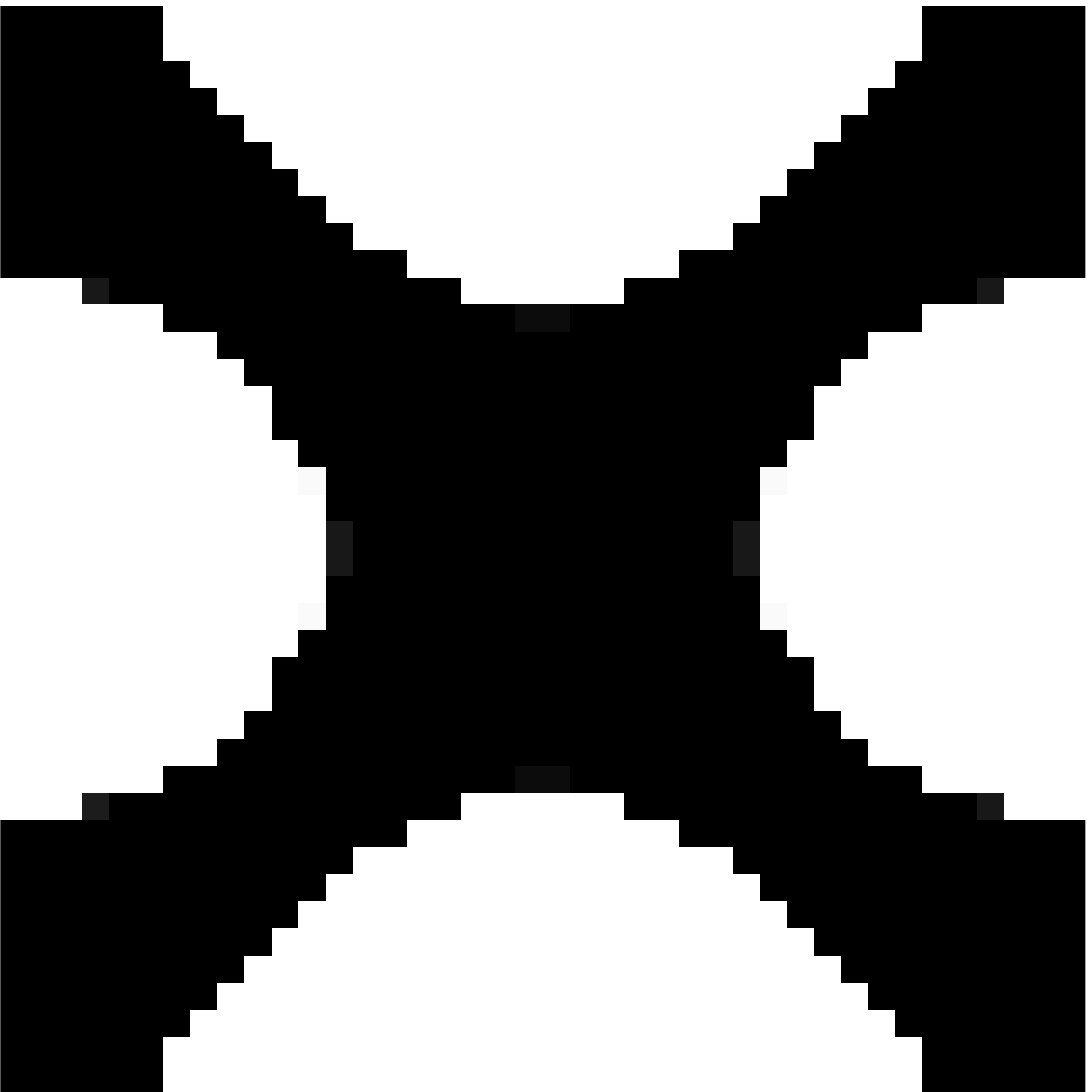}	\label{fig:doubleclamped_robust-b}}
	\hspace*{0.04\textwidth}
	\subfloat[$M_{x} = 16$]{\includegraphics[clip, width=0.16\textwidth]{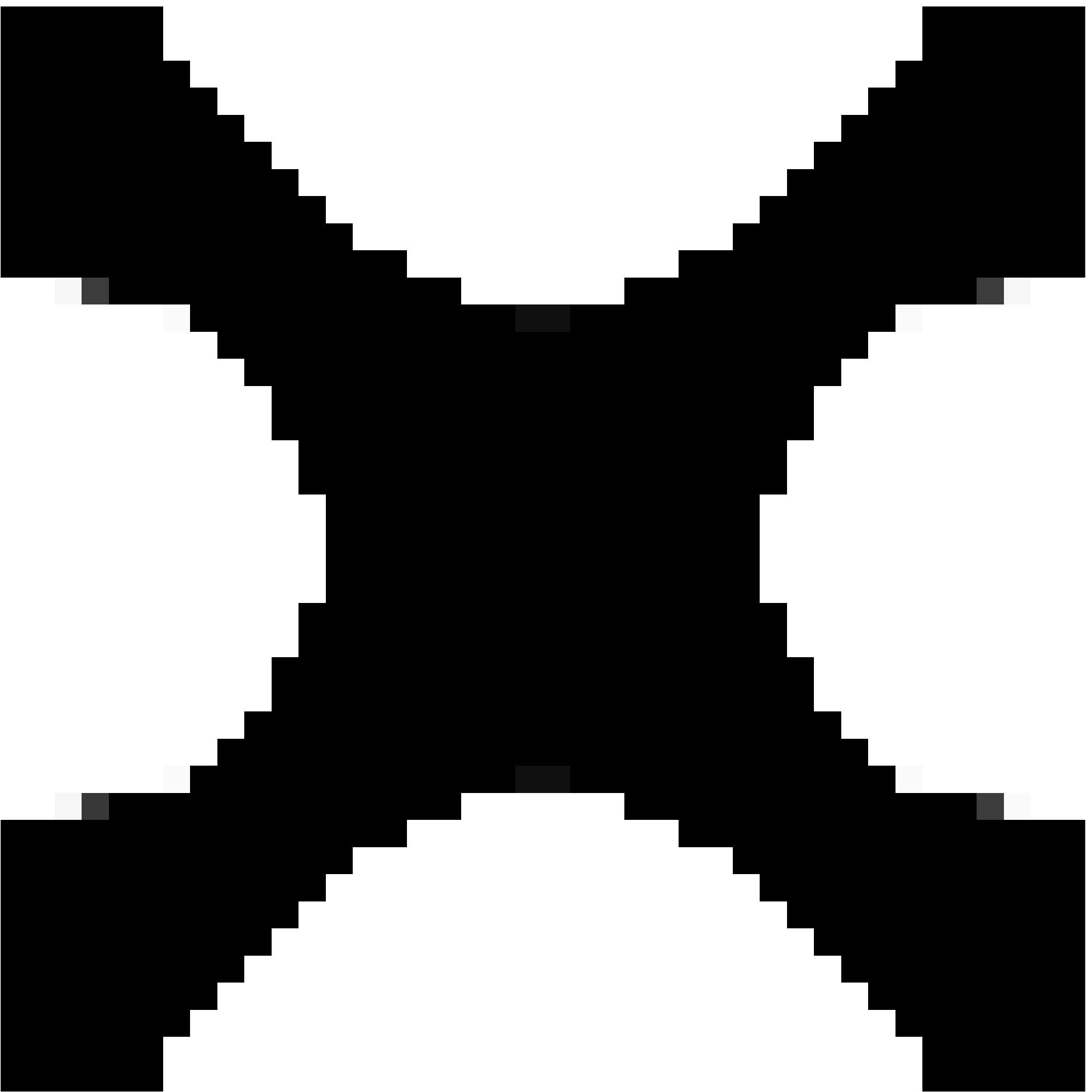}	\label{fig:doubleclamped_robust-f}}
	\caption{Optimised robust microstructures for the double-clamped beam with a concentrated load. The microstructures are shown for varying number of cells across the height, $M_{x} = \{ 2,4,16 \} $.} \label{fig:doubleclamped_robust}
	\end{figure}
Figure \ref{fig:doubleclamped_robust} shows the optimised robust periodic microstructures for the double-clamped beam with a concentrated load in the centre. As for the single design case, it can be seen that the microstructure converges very fast and the overall topology does not qualitatively change for $M_{x} > 4$, therefore only a select few have been shown out of the investigated set, $M_{x} = \{ 2,4,6,8,12,16 \} $. It is interesting to note, that the robust topologies are very similar to the optimised topologies presented in \citep{Zuo2013}.
	\begin{figure}
	\centering
	\subfloat[Eroded]{\includegraphics[clip, width=0.16\textwidth]{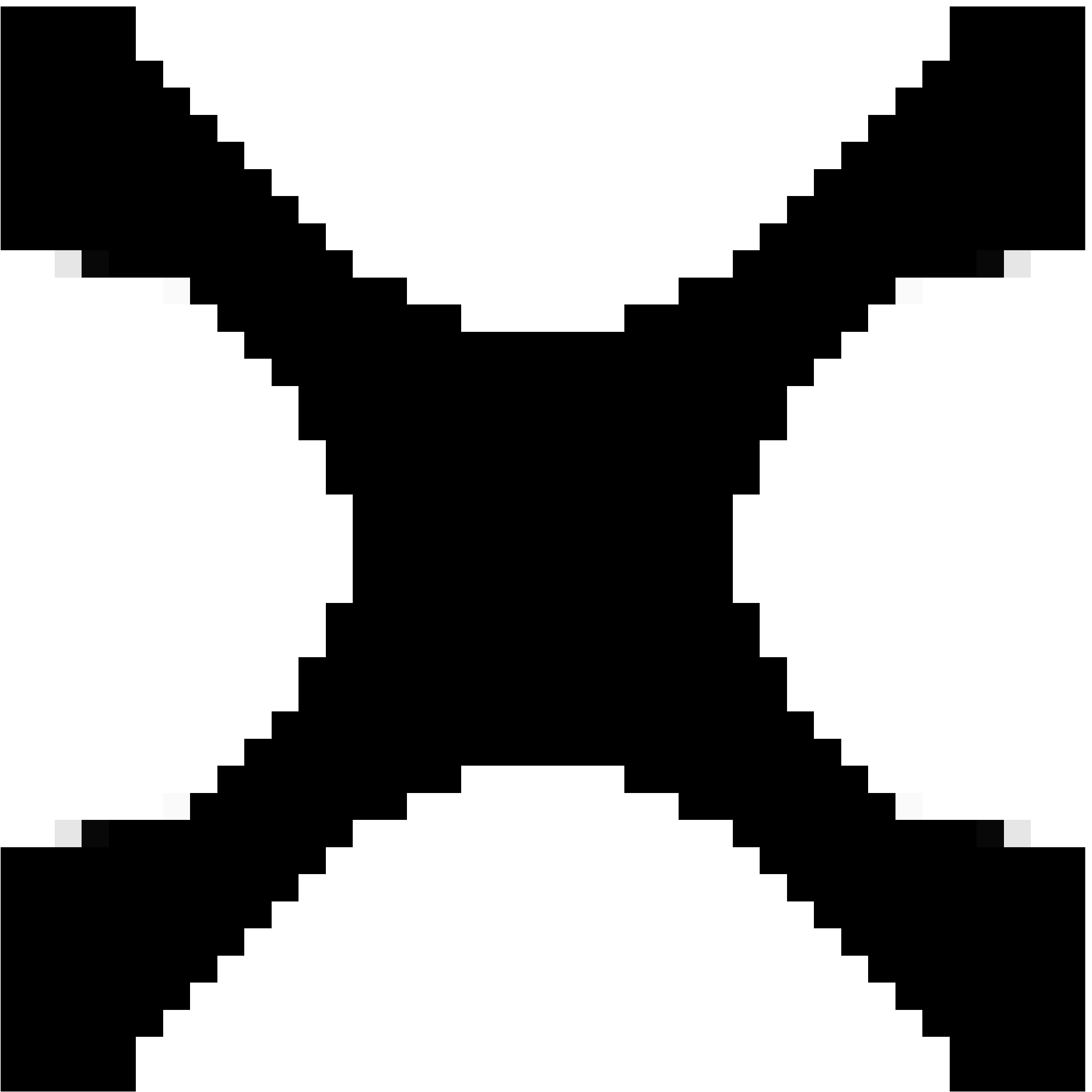}	\label{fig:doubleclamped_robust_realisations-a}}
	\hspace*{0.04\textwidth}
%\\
	\subfloat[Intermediate]{\includegraphics[clip, width=0.16\textwidth]{doubleclamped_robust_Mx16_PNx40_microInt_bw}	\label{fig:doubleclamped_robust_realisations-d}}
	\hspace*{0.04\textwidth}
%	\\
	\subfloat[Dilated]{\includegraphics[clip, width=0.16\textwidth]{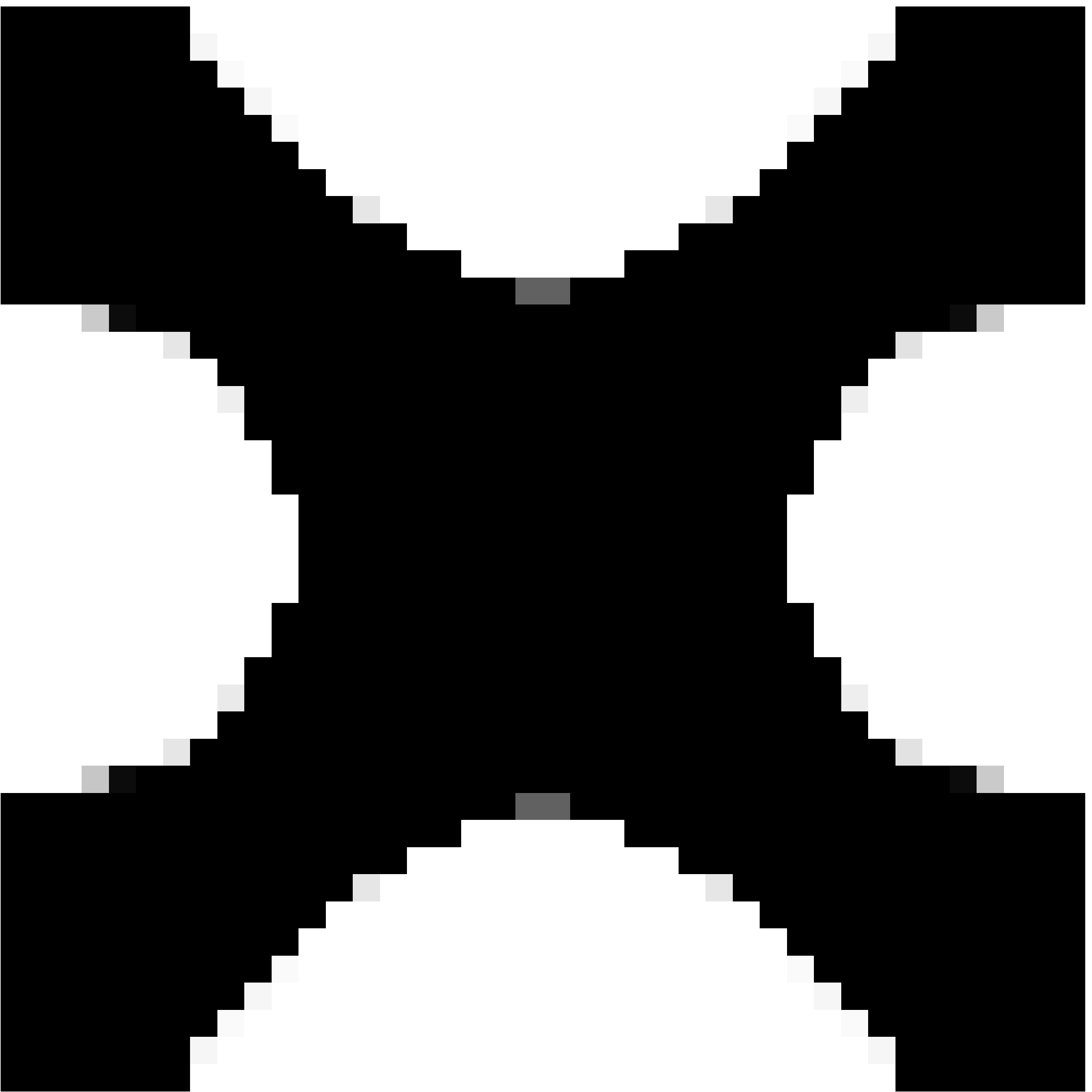}	\label{fig:doubleclamped_robust_realisations-c}}
							
	\caption{The three realisations of the optimised robust microstructures for the double-clamped beam with a concentrated load and $M_{x} = 16$.} \label{fig:doubleclamped_robust_realisations}
	\end{figure}
The main difference is that the topologies presented in figure \ref{fig:doubleclamped_robust} have been optimised to be robust with respect to erosion and dilation of the design, which can be seen in figure \ref{fig:doubleclamped_robust_realisations} for $M_{x} = 16$. Comparing the non-robust and robust topologies, figures \ref{fig:doubleclamped_singleproj} and \ref{fig:doubleclamped_robust} respectively, it can be observed that the small horisontal and vertical bars in the left- and right-hand side of the unit cell of the non-robust design are not present in the robust designs. This is due to the length scale imposed by the robust formulation, as opposed to the lack of length scale for the non-robust result.

As for the non-robust case, it is observed that the compliance increases as the number of unit cells is increased. 
It is also seen that in terms of average time (total for all three realisations) per design iteration, the direct solver approach is faster for smaller problems, but the MsFEM-GMRES approach becomes faster for larger problems, $M_{x} > 8$. 
For $M_{x} = 16$ the observed average times were 188.91 and 246.68 seconds for the MsFEM-GMRES and direct approaches, respectively. The relative difference is $23.4\%$ and thus, the savings in time are also quite substantial for the robust case. Using a linear fit, the projected reduction in time for $M_{x} = 32$ and $M_{x} = 64$ is $36.8\%$ and $49.5\%$, respectively. 

	\begin{figure}
	\centering
	\includegraphics[clip, width=0.9\textwidth]{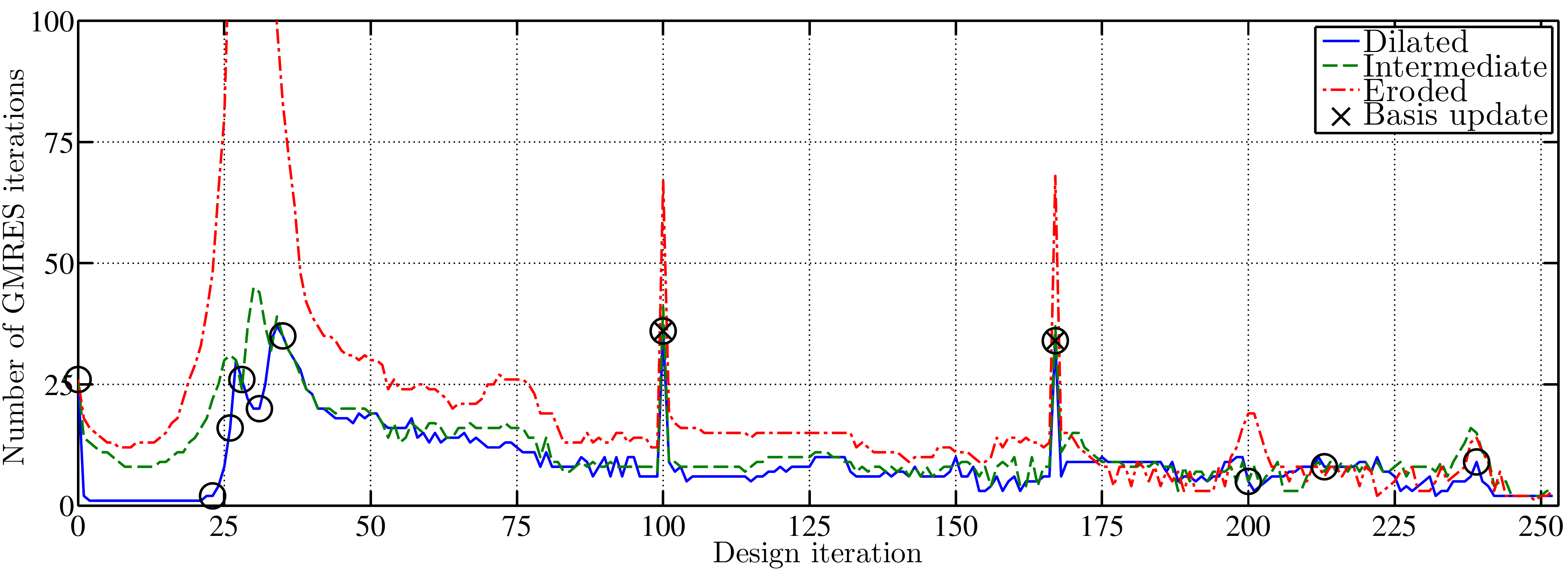}	
	\caption{The number of GMRES iterations as a function of design iteration for the double clamped beam with a concentrated load using the robust formulation and $M_{x} = 16$.} \label{fig:doubleclamped_robust_itervsGMRES}
	\end{figure}
The observed cross-over point is higher for the robust case, which likely is due to a higher computational cost per design iteration. This higher computational cost is associated with the fact that the spectral coarse basis is formed for the dilated design and applied to all three realisations of the design. As expected, the iterative solution of the intermediate and eroded designs, using the spectral coarse basis based on the dilated design, needs more iterations to converge to the specified tolerance.
For the adopted approach of using a moderately high $\beta_{0}$, the three realisations are significantly different in the initial phase of the optimisation process when the topology is being formed. It is in this phase, where large differences in iteration numbers are observed, especially for the eroded case.
Figure \ref{fig:doubleclamped_robust_itervsGMRES} shows the number of GMRES iterations during the optimisation process for $M_{x} = 16$. It can be seen that around design iterations 25-35, the number of GMRES iterations for the eroded design peaks at over 100. It is observed that the peak in GMRES iterations occurs at the same time as the designs are changing very quickly. 
When the topology has formed and the three realisations are similar, the performance can be seen to become much better and the difference in the number of GMRES iterations is relatively small. One could argue that it would be better to adopt the usual $\beta$-continuation, since these problems are not observed for this approach, because the realisations are very similar throughout the optimisation process. However, it has been observed that the presented approach needs significantly fewer design iterations to produce a competitive design and the approach is thus still faster, even with the increased computational time associated with the mismatch of the realisations. Improvement of the spectral basis for robust topology optimisation is a subject of future research.

\subsection{Cantilever beam with distributed load} \label{sec:results_cantidist}
	
\begin{figure}[t]
\centering
\includegraphics[width=0.6\textwidth]{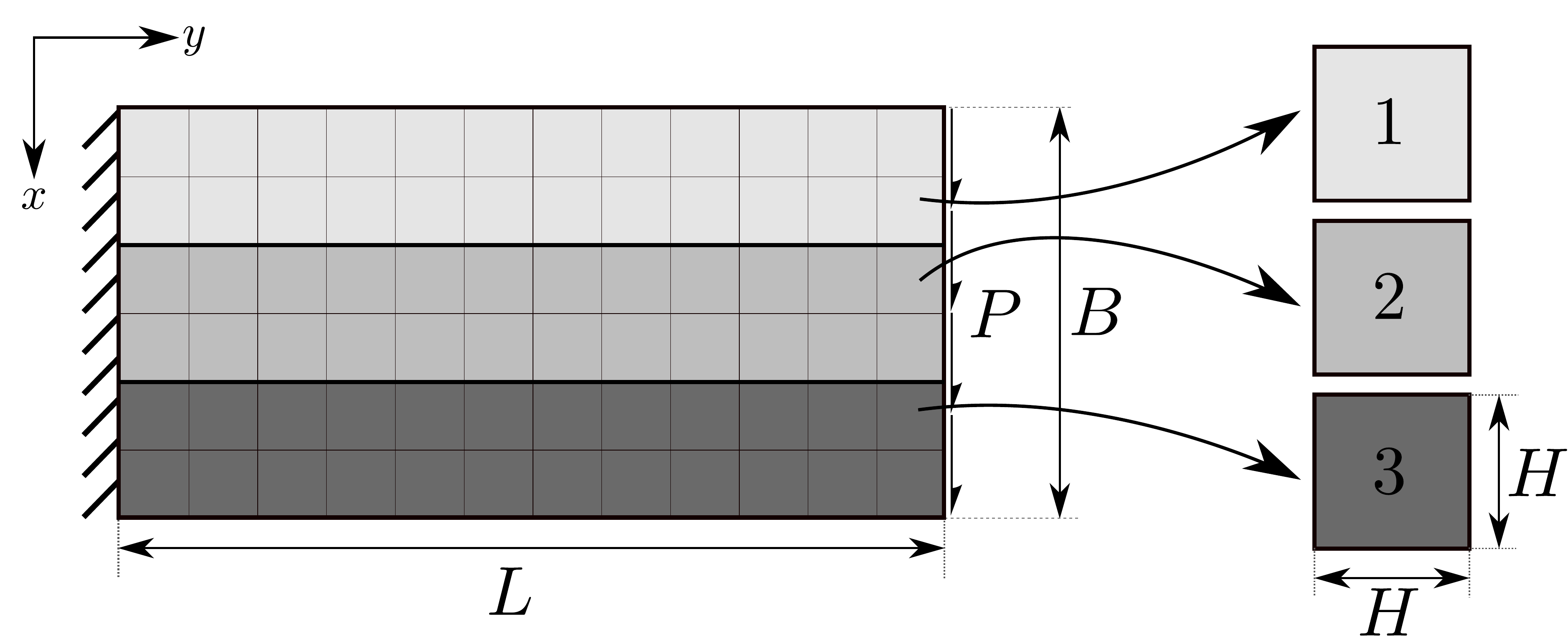}
\caption{Illustration and dimensions of a cantilever beam subjected to a distributed force along the right-hand side. The beam is made up of three layers containing separate periodic microstructures with square unit cells.} \label{fig:cantilever_numexp_illustration}
\end{figure}
In order to demonstrate the methodology for design domains with multiple microstructural layers, a cantilever, as shown in figure \ref{fig:cantilever_numexp_illustration}, is investigated. The cantilever beam is subjected to a distributed force in the positive $x$-direction along the right-hand side. The height of the beam is set to $B = 1$, the length to $L=2$ and the size of the force is set to $P = 0.01$ per unit length. The beam is made up of three layers with equal thickness, each subdivided into square coarse cells, with an edge length of $H$, containing a locally periodic microstructure. Each coarse cell is discretised with 40 by 40 linear finite elements. 
The following parameters are used: $r_{min} = 4h$, $v_{f} = 0.5$, $\beta_{0} = 16$, $\beta_{1} = 64$, $p_{1} = 5$, $\lambda_{\Omega} = 6.5\times10^{-4}$, $\varepsilon_{rel} = 10^{-6}$.

The problem is first investigated for a single periodic microstructure and afterwards for a varying number of coarse cells across the thickness of the three layers. Finally, the problem is investigated for varying thickness of the outer two layers.

\subsubsection{Single periodic robust microstructure}
	
This problem was also investigated in \citep{Zuo2013}, however, in the present work, the distributed load is in direct contact with the design domain, whereas a solid non-design region was included in \citep{Zuo2013} in order to allow for a direct comparison with results obtained through homogenisation. This is not done in the current work, in order to highlight the fact that by considering the full macroscopic structure, one actually obtains a design that automatically takes the boundary and loading conditions into account. 

	\begin{figure}
	\centering
	\subfloat[$M_{x} = 2$]{\includegraphics[clip, width=0.145\textwidth]{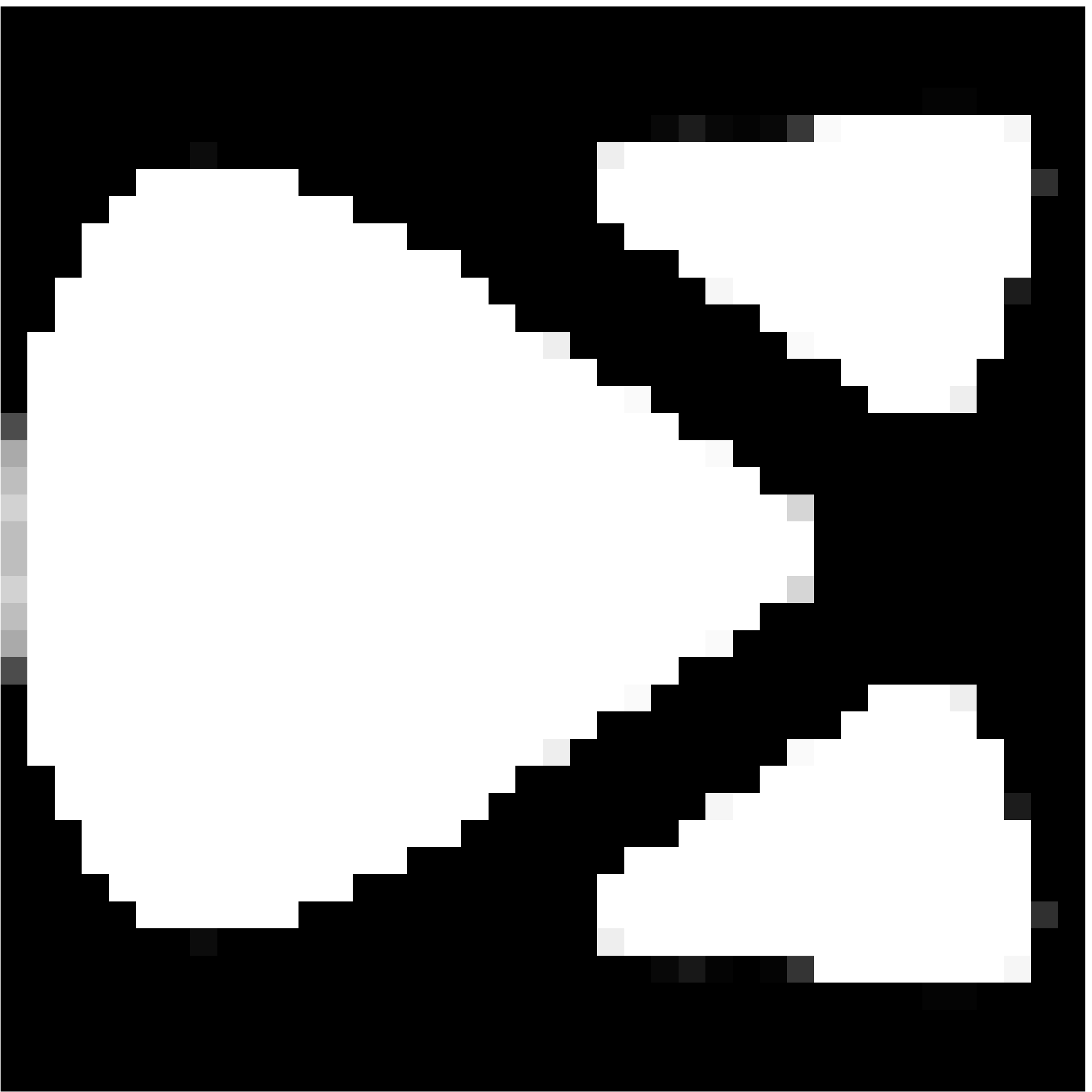}	
	\label{fig:cantidist_robust_single-a}}
	\hspace*{0.015\textwidth}
	\subfloat[$M_{x} = 4$]{\includegraphics[clip, width=0.145\textwidth]{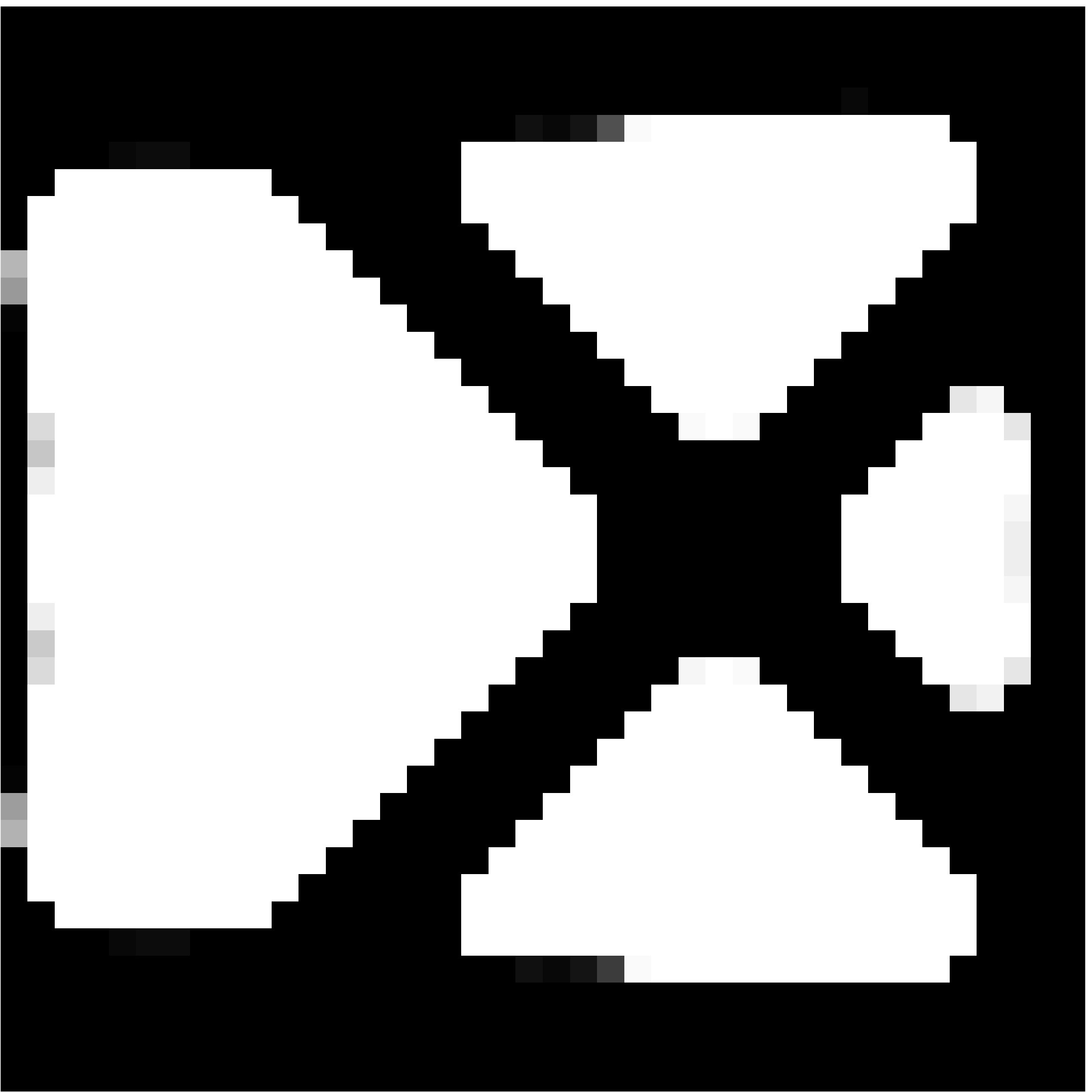}	\label{fig:cantidist_robust_single-b}}
	\hspace*{0.015\textwidth}
	\subfloat[$M_{x} = 8$]{\includegraphics[clip, width=0.145\textwidth]{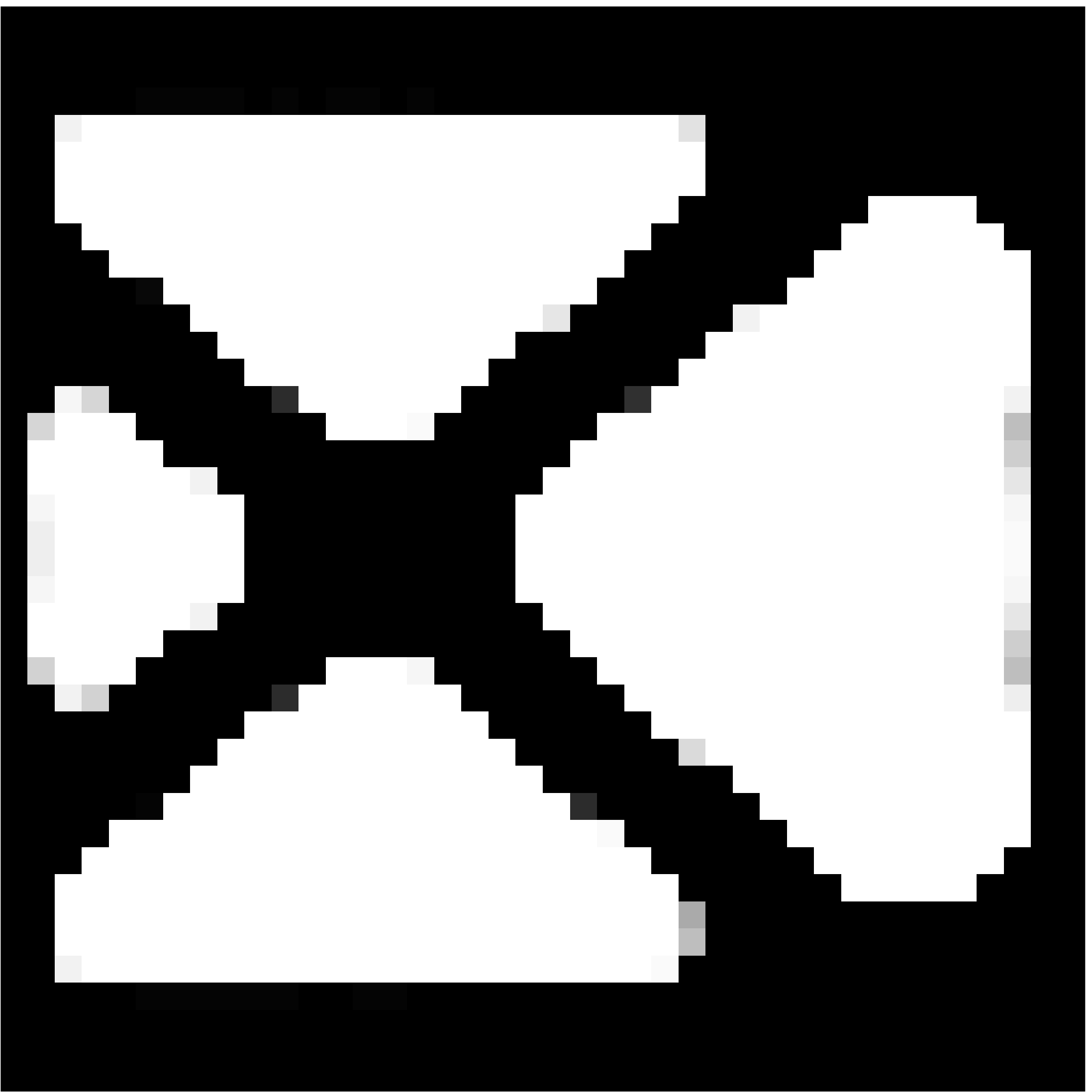}	\label{fig:cantidist_robust_single-c}}
	\hspace*{0.015\textwidth}
	\subfloat[$M_{x} = 12$]{\includegraphics[clip, width=0.145\textwidth]{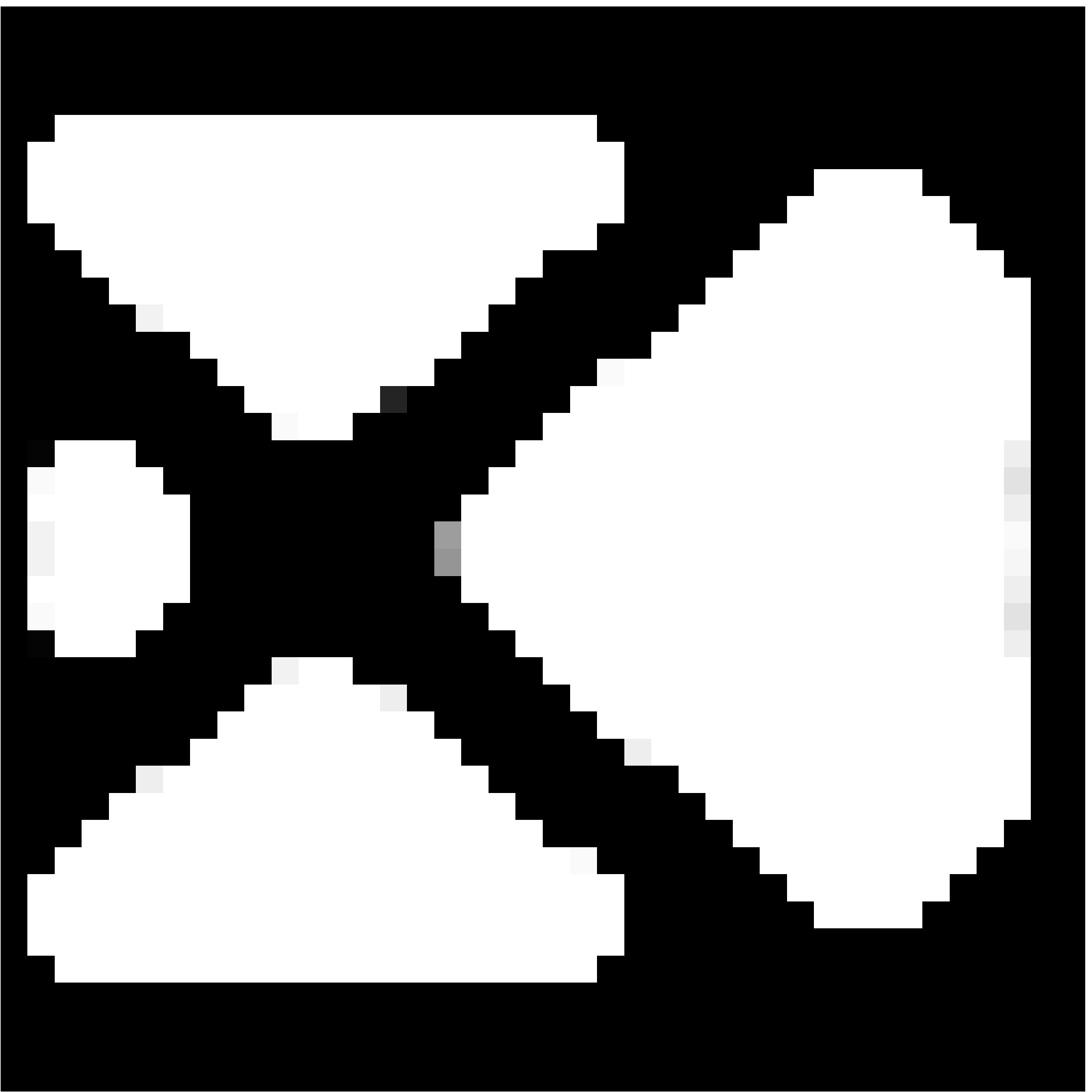}	\label{fig:cantidist_robust_single-d}}
	\hspace*{0.015\textwidth}
	\subfloat[$M_{x} = 16$]{\includegraphics[clip, width=0.145\textwidth]{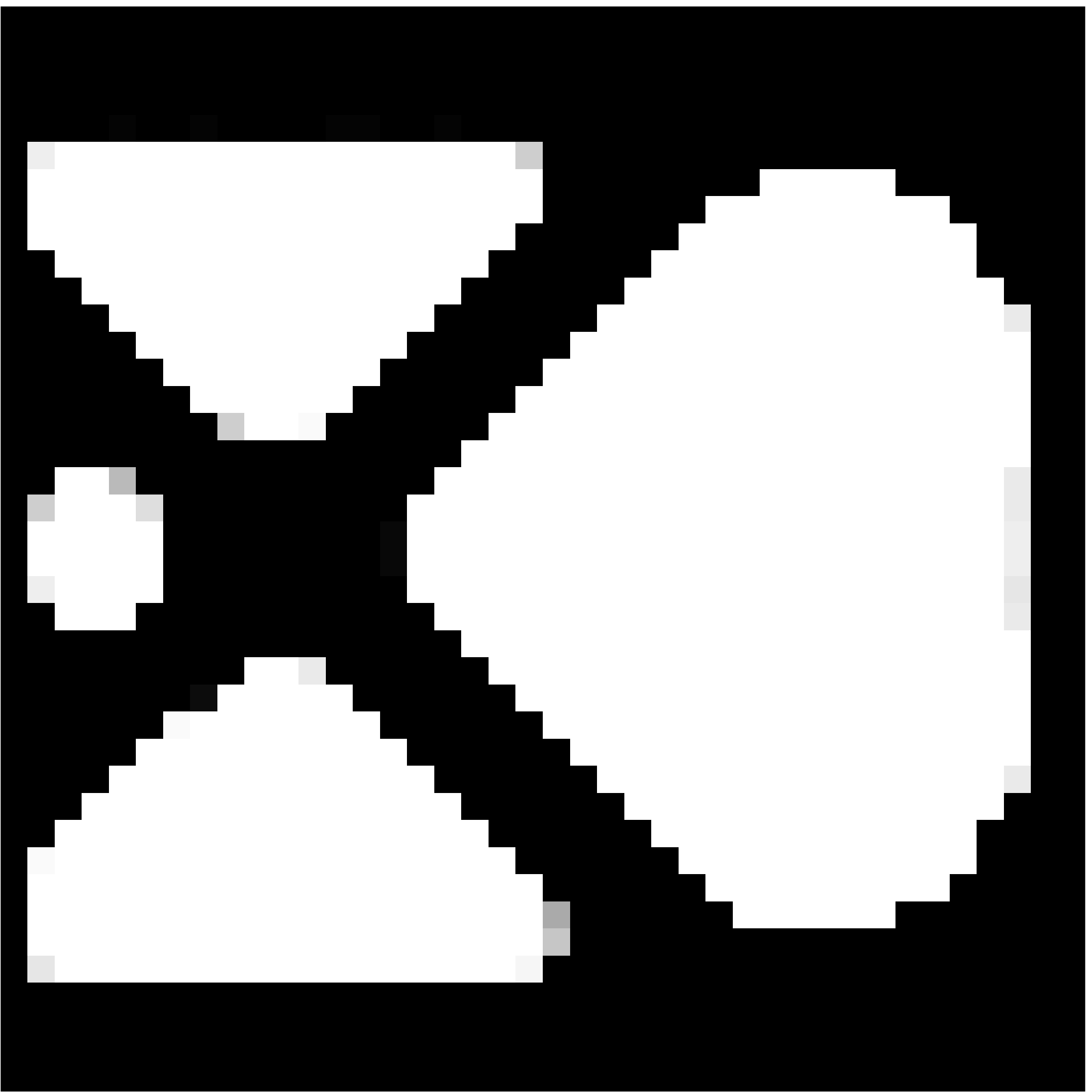}	
	\label{fig:cantidist_robust_single-e}}	
	\\
	\subfloat[$M_{x} = 2$]{
	\includegraphics[clip, width=0.40\textwidth]{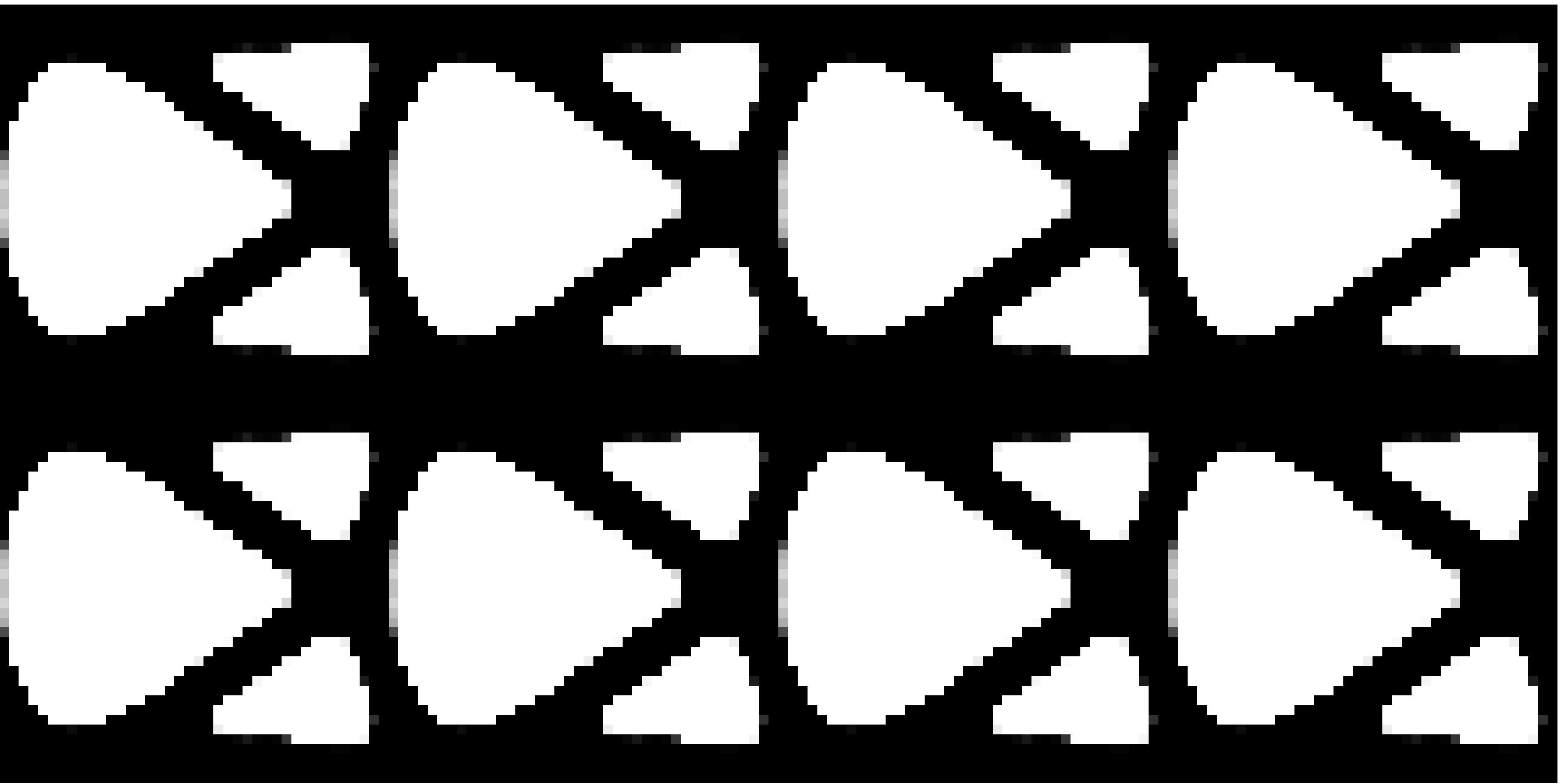}	
	\label{fig:cantidist_robust_single-f}}
	\hspace*{0.02\textwidth}
	\subfloat[$M_{x} = 16$]{
	\includegraphics[clip, width=0.40\textwidth]{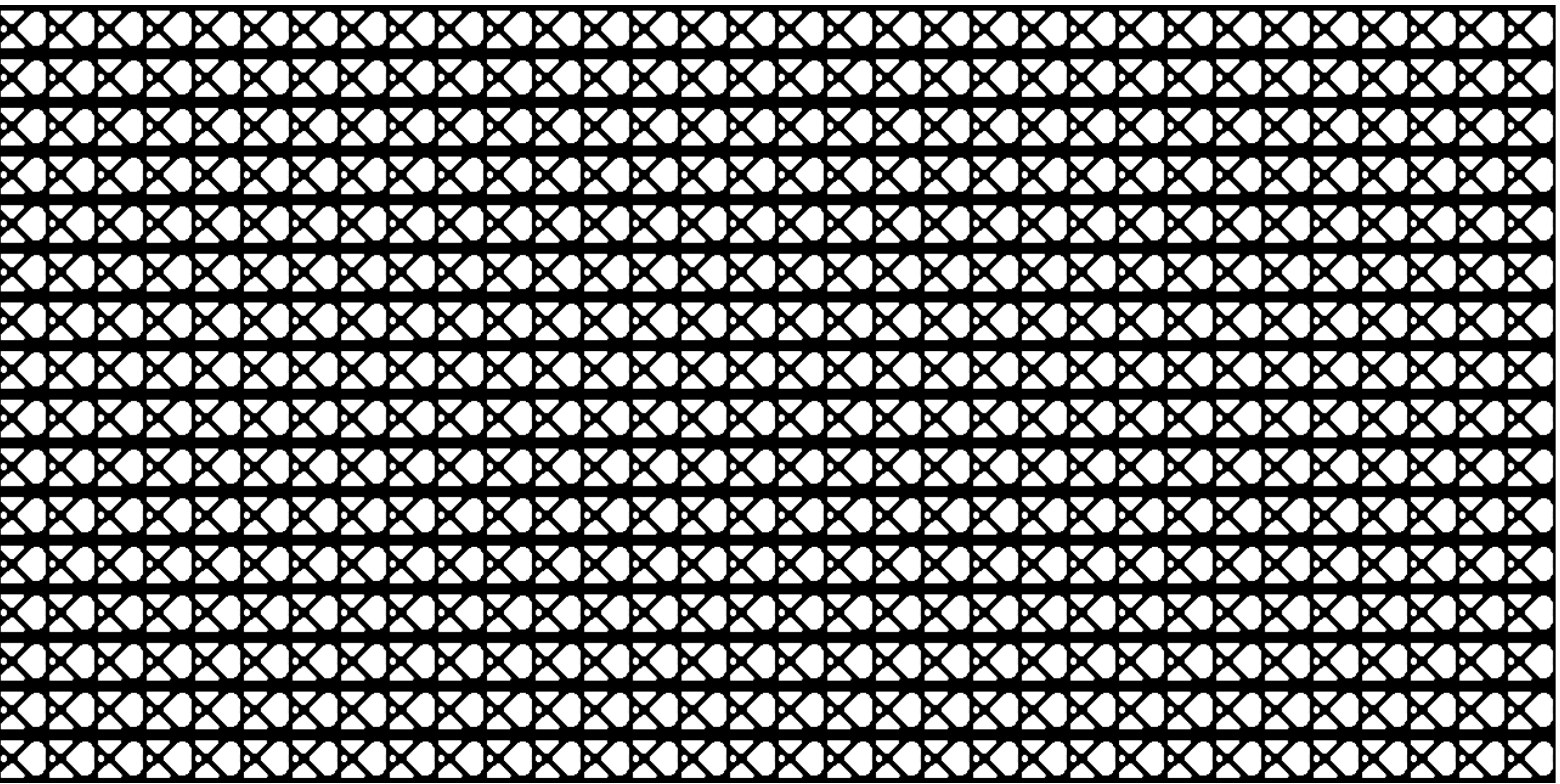}	
	\label{fig:cantidist_robust_single-g}}		
	\caption{Optimised robust microstructures for the cantilever beam with a distributed load. The microstructures are shown for varying number of cells across the height, $M_{x} = \{ 2,4,8,12,16 \} $. The full macrostructures are shown for $M_{x} = 2$ and $M_{x} = 16$.} \label{fig:cantidist_robust_single}
	\end{figure}
Figure \ref{fig:cantidist_robust_single} shows the optimised robust microstructures for the cantilever beam with a distributed load along the right-hand edge. It can be observed that, unlike for the double clamped beam problem of section \ref{sec:results_doubleclamp}, the microstructure does not converge within the observed range of unit cell refinement. It can be seen that the microstructure has the same overall topology from $M_{x} = 4 - 16$, but that the size of the holes are changing. By performing a cross-check analysis, this development in the topology appears to be due to local minima. This is not considered as important for the current study, as the aim of this paper is not to investigate the convergence of the unit cell design with respect to unit cell size .

Comparing the microstructures in figure \ref{fig:cantidist_robust_single}  to those presented in \citep{Zuo2013}, an important difference is the vertical bar at the right-hand side of the unit cell. This vertical bar is needed for the unit cells in contact with the distributed load and due to the imposed periodicity it exist in all unit cells within the beam, as seen in figures \ref{fig:cantidist_robust_single-f} and \ref{fig:cantidist_robust_single-g}. Of course, this leads to an increasingly sub-optimal design within the design domain, as the vertical bar is only needed at the very edge. But as already stated, it is seen as important to illustrate that the presented methodology, of taking the entire macrostructure into account, ensures a physically valid design with respect to boundary and loading conditions. The optimised microstructures can be seen to exhibit a combination of bending and shear stiffness, the first characterised by the horisontal bars and the latter characterised by the cross-like members. This is as expected, as forcing the microstructure to be the same all over the short beam, ensures that the microstructure needs to exhibit both characteristics.
		
	\begin{figure}
	\centering
	\subfloat[Eroded]{\includegraphics[clip, width=0.16\textwidth]{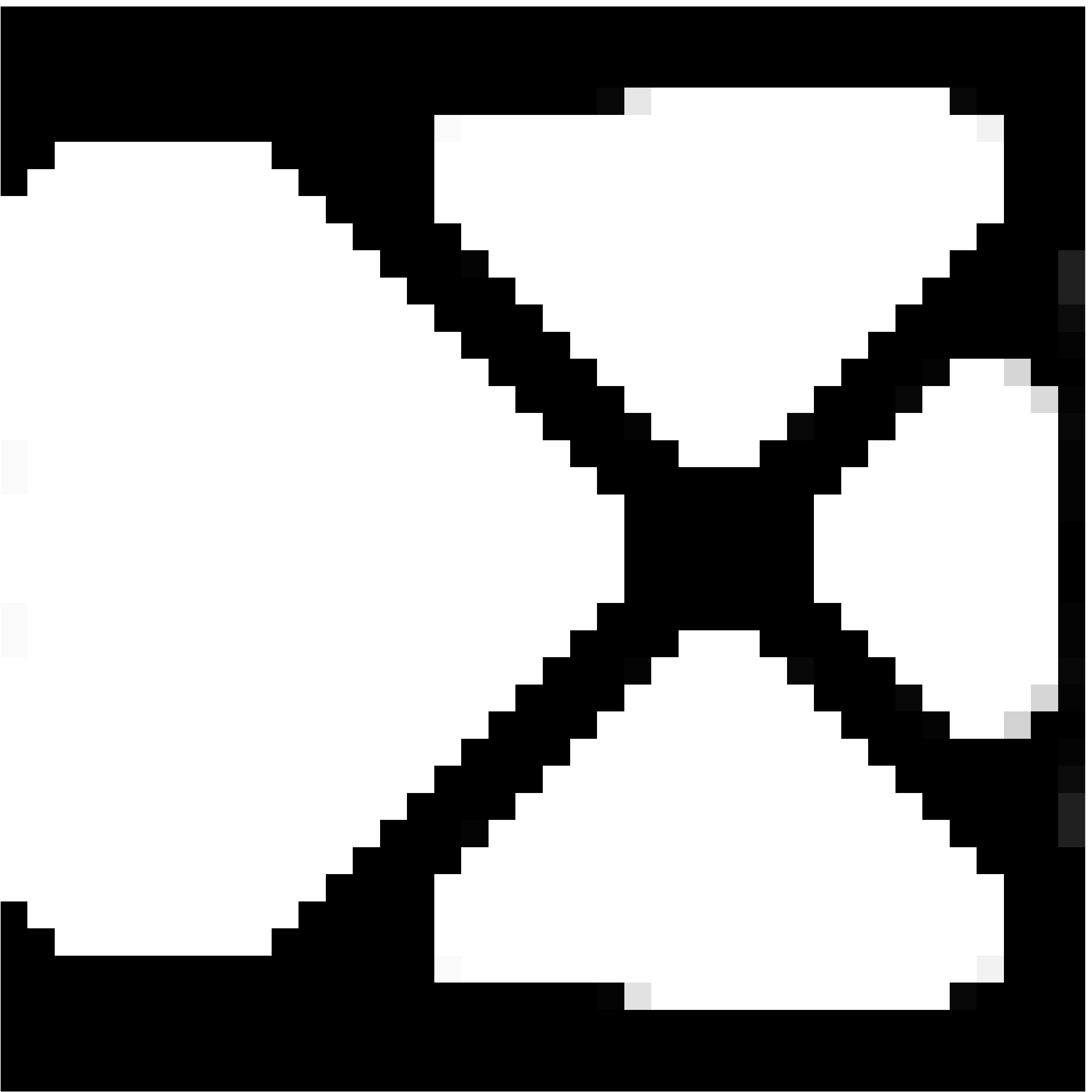}	\label{fig:cantidist_robust_single_realisations-a}}
	\hspace*{0.04\textwidth}
	\subfloat[Intermediate]{\includegraphics[clip, width=0.16\textwidth]{cantidist_singlelayer_Mx4_PNx40_microInt_bw}	\label{fig:cantidist_robust_single_realisations-d}}
	\hspace*{0.04\textwidth}
	\subfloat[Dilated]{\includegraphics[clip, width=0.16\textwidth]{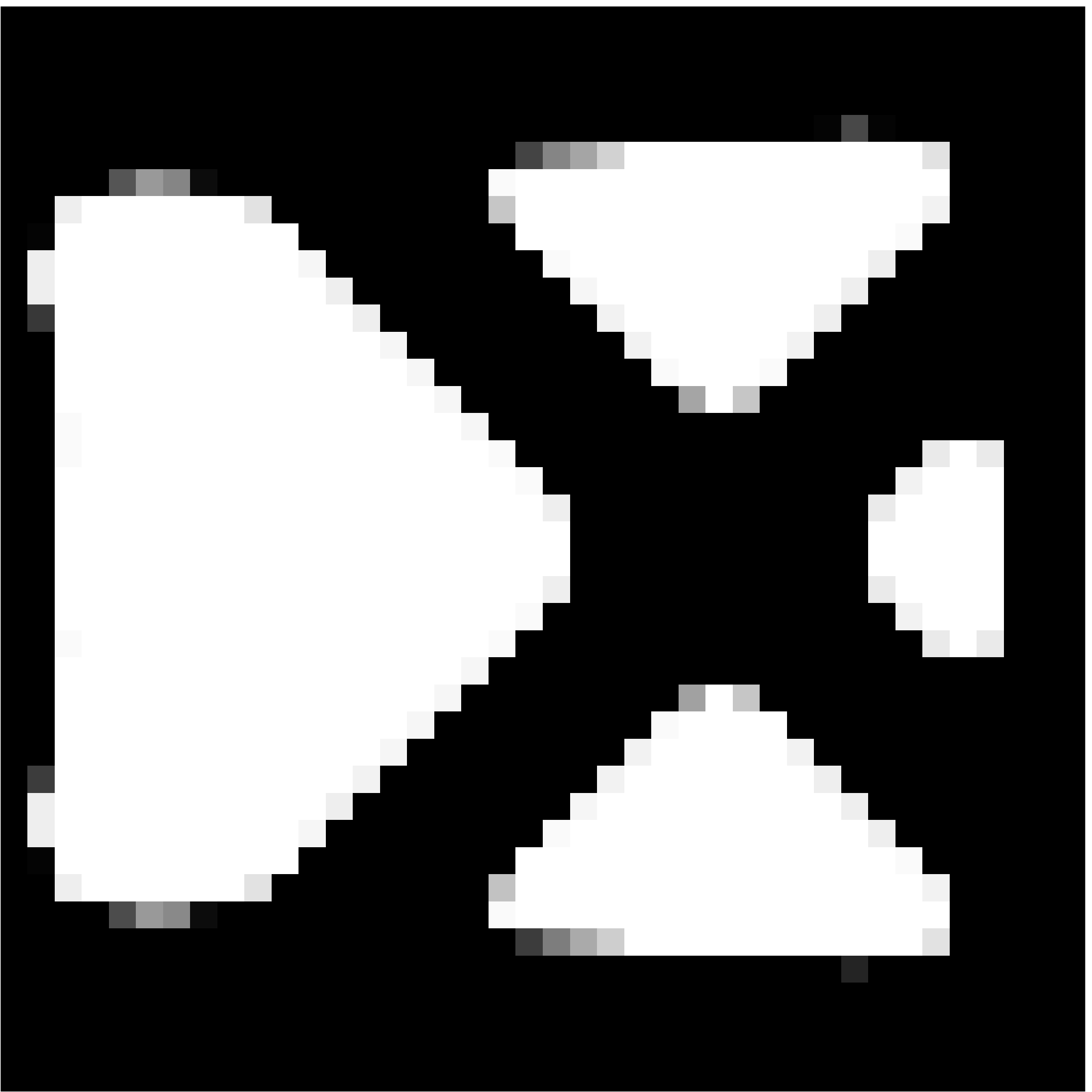}	\label{fig:cantidist_robust_single_realisations-c}}
							
	\caption{The three realisations of the optimised robust microstructures for the cantilever beam with a distributed load and $M_{x} = 4$.} \label{fig:cantidist_robust_single_realisations}
	\end{figure}
	Figure \ref{fig:cantidist_robust_single_realisations} shows the three realisations of the final optimised design for $M_{x} = 4$. Here it can be clearly seen, that the design is robust with respect to erosion and dilation. It is important to point out that the vertical bar ensuring connection to the distributed load is ensured for the eroded design also, as seen in figure \ref{fig:cantidist_robust_single_realisations-a}.

\subsubsection{Three-layered robust microstructure}

One of the main benefits of analysing the full macrostructure with all microstructural details resolved, is the ability to design structures with varying microstructure and ensuring that these microstructures are connected. The presented implementation has the ability to optimise layered microstructures. The cantilever beam with a distributed load along the right-hand edge is split into three layers of equal thickness each made up of a number of periodic cells, as illustrated in figure \ref{fig:cantilever_numexp_illustration}.
	
\begin{figure}
	\centering
	\includegraphics[clip, width=0.9\textwidth]{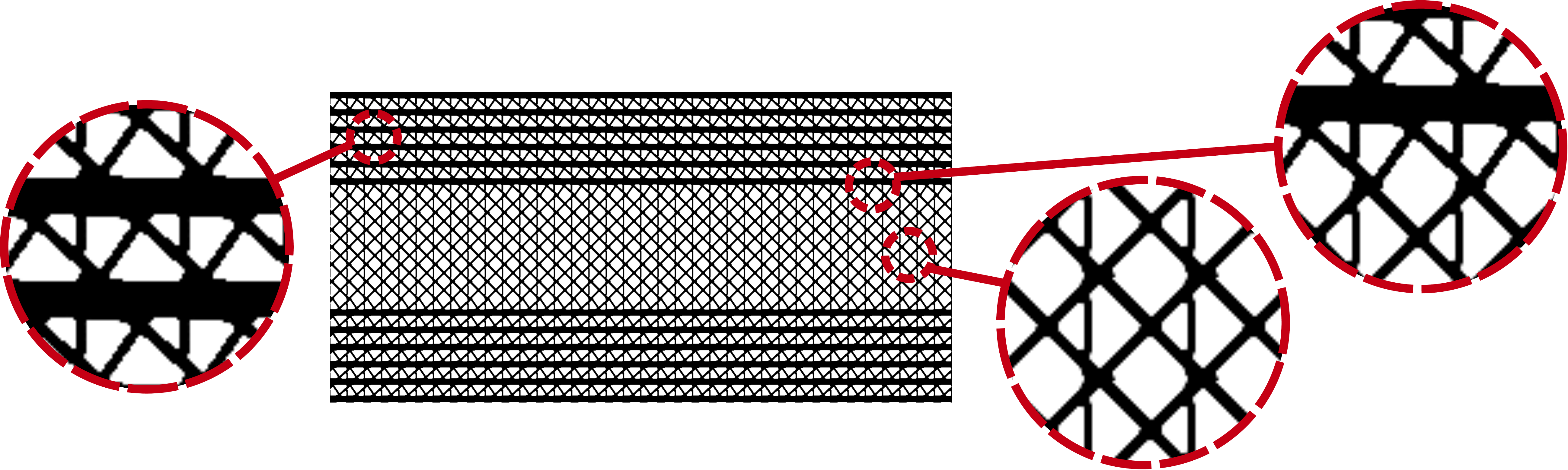}
	\caption{Optimised robust layered macrostructure for the cantilever beam with a distributed load for $M_{x} = 18$ and layer thickness $\left\lbrace \frac{6}{18} , \frac{6}{18} , \frac{6}{18} \right\rbrace$.} \label{fig:cantidist_robust_three_equal_macro}
\end{figure}
Figure \ref{fig:cantidist_robust_three_equal_macro} shows the full macrostructure with optimised robust layered microstructures for $M_{x} = 18$ with layer thickness $\left\lbrace \frac{6}{18} , \frac{6}{18} , \frac{6}{18} \right\rbrace$, where the first entry denotes the relative thickness of the top layer, the second denotes the middle layer and the third denotes the bottom layer. It can clearly be seen that different microstructures exist in the three layers. As expected, the top and bottom layers exhibit microstructures with a high bending stiffness, characterised by the thick horisontal bars, whereas the middle layer exhibits a microstructure with a high shear stiffness, characterised by the cross structure. This is perfectly in line with the stress distribution of a short beam under bending; high axial stresses near the top and bottom and high shear stresses in the middle. Furthermore, due to the skew-symmetry of the stress distribution, the optimal topology should be symmetric \citep{Rozvany2011} which is also observed for the optimised designs. It is important to note that no symmetry has been imposed during the optimisation process. It is observed that the average compliance does not change significantly with respect to increasing the number of coarse cells. Also, the microstructural topology does not change significantly and these are therefore not shown.

%\paragraph{Convergence to the homogenised case}
%
%\red{Comparison of the three-layered homogenisation case - pure analysis only!}

\subsubsection{Three-layered robust microstructure of variable thickness}

The three-layered cantilever beam with a distributed load along the right-hand edge is now investigated for varying thickness of the outer layers.
	\begin{figure}
	\centering
	\begin{tabular}{cc|c|c|c|c}

	&
	$\left\lbrace \frac{6}{18} , \frac{6}{18} , \frac{6}{18} \right\rbrace$
	&
	$\left\lbrace \frac{5}{18} , \frac{8}{18} , \frac{5}{18} \right\rbrace$
	&
	$\left\lbrace \frac{4}{18} , \frac{10}{18} , \frac{4}{18} \right\rbrace$
	&
	$\left\lbrace \frac{3}{18} , \frac{12}{18} , \frac{3}{18} \right\rbrace$
	&
	$\left\lbrace \frac{2}{18} , \frac{14}{18} , \frac{2}{18} \right\rbrace$
	
	\\
	
	&	&	&	&	&	\\
	
	\rotatebox{90}{\hspace*{0.5cm}top}
	&
	\includegraphics[clip, width=0.10\textwidth]{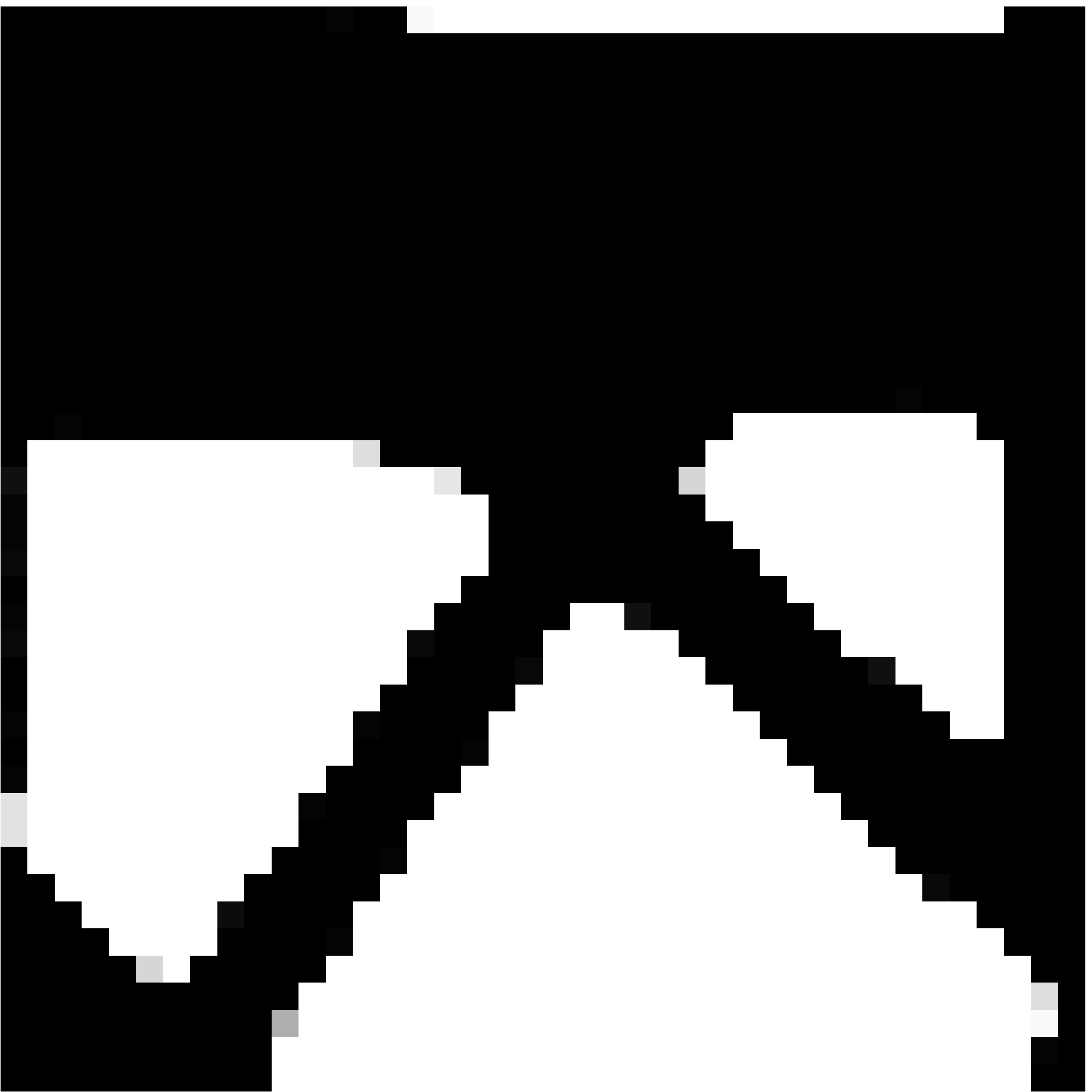}
	&
	\includegraphics[clip, width=0.10\textwidth]{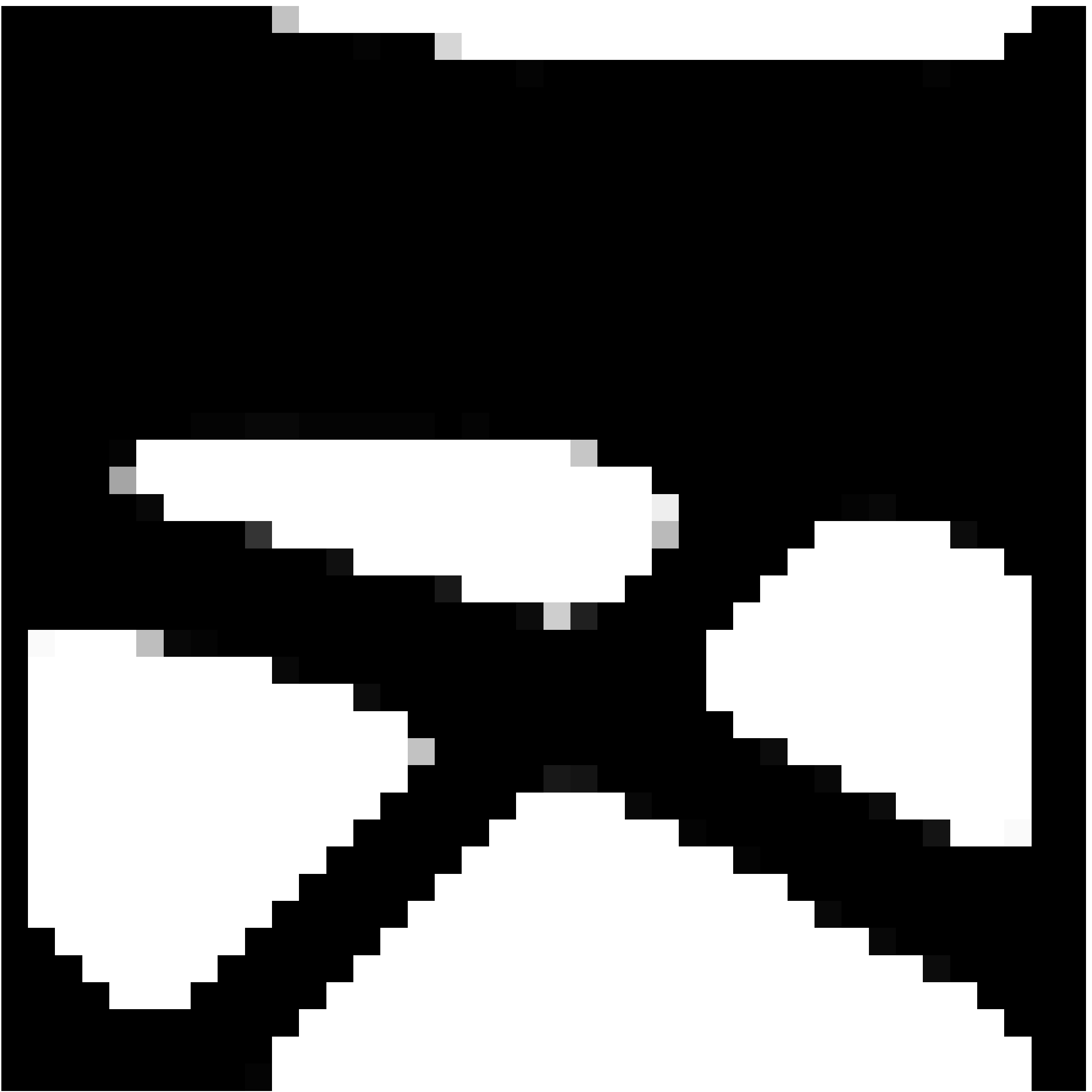}	
	&
	\includegraphics[clip, width=0.10\textwidth]{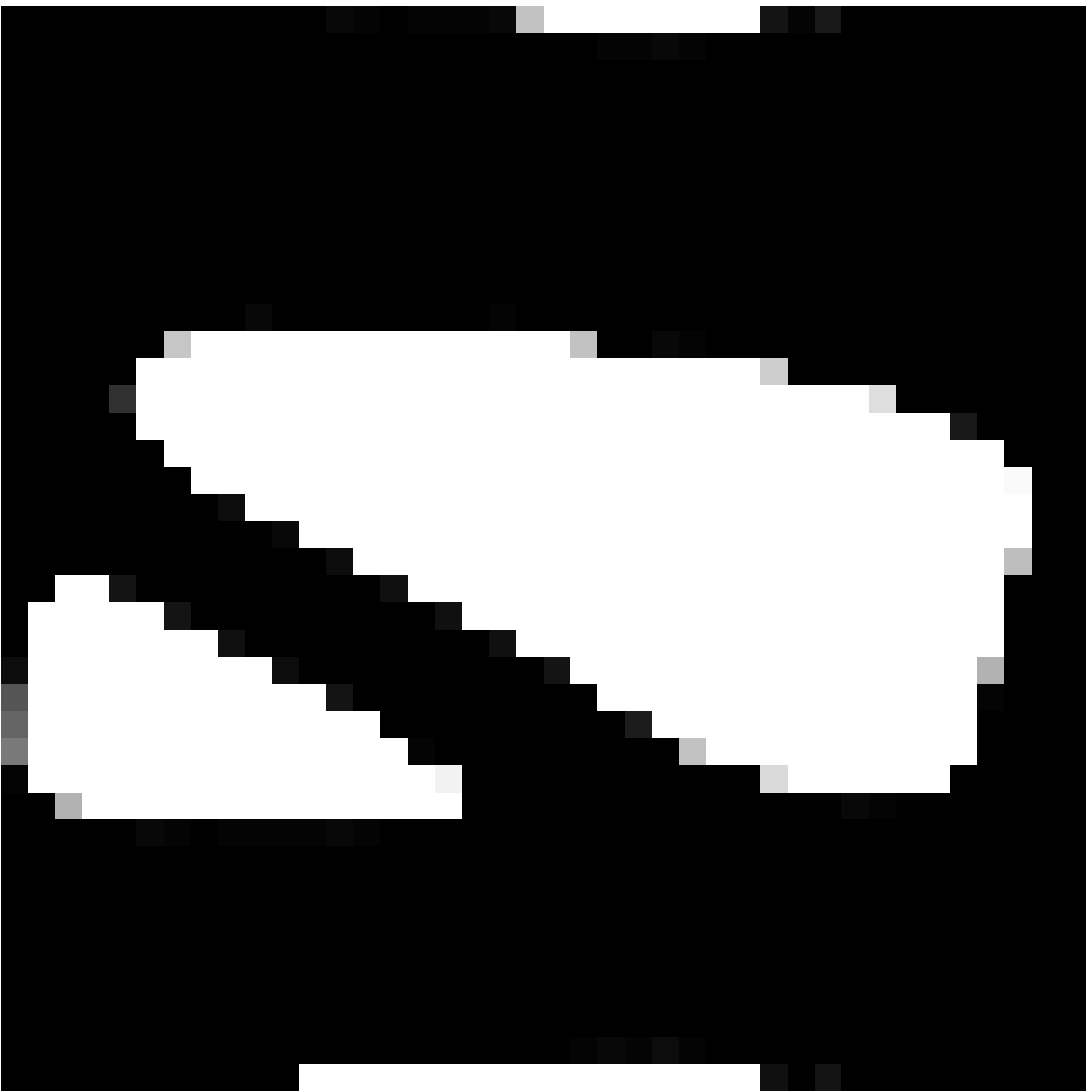}	
	&
	\includegraphics[clip, width=0.10\textwidth]{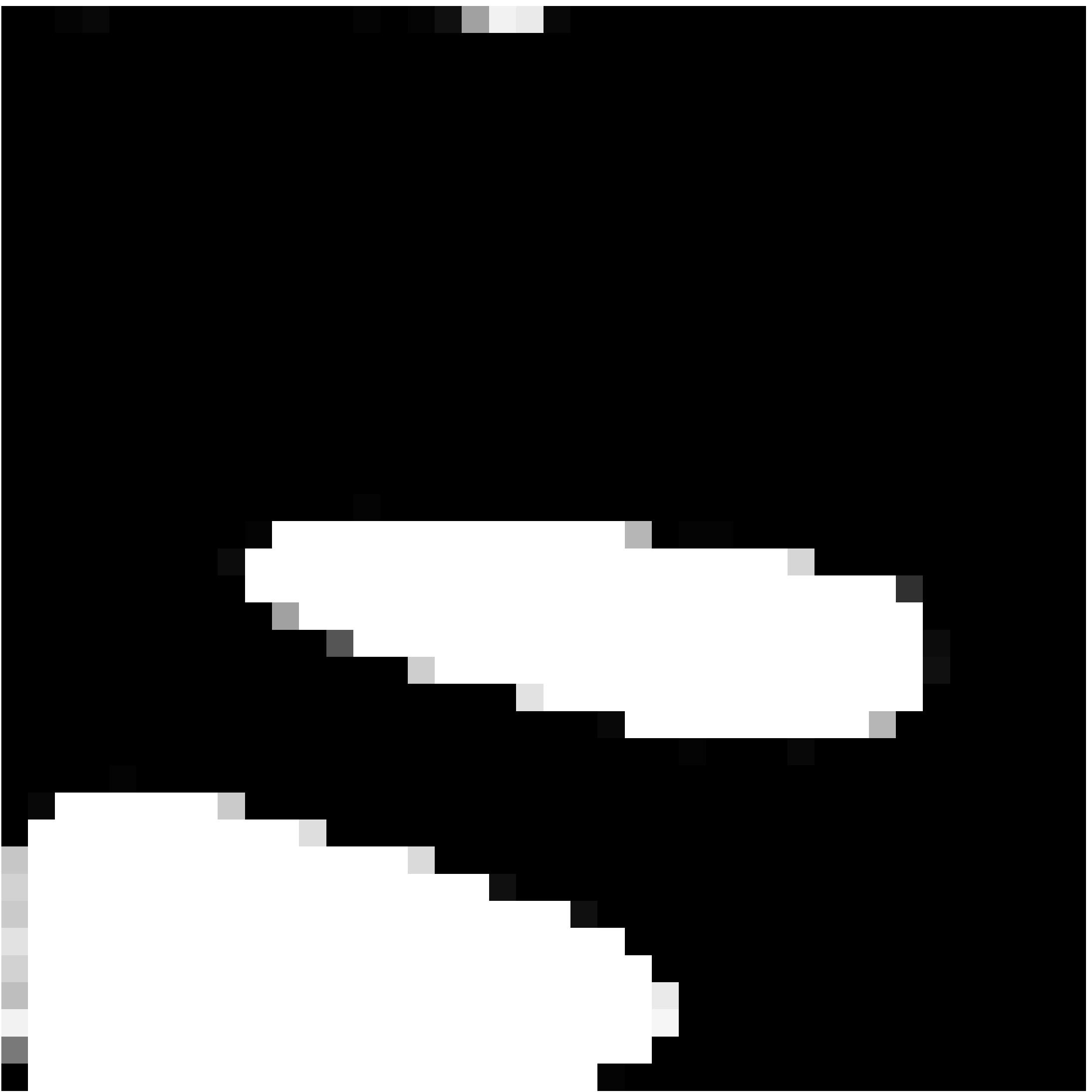}	
	&
	\includegraphics[clip, width=0.10\textwidth]{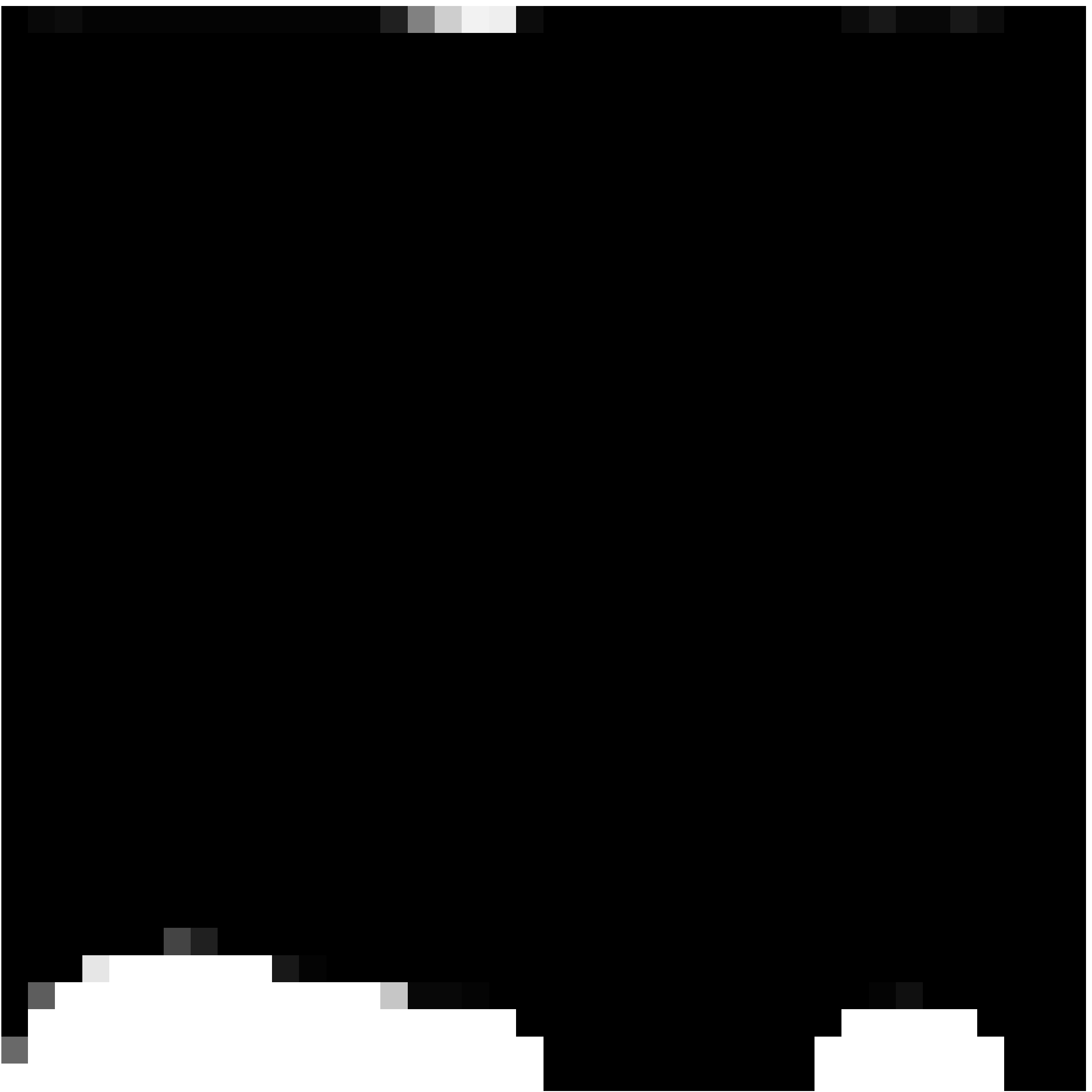}	
	\\
	\rotatebox{90}{\hspace*{0.25cm}middle}
	&
	\includegraphics[clip, width=0.10\textwidth]{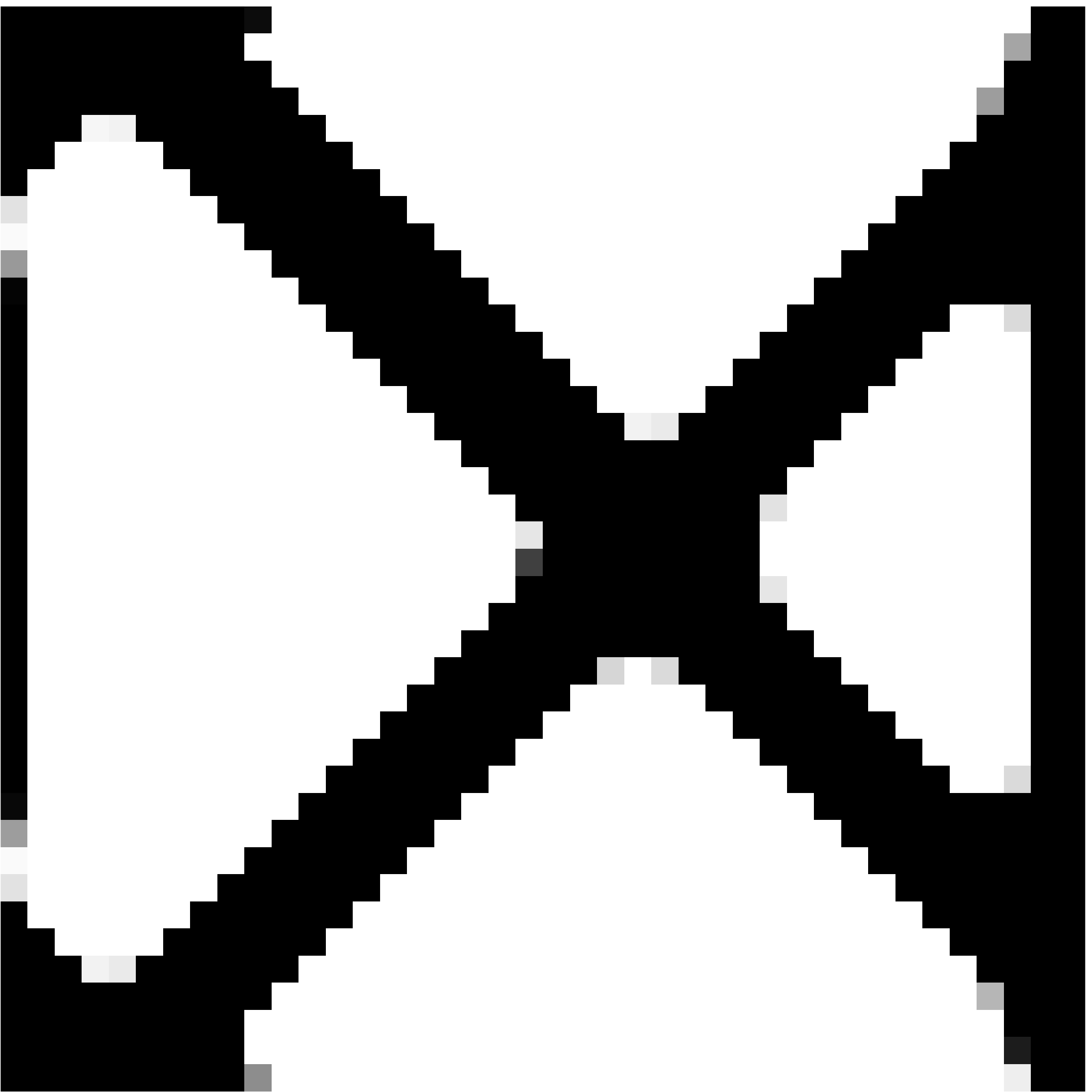}	
	&
	\includegraphics[clip, width=0.10\textwidth]{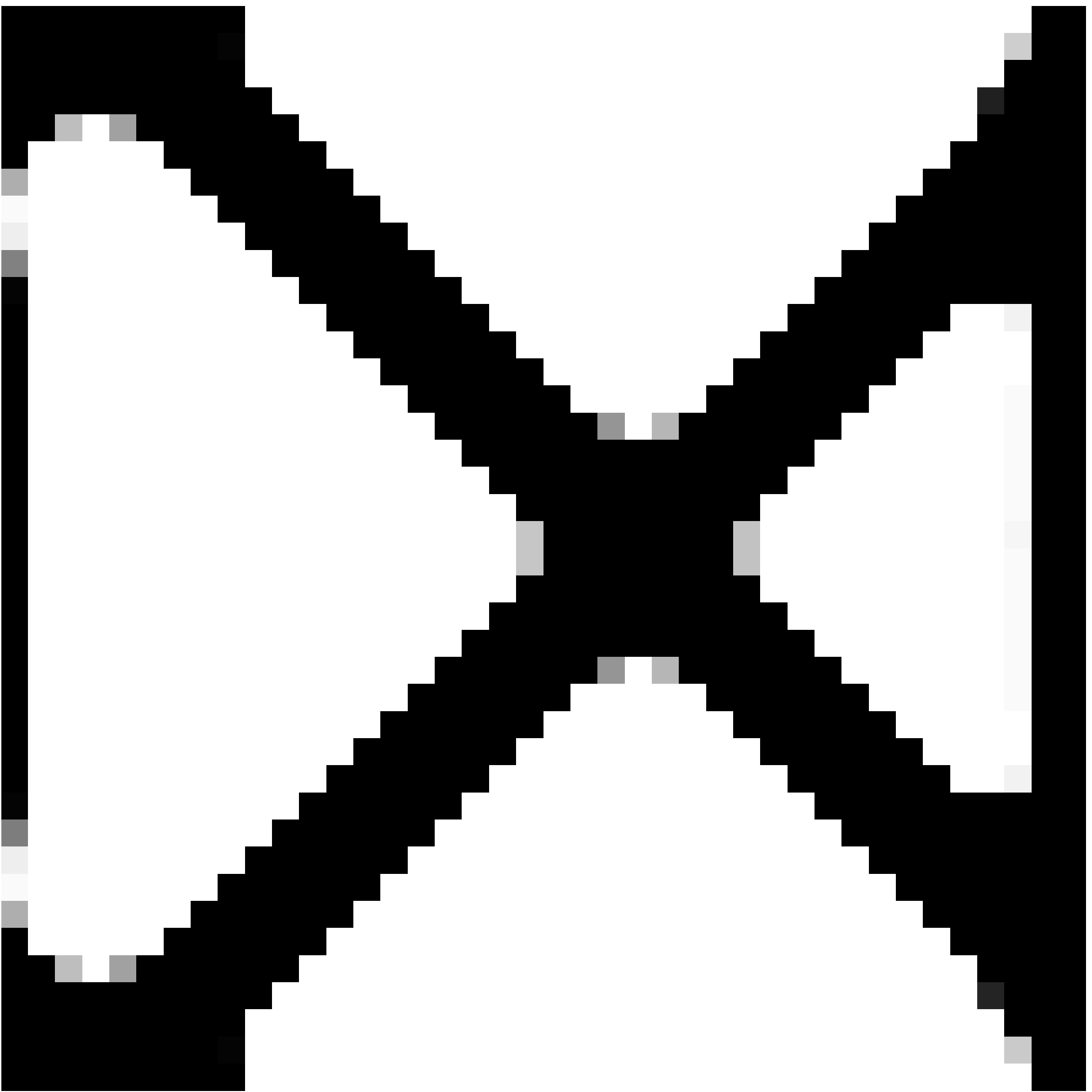}	
	&
	\includegraphics[clip, width=0.10\textwidth]{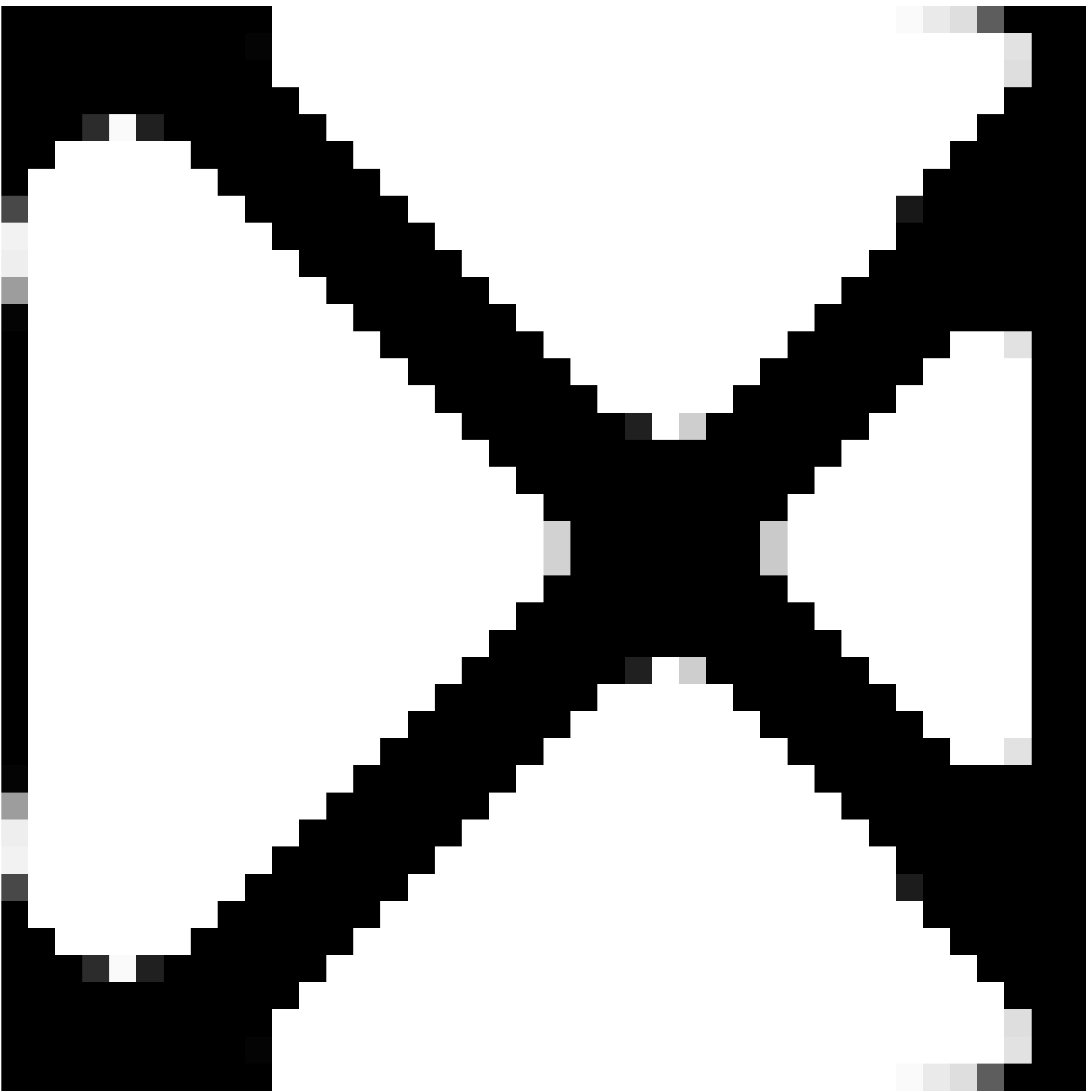}	
	&
	\includegraphics[clip, width=0.10\textwidth]{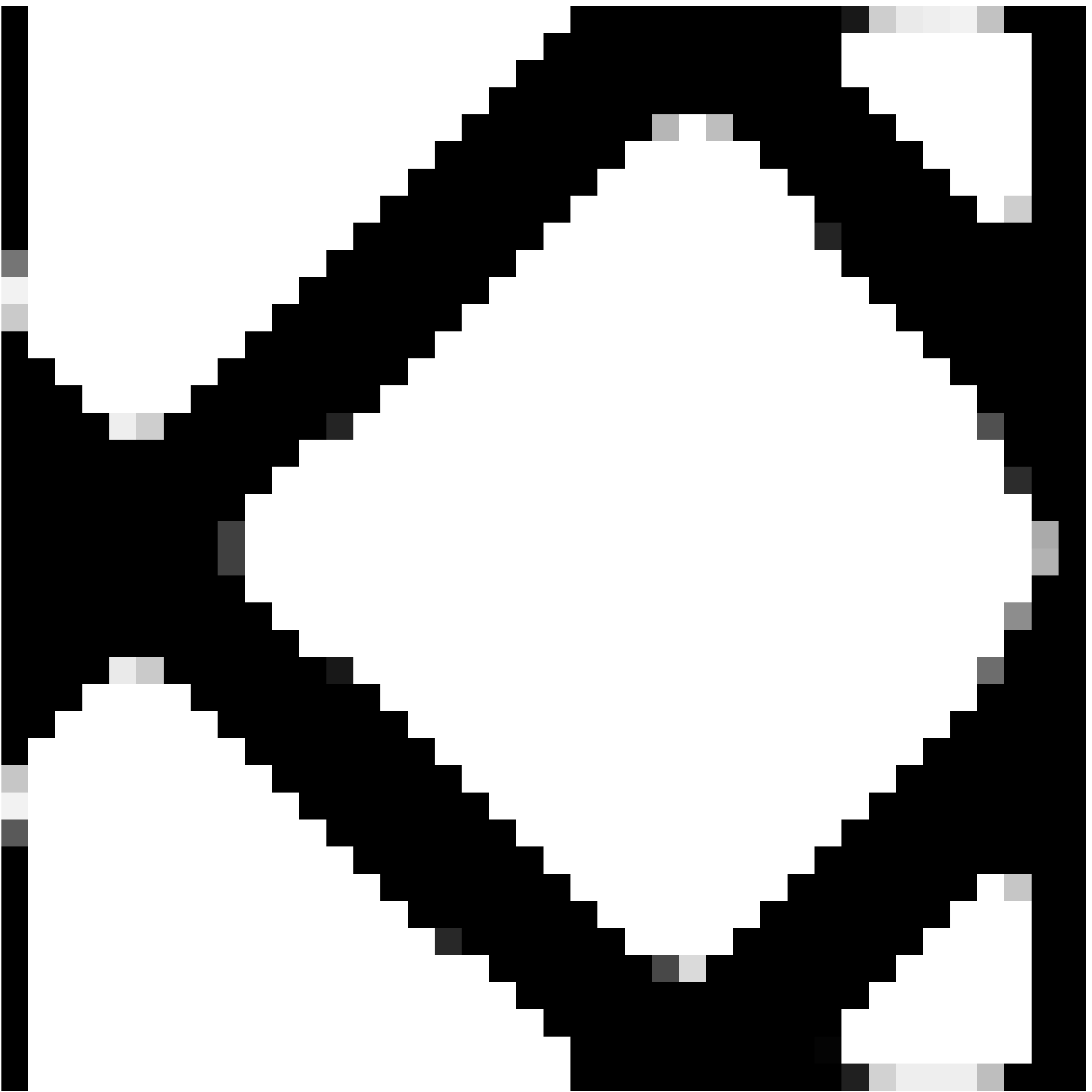}	
	&
	\includegraphics[clip, width=0.10\textwidth]{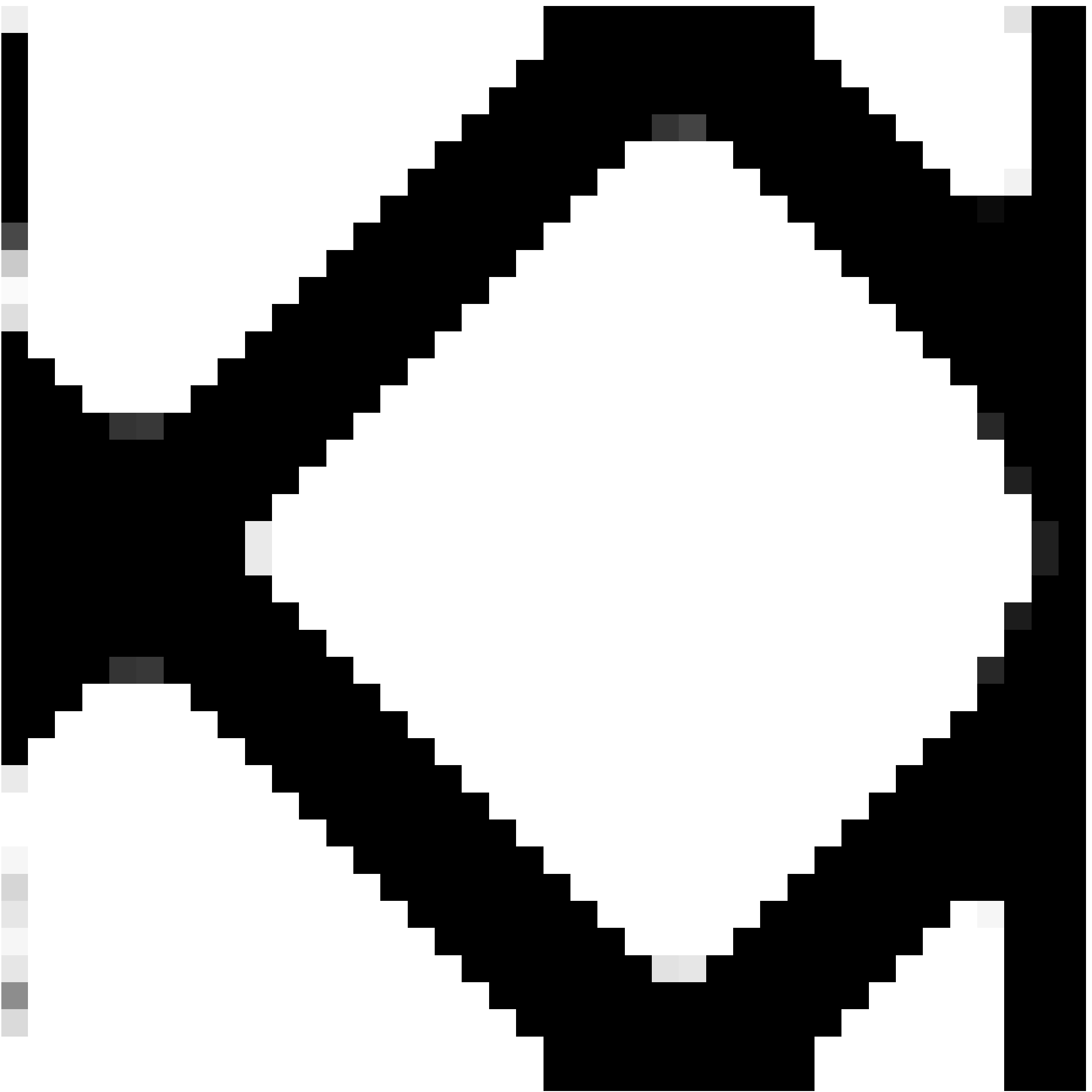}	
	\\
	\rotatebox{90}{\hspace*{0.25cm}bottom}
	&
	\includegraphics[clip, width=0.10\textwidth]{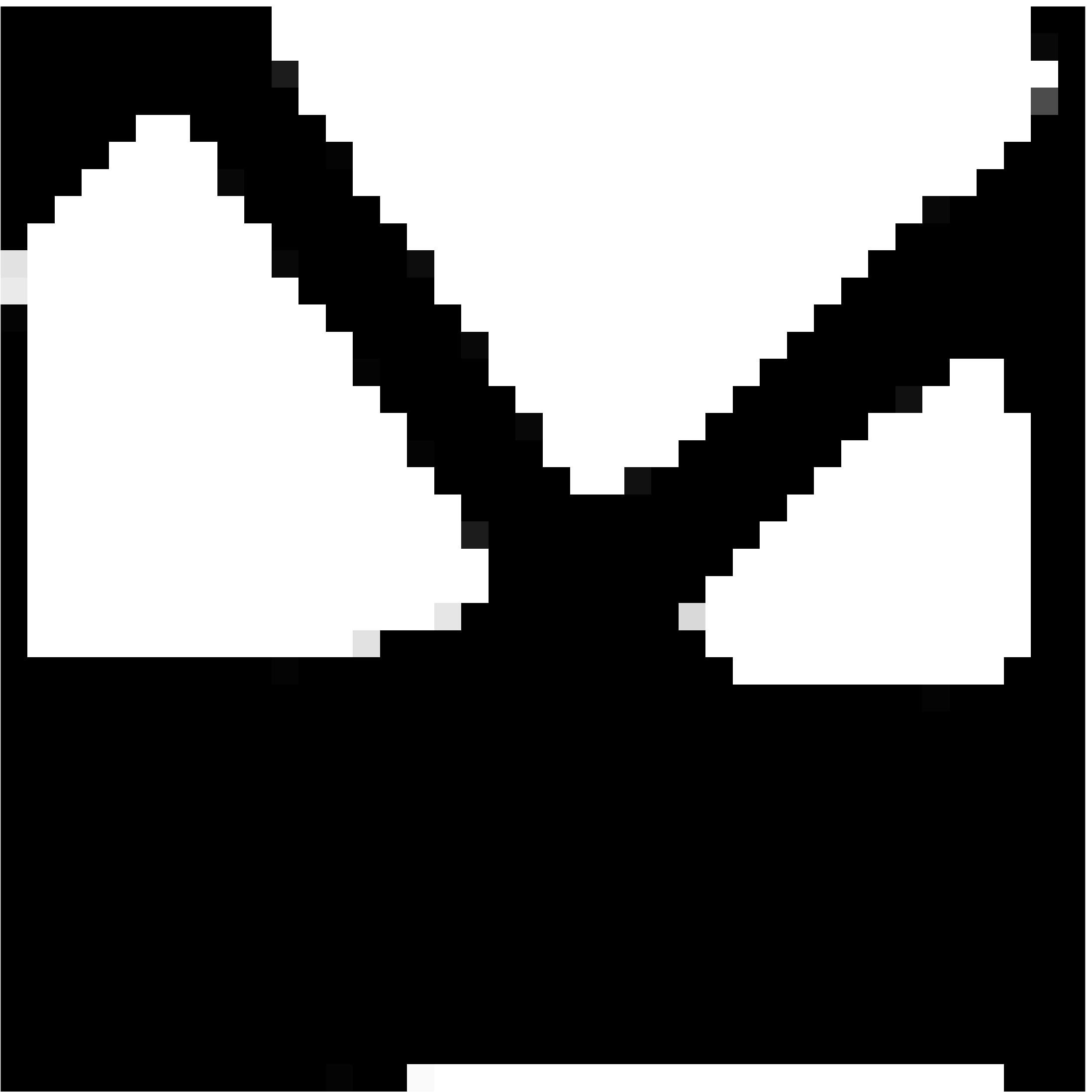}	
	&
	\includegraphics[clip, width=0.10\textwidth]{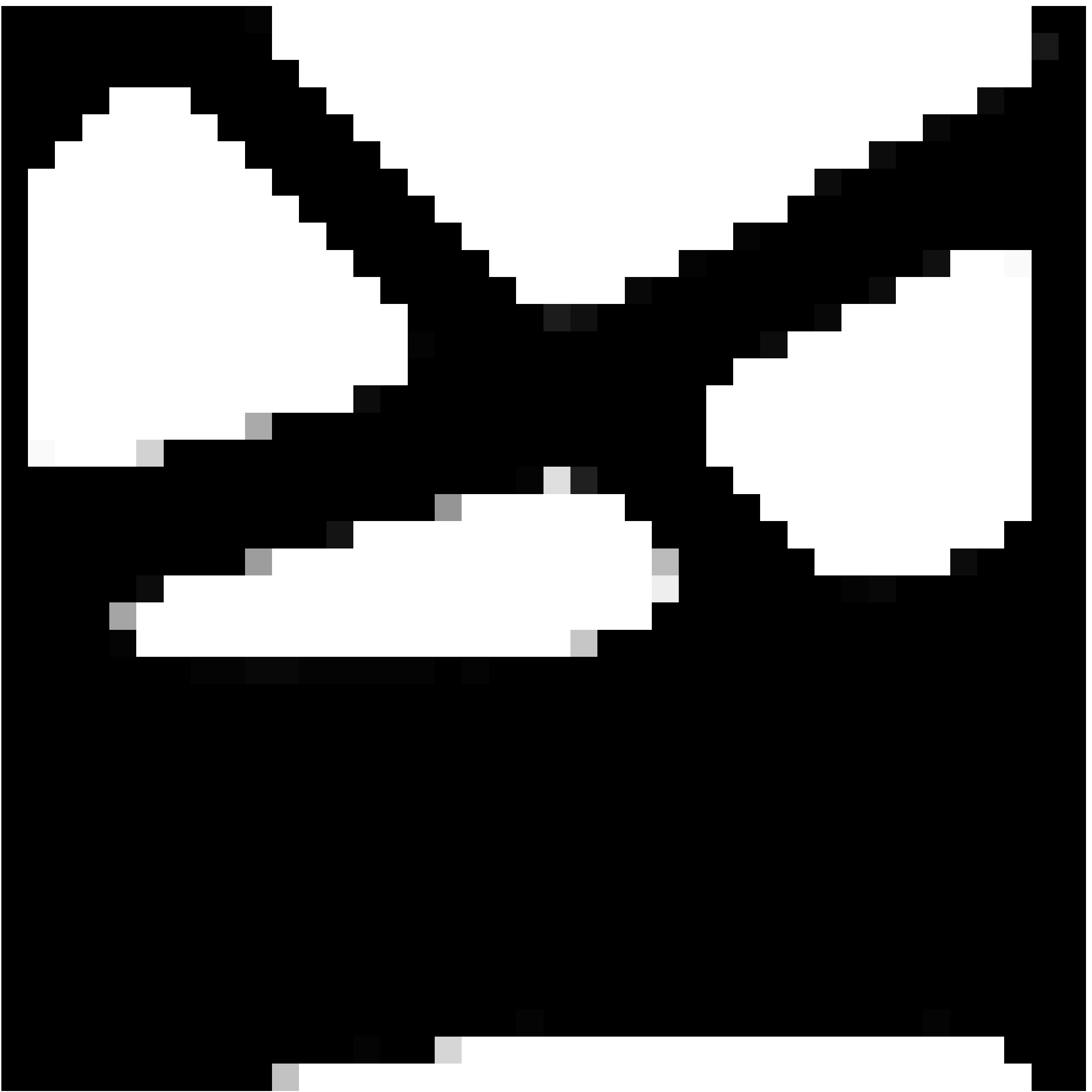}	
	&
	\includegraphics[clip, width=0.10\textwidth]{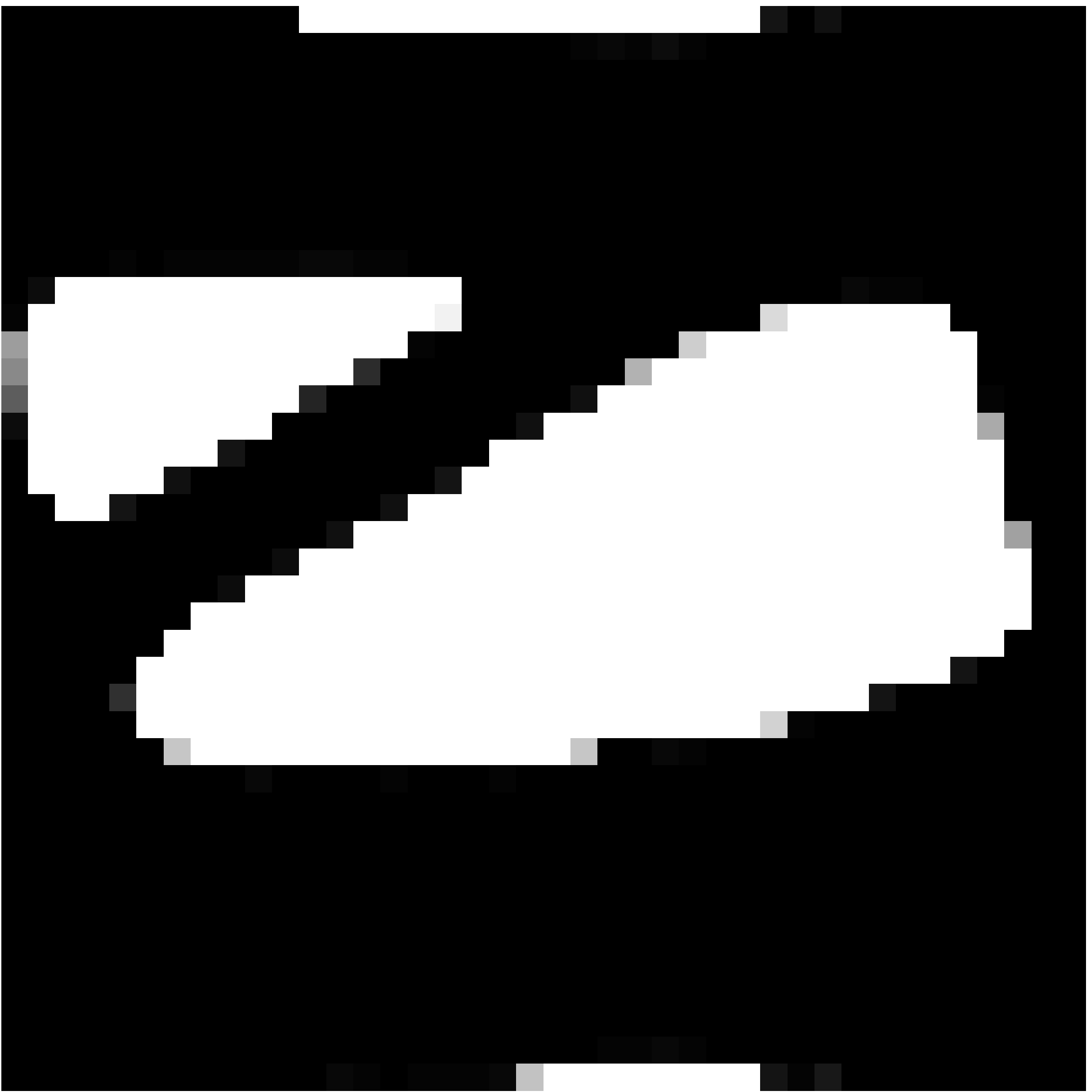}	
	&
	\includegraphics[clip, width=0.10\textwidth]{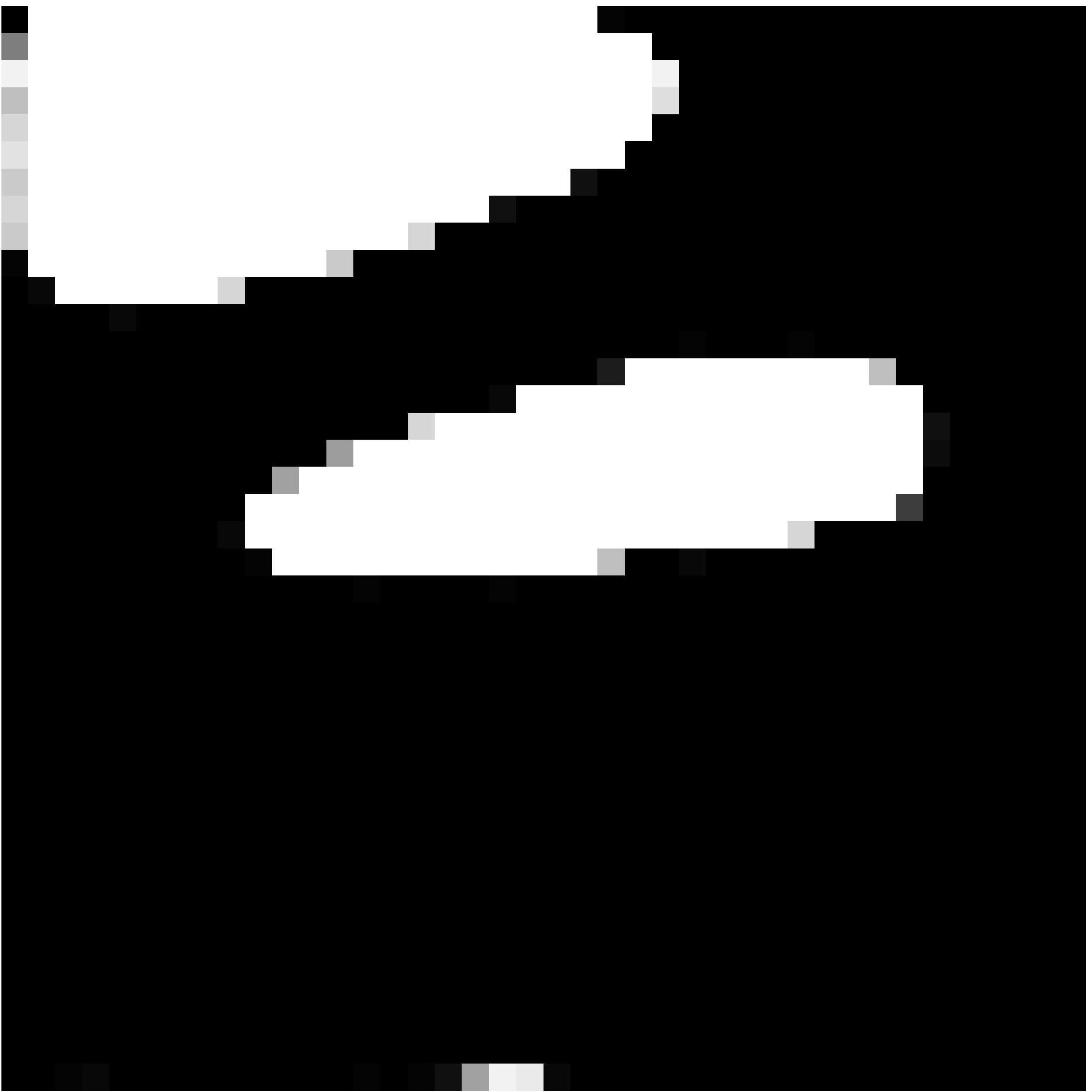}	
	&
	\includegraphics[clip, width=0.10\textwidth]{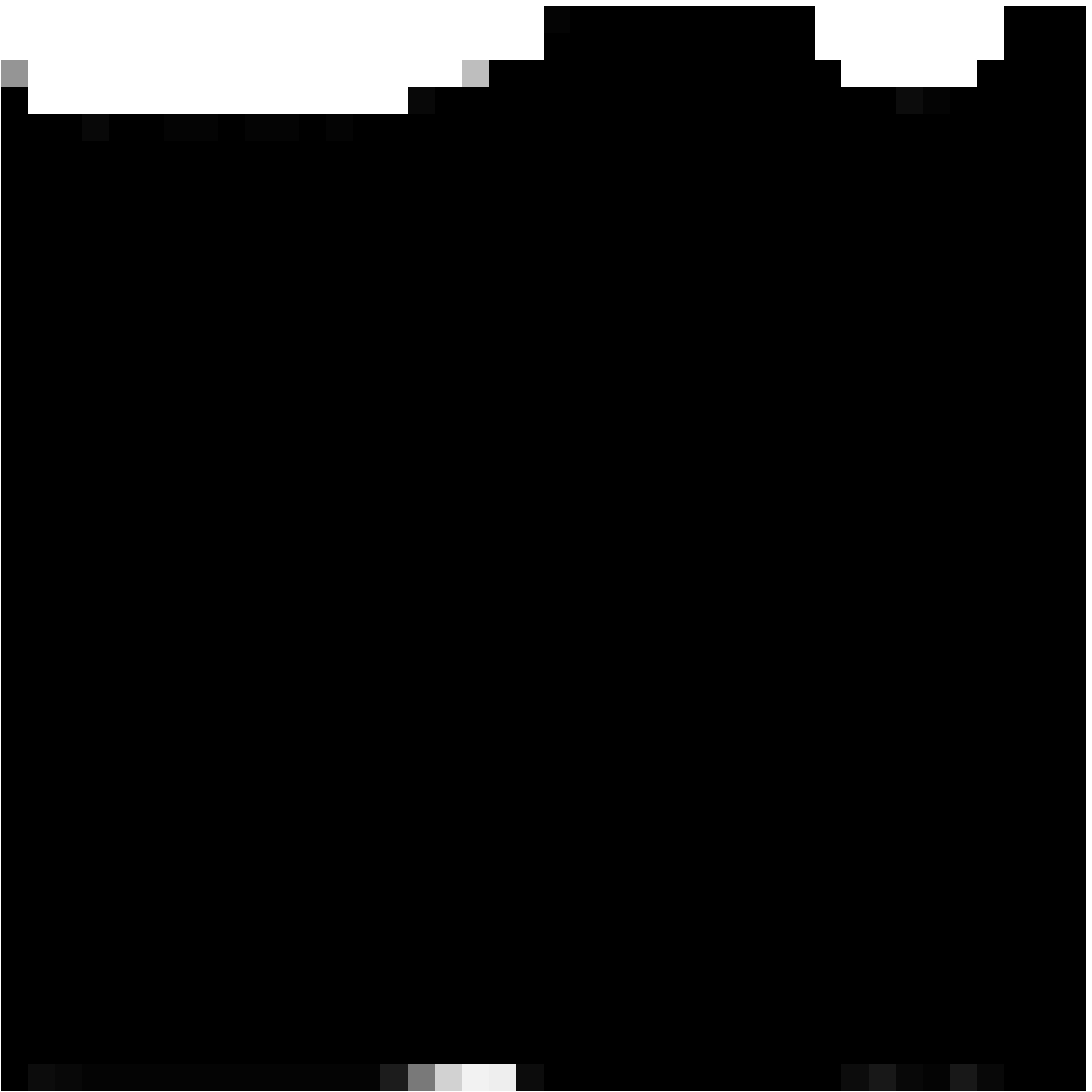}	
	
	\end{tabular}
		
	\caption{Optimised robust microstructures for the cantilever beam with a distributed load for three layers of varying outer layer thickness.} \label{fig:cantidist_robust_three_var_micro}
	\end{figure}
Figure \ref{fig:cantidist_robust_three_var_micro} shows the optimised robust microstructures for the three-layered cantilever beam with a distributed load for varying outer layer thickness. It can be seen that the microstructure of the middle layer does not change significantly when the thickness of the outer layers is reduced, in turn leading to an increase in the thickness of the middle layer. It should be noted, that although the topologies for $\left\lbrace \frac{3}{18} , \frac{12}{18} , \frac{3}{18} \right\rbrace$ and $\left\lbrace \frac{2}{18} , \frac{14}{18} , \frac{2}{18} \right\rbrace$ appear significantly different than for the others, they are merely the same topology as the others but shifted vertically half a coarse cell. However, the topology of the outer layers does change quite significantly as the thickness of the outer layers is reduced. It appears that the outer cells are becoming more dense and this is supported by looking at the local volume fraction of the outer coarse cells.
	\begin{figure}
	\centering
	\subfloat[Average compliance]{\includegraphics[clip, width=0.45\textwidth]{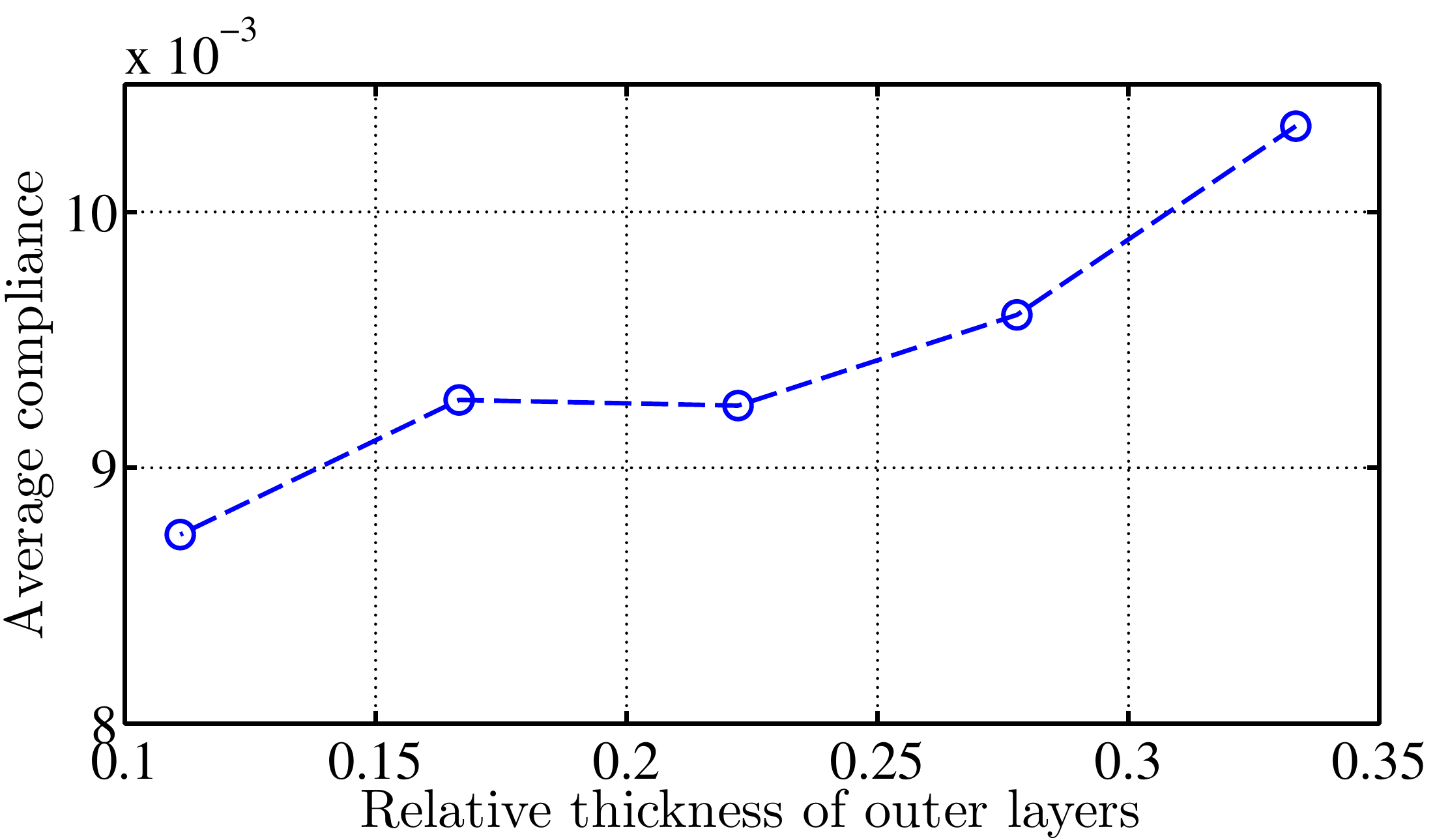}	\label{fig:doubleclamped_robust_cellconv-a}}
	\hspace*{0.04\textwidth}
	\subfloat[Local volume fraction]{\includegraphics[clip, width=0.45\textwidth]{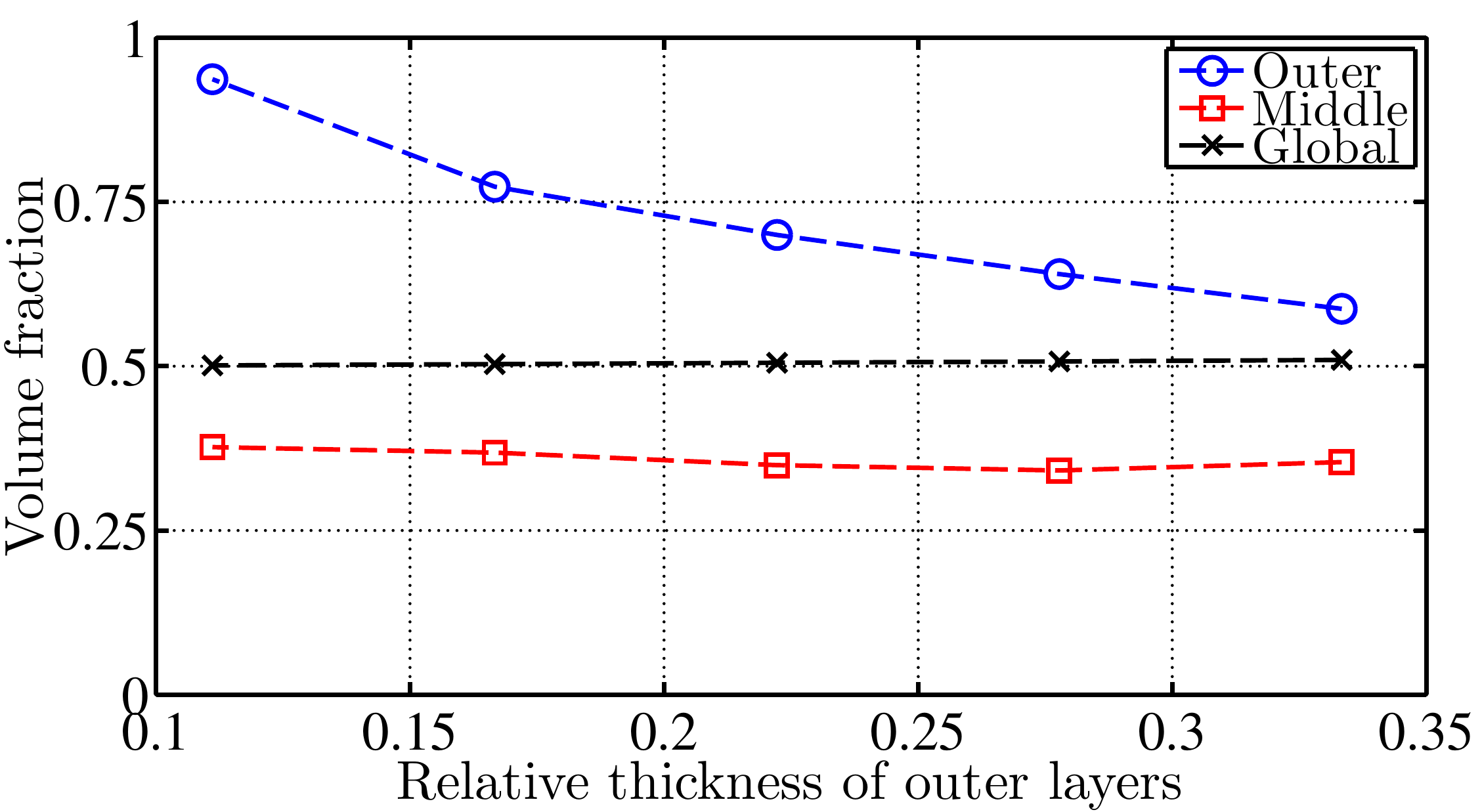}	\label{fig:doubleclamped_robust_cellconv-b}}	
	\caption{Average compliance and local volume fractions as a function of the relative thickness of the outer layers for the three-layered cantilever beam with a distributed load.} \label{fig:cantidist_robust_three_varconv}
	\end{figure}
Figure \ref{fig:doubleclamped_robust_cellconv-b} clearly shows that as the thickness of the outer layers is reduced, the local volume fraction of material is increased in the outer layers. On the contrary, the local volume fraction of material in the middle layer is more or less constant.
Figure \ref{fig:doubleclamped_robust_cellconv-a} shows the average compliance and it can be seen that as the thickness of the outer layers is reduced, the compliance decreases. Despite the compliance anomaly for $\left\lbrace \frac{3}{18} , \frac{12}{18} , \frac{3}{18} \right\rbrace$, likely due to a local minimum of the optimisation problem, it is concluded that it is beneficial to have thin outer layers with very high volume fractions and a thick middle layer with a relatively low volume fraction. This is perfectly in line with sandwich theory, where very thin and axially-stiff plates are joined to a low-density shear-stiff core in order to best resist the high axial stresses near the top and bottom and high shear stresses in the middle.

	\begin{figure}
	\centering
	\subfloat[$\left\lbrace \frac{4}{18} , \frac{10}{18} , \frac{4}{18} \right\rbrace$]{\includegraphics[clip, width=0.45\textwidth]{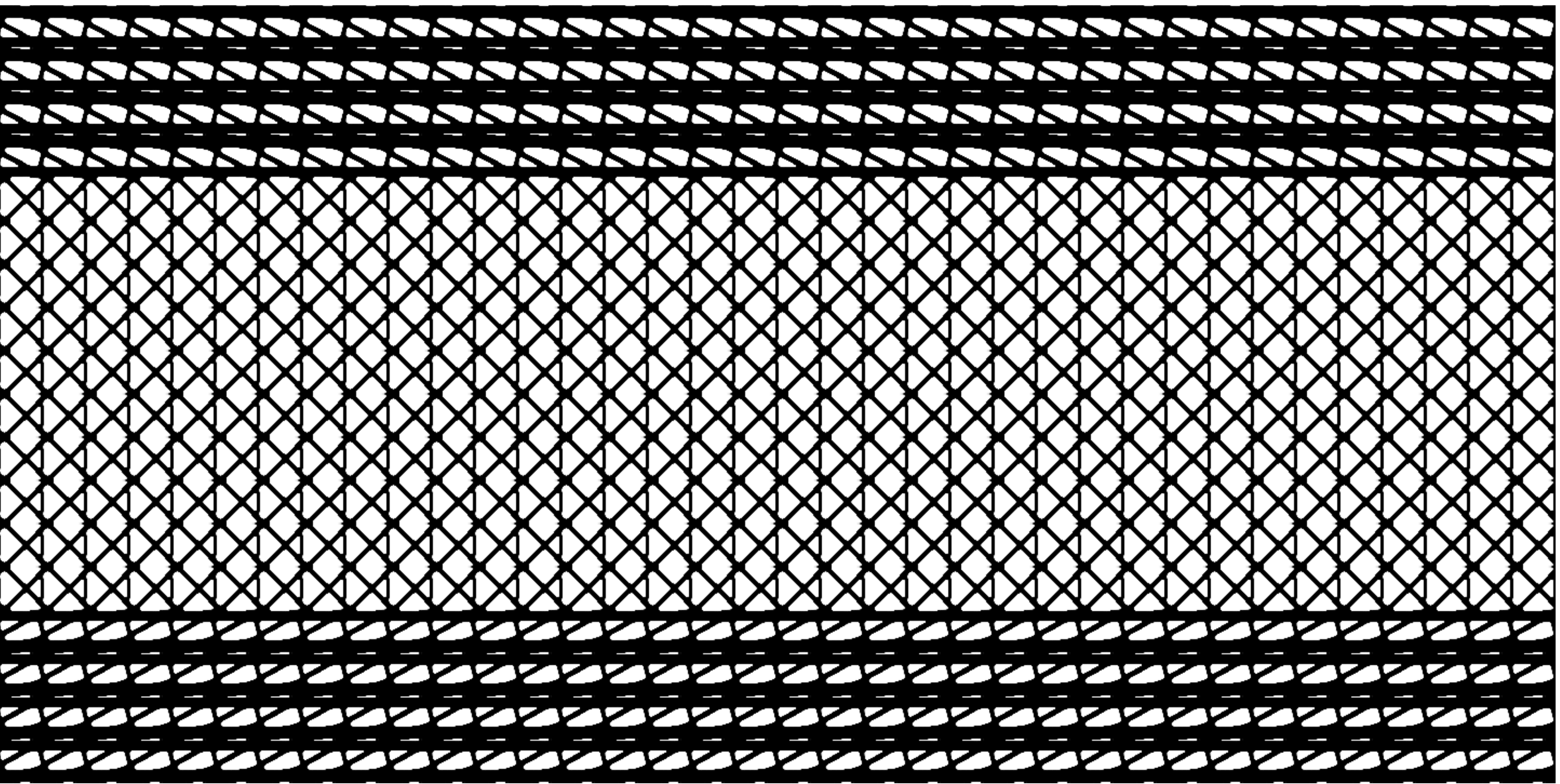}
	\label{fig:cantidist_robust_three_var_macro_a}}
	\hspace*{0.01\textwidth}
	\subfloat[$\left\lbrace \frac{2}{18} , \frac{14}{18} , \frac{2}{18} \right\rbrace$]{\includegraphics[clip, width=0.45\textwidth]{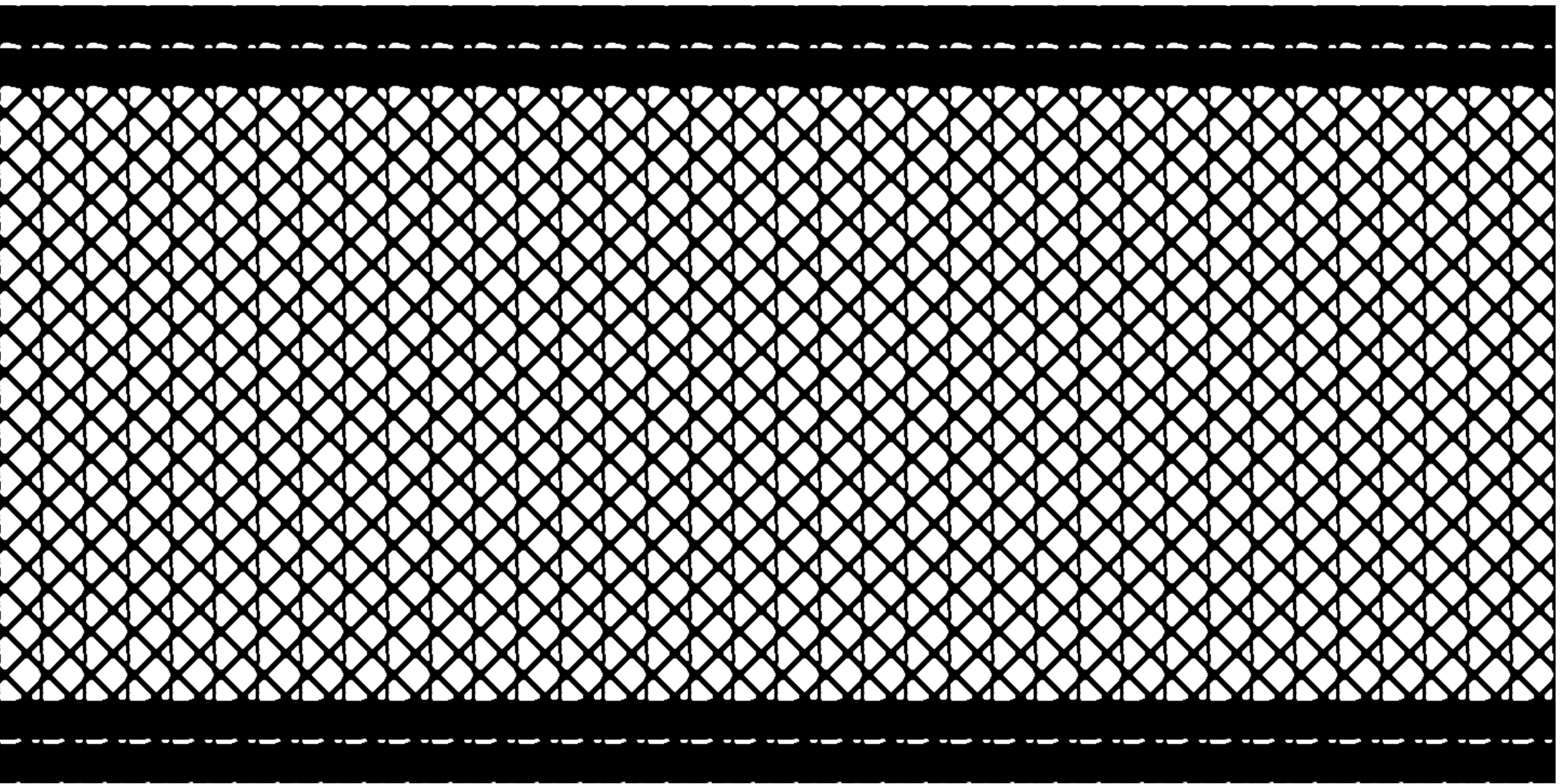}
	\label{fig:cantidist_robust_three_var_macro_b}}
	\caption{Optimised robust layered macrostructure for the cantilever beam with a distributed load for two selected layer thicknesses.} \label{fig:cantidist_robust_three_var_macro}
	\end{figure}
In order to fully appreciate the adaptivity of the approach to the varying outer layer thickness and the relation to sandwich theory, figure \ref{fig:cantidist_robust_three_var_macro} shows the full macrostructure for two select layer thicknesses: $\left\lbrace \frac{4}{18} , \frac{10}{18} , \frac{4}{18} \right\rbrace$ and  $\left\lbrace \frac{2}{18} , \frac{14}{18} , \frac{2}{18} \right\rbrace$. Here it can be clearly seen that the middle microstructure topology is essentially the same and that the outer layers become increasingly dense.

\subsection{Cantilever beam with concentrated load} \label{sec:results_canticonc}

In closing, in order to showcase the capabilities of the proposed methodology, a cantilever beam with a concentrated load at the lower right-hand corner is optimised where every layer of microstructure is different.
\begin{figure}[t]
	\centering
	\includegraphics[clip, width=0.44\textwidth]{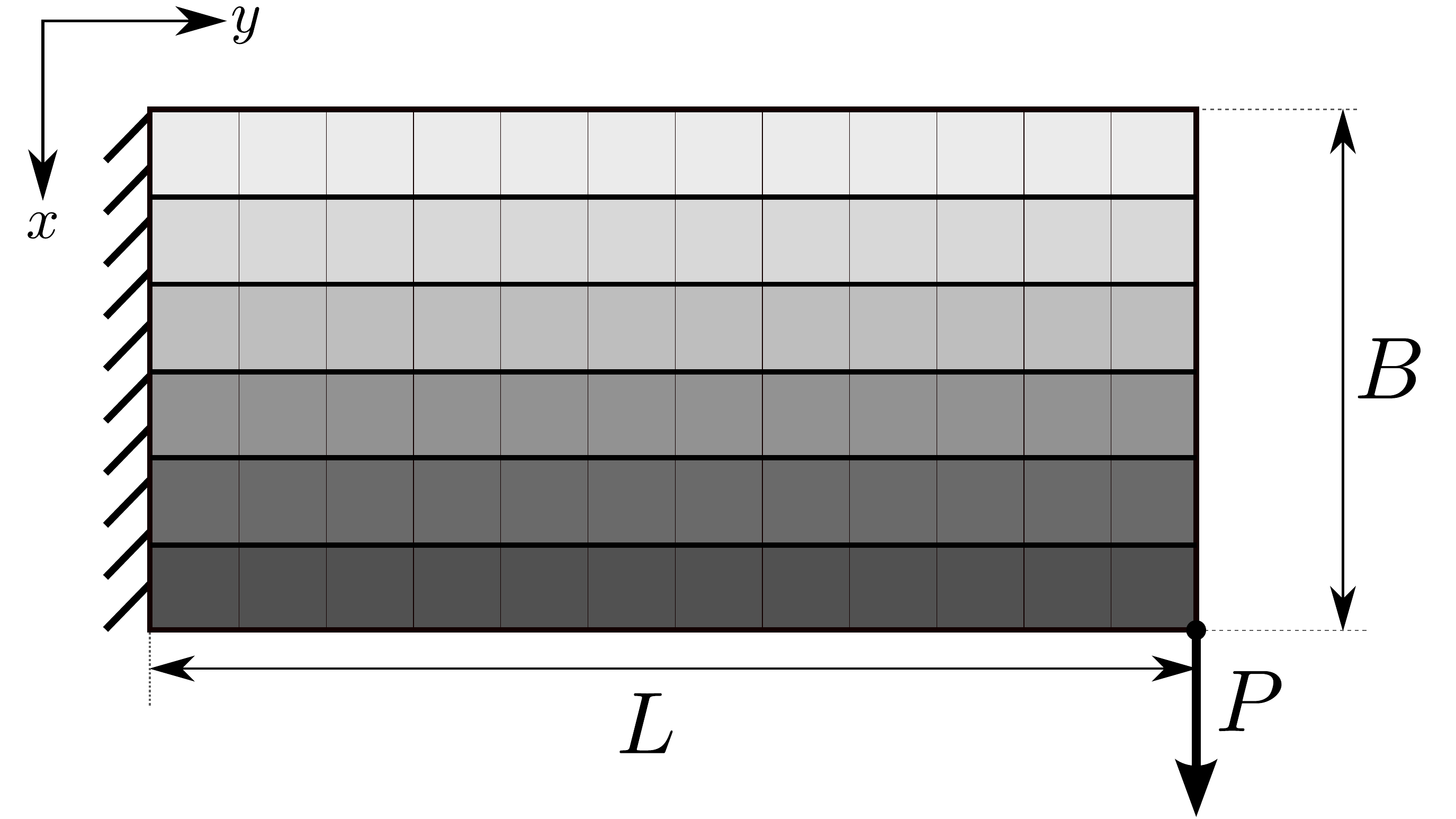}	
	\caption{Illustration and dimensions of a cantilever beam subjected to a concentrated force at the bottom rightmost corner. Every layer has a unique microstructure which is periodic in the horisontal direction only.} \label{fig:canticonc_illustr}
	\end{figure}
	\begin{figure}
	\centering
	\includegraphics[clip, width=0.8\textwidth]{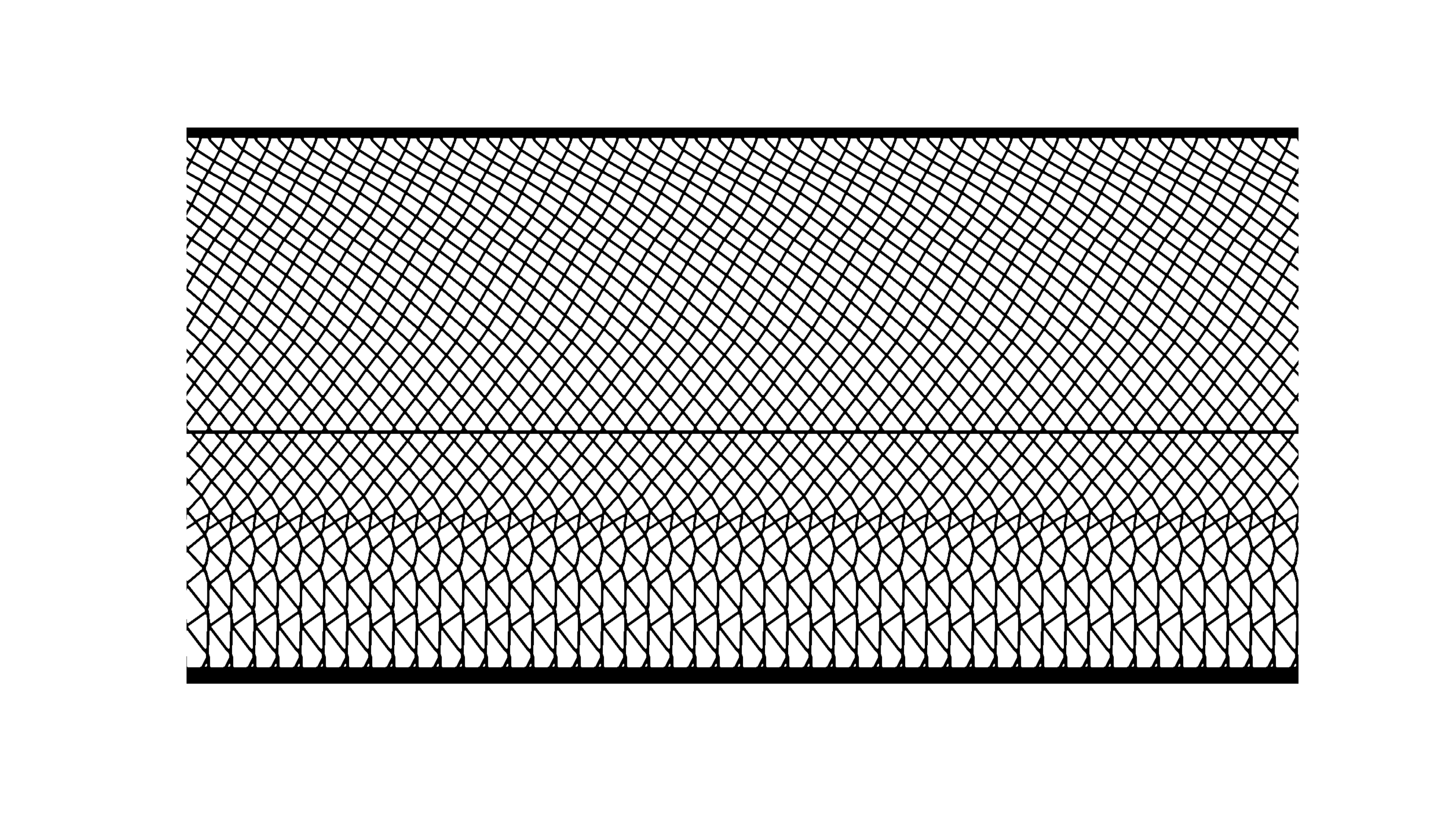}	
	\caption{Optimised robust layered macrostructure for the cantilever beam with a concentrated load for $M_{x} = 24$.} \label{fig:canticonc_robust_multilayer}
\end{figure}
Figure \ref{fig:canticonc_illustr} shows the problem layout, where it can be seen that every layer has a unique microstructure which is periodic in the horisontal direction only. A small change is made to the filtering procedure, described in section \ref{sec:topopt_multiblock}, which is that internal periodicity in the vertical direction is no longer required. That is, the constraints applied on the horisontal edges of the microstructures, see figure \ref{fig:filtering_threeblock}, are no longer imposed and the filtering domain essentially becomes a single vertical slice of the design domain with periodic boundary conditions in the horisontal direction. The following parameters are used: $r_{min} = 6h$, $v_{f} = 0.3$, $\beta_{0} = 8$, $\beta_{1} = 32$, $\eta \in \left\lbrace 0.3,0.5,0.7 \right\rbrace$, $\lambda_{\Omega} = 5\times10^{-4}$, $\varepsilon_{rel} = 10^{-6}$.

Figure \ref{fig:canticonc_robust_multilayer} shows the optimised robust macrostructure for the cantilever beam with a concentrated load and twenty-four layers, $M_{x} = 24$. It can be seen that the microstructural details vary continuously throughout the thickness. Many layers are similar with small variations, due to gradual change of the principal strain directions. The top and bottom layers exhibits bending stiffness, where most of the inner layers exhibit shear stiffness. It can be noticed that there is a horisontal bar of material in a layer just below the middle. This is likely due to convergence to local minima, due to the rather aggressive continuation approach used. It has been observed that these features are less likely to appear when robustness is not required, however, they do still appear at times and the problem increases with an increasing number of layers. The problem appears to be due to the propagation of the design information during the optimisation procedure, where the topologies of the local microstructural details are determined at very different speeds. Thus, the topologies of some layers are uniquely determined by the surrounding, already formed, layers.

\section{Discussion}

\subsection{Need for manufacturable microstructures}

The need for manufacturable microstructures of finite size is rather important. Current manufacturing processes, e.g. additive manufacturing, two photon polymerisation and photolithography, do not allow for the manufacture of microstructural details with length scales far below that of the specimen size. Furthermore, if the microstructural details are spatially-varying, it is crucial that connectivity is ensured throughout the manufactured structure in order to ensure the expected performance. It can be argued that going from a structure with disconnected, but optimal, microstructural details, as in \cite{Schury2012} in the context of FMO, to modified connected microstructural details ensures a manufacturable and well-performing structure. But when such modifications are made after the optimisation procedure itself, with no regard to the optimisation objective, the final structure may perform far from what was originally predicted. This may work to some extent for the forgiving compliance objective, but when moving to more advanced objectives or physics, the only rational approach is to include the requirement for connectivity in the original optimisation itself.

\subsubsection{Connectivity constraints}

It is important to note that small seemingly useless features can appear at the interfaces of the microstructural cells when dealing with multiple blocks. These small features are artefacts resulting from the very tough connectivity constraints imposed through the filtering procedure. In effect one imposes that all horisontal interfaces between any two given coarse cells should be the same. Generally it is observed that the microstructures that develop a clear topology first, that is the ones with the largest sensitivities, have a dominant influence on the interfaces. This is as expected, but the filtering procedure ensures connected microstructures that have as smoothly connected interfaces as possible in order to ensure manufacturability. When the design is further relaxed into multiple layers with unique microstructures, these artefacts do not occur because internal periodicity is not required in the vertical direction.

\subsection{Microstructure restriction}

In this article, the microstructural details have been forced to be periodic or in layers. The ideal situation is, of course, the ability of the microstructural details to change in all directions. Forcing the design to have microstructural details everywhere, essentially imposes a maximum length scale on the resulting members. Providing the possibility of unrestricted microstructural details is trivial for the presented approach, as the optimisation problem reverts back to the original unrestricted topology optimisation problems. In this case, all length scales above the minimum length scale imposed by the robust approach are allowed in the design space and thus a significantly better performing structure is to be expected than when restricting the design space using forced microstructural details. If one for instance wants microstructural details everywhere, one can imposed a maximum length scale on the design using local lower-bound constraints on the void volume fraction, similar to the approach presented in \cite{Guest2009}.

\subsection{Spectral MsFEM as a solver}

Most multiscale finite element methods have been developed for use as an approximate solver, rather than as a preconditioner. It is also possible to use the presented spectral MsFEM as a solver for high-contrast linear elasticity problems. As the error convergence studies presented in section \ref{sec:msfem_numexp} suggest, a lot of eigenmodes are needed to approximate the true solution well. Using only a few modes yields results similar to homogenisation and thus designs can end up being massively disconnected, as discussed in the original work \cite{Lazarov2014}. By increasing the number of modes slightly, one can increase the sensitivity of the structural response to disconnected members. However, for microstructural design problems, these low-dimension bases require eigenfunction derivatives for the optimisation to proceed correctly, which is computationally expensive. Hence, using the presented spectral MsFEM as a solver is left as a subject for future research.

\subsection{Future work}

For the presented problems, the addition of local constraints on the minimum void volume fraction in each unique microstructure cell is a cheap way to ensure a maximum length scale, similar to \cite{Guest2009}. Of course, the number of local constraints increases with the amount of unique microstructures, that is the number of layers. This approach has been tested successfully, however, it does not show significant differences for the presented problems with relatively low volume fractions. The extension to other local constraints will be investigated in future work. 

A full comparison of the presented approach to homogenisation for varying microstructural details is part of current research, which will include investigations on the possibility of using the presented methodology to find the optimal design for cores of sandwich beam, plate and shell structures. The extension of the MsFEM-GMRES solver to three-dimensional general topology optimisation problems in a large scale parallel framework \cite{Aage2013,Aage2014} is currently being pursued. Here the computational performance of the method is expected to excel due to the inherent parallisation properties of the decoupled local problems.

\section{Conclusion}

The presented approach treats the optimisation of microstructural details with respect to the macroscopic response, while resolving all microstructural details as compared to the homogenisation approach. This results in an optimisation that takes localised behaviour, such as boundary conditions and forces, into account and ensures microstructures tailored for a specific application rather than specific properties. The approach is fully flexible and can treat microstructural details with and without clear separation of length scale compared to macrostructure. Finally, the approach yields robust and manufacturable solutions, with smoothly connected and continuously varying microstructural details in one direction.
A further advantage of the proposed methodology, is that the application of the MsFEM-GMRES solver allows for huge problems to be solved in \textsc{Matlab} and that it exhibits contrast-independent behaviour ensuring fast computations for topology optimisation problems.

\section{Acknowledgements} \label{sec:acknow}
The authors would like to thank Erik Andreassen for providing \textsc{Matlab} subroutines for homogenisation and the TopOpt group for useful comments and discussions during the preparation of the work.

Both authors were funded by Villum Fonden through the NextTop project, as well as the EU FP7-MC-IAPP programme LaScISO.

\bibliographystyle{elsarticle-num}
\bibliography{bib}

\end{document}